\documentclass[a4paper,11pt]{article}

\usepackage{jheppub}

\usepackage{hyperref}
\usepackage{graphicx}
\usepackage{amsmath}
\usepackage{amssymb}
\usepackage{xspace}
\usepackage{mathrsfs}
\usepackage{subfigure}
\usepackage{slashed}
\usepackage[normalem]{ulem}
\usepackage{xcolor} 
\usepackage{epstopdf}
\usepackage{soul}

\DeclareRobustCommand{\eq}[1]{eq.~\eqref{eq:#1}}
\DeclareRobustCommand{\eqs}[2]{eqs.~\eqref{eq:#1} and \eqref{eq:#2}}
\DeclareRobustCommand{\fig}[1]{fig.~\ref{fig:#1}}

\DeclareRobustCommand{\sec}[1]{sec.~\ref{sec:#1}}
\DeclareRobustCommand{\secs}[2]{secs.~\ref{sec:#1} and \ref{sec:#2}}
\DeclareRobustCommand{\subsec}[1]{sec.~\ref{subsec:#1}}
\DeclareRobustCommand{\subsecs}[2]{secs.~\ref{subsec:#1} and \ref{subsec:#2}}
\DeclareRobustCommand{\app}[1]{app.~\ref{app:#1}}
\DeclareRobustCommand{\refcite}[1]{ref.~\cite{#1}}
\DeclareRobustCommand{\refcites}[1]{refs.~\cite{#1}}

\usepackage{marginnote}

\newcommand{\abs}[1]{\lvert#1\rvert}

\newcommand{\ord}[1]{\mathcal{O}(#1)}

\newcommand{\bra}[1]{\langle#1\rvert}
\newcommand{\ket}[1]{\lvert#1\rangle}

\newcommand{\df}{\mathrm{d}}

\newcommand{\sdt}{\!\cdot\!}

\newcommand{\eps}{\epsilon}

\newcommand{\cI}{{\mathcal I}}

\newcommand{\nn}{\nonumber}

\newcommand{\lqcd}{\Lambda_\mathrm{QCD}}

\newcommand{\Ecm}{E_\mathrm{cm}}

\newcommand{\cusp}{\mathrm{cusp}}

\newcommand{\SCETa}{\ensuremath{{\rm SCET}_{\rm I}}\xspace}
\newcommand{\SCETb}{\ensuremath{{\rm SCET}_{\rm II}}\xspace}
\newcommand{\SCETp}{\ensuremath{{\rm SCET}_+}\xspace}

\newcommand{\e}[0]{\epsilon}

\usepackage{suffix}
\usepackage{mathtools}
\DeclarePairedDelimiterX\MeijerM[3]{\lparen}{\rparen}%
{\begin{smallmatrix}#1 \\ #2\end{smallmatrix}\delimsize\vert\,#3}

\newcommand\MeijerG[8][]{%
  G^{\,#2,#3}_{#4,#5}\MeijerM[#1]{#6}{#7}{#8}}

\WithSuffix\newcommand\MeijerG*[7]{%
  G^{\,#1,#2}_{#3,#4}\MeijerM*{#5}{#6}{#7}}

\def\Li{\textrm{Li}}
\def\nn{\nonumber}

\def\df{\textrm{d}}

\def\MS{\overline{\rm MS}}

\newcommand{\Tau}{{\mathcal T}}

\allowdisplaybreaks[2]

\preprint{\begin{flushright}
DESY  16-199 \\
UWTHPH 2017-3 \\
HU-EP-16/42
\end{flushright}}

\title{Factorization and Resummation for Massive Quark Effects in Exclusive Drell-Yan}

\author[a]{Piotr Pietrulewicz,}
\author[b]{Daniel Samitz,}
\author[c]{Anne Spiering,}
\author[a]{and Frank J.~Tackmann}

\affiliation[a]{Theory Group, Deutsches Elektronen-Synchrotron (DESY), D-22607 Hamburg, Germany}
\affiliation[b]{University of Vienna, Faculty of Physics, Boltzmanngasse 5, A-1090 Wien, Austria}
\affiliation[c]{Institut f\"{u}r Physik, Humboldt-Universit\"{a}t zu Berlin, Zum Gro\ss en Windkanal 6, D-12489 Berlin, Germany}

\emailAdd{piotr.pietrulewicz@desy.de}
\emailAdd{daniel.samitz@univie.ac.at}
\emailAdd{spiering@physik.hu-berlin.de}
\emailAdd{frank.tackmann@desy.de}

\abstract{
Exclusive differential spectra in color-singlet processes at hadron colliders are
benchmark observables that have been studied to high precision in theory and experiment.
We present an effective-theory framework utilizing soft-collinear effective theory
to incorporate massive (bottom) quark effects into resummed differential distributions,
accounting for both heavy-quark initiated primary contributions to the hard scattering process
as well as secondary effects from gluons splitting into heavy-quark pairs.
To be specific, we focus on the Drell-Yan process and consider the vector-boson transverse momentum,
$q_T$, and beam thrust, $\Tau$, as examples of exclusive observables.
The theoretical description depends on the hierarchy between the hard, mass, and
the $q_T$ (or $\Tau$) scales,
ranging from the decoupling limit $q_T \ll m$ to the massless limit $m \ll q_T$.
The phenomenologically relevant intermediate regime $m \sim q_T$ requires in particular
quark-mass dependent beam and soft functions.
We calculate all ingredients for the description of primary and secondary mass effects
required at NNLL$'$ resummation order (combining NNLL evolution with NNLO boundary conditions)
for $q_T$ and $\Tau$ in all relevant hierarchies. For the $q_T$ distribution
the rapidity divergences are different from the massless case and we discuss features of the
resulting rapidity evolution.
Our results will allow for a detailed investigation of quark-mass effects in the ratio of $W$
and $Z$ boson spectra at small $q_T$, which is important for the precision measurement
of the $W$-boson mass at the LHC.
}

\begin{document}
\maketitle

\section{Introduction}
\label{sec:Intro}

Differential cross sections for the production of color-singlet states (e.g.~electroweak vector bosons or the Higgs boson) represent benchmark observables at the LHC. For the Drell-Yan process, the measurements of the transverse momentum ($q_T$) spectrum of the vector boson (and related variables) have reached uncertainties below the percent level~\cite{Aad:2011fp, Chatrchyan:2011wt, Aad:2014xaa, Khachatryan:2015oaa, Aad:2015auj, Khachatryan:2016nbe}, allowing for stringent tests of theoretical predictions from both analytic resummed calculations and parton-shower Monte-Carlo programs.
An accurate description of the $q_T$ spectrum is also a key ingredient for a precise measurement of the $W$-boson mass at the LHC, which requires a thorough understanding of the $W$-boson and $Z$-boson spectra and in particular their ratio~\cite{Buge:2006dv, Besson:2008zs, CMS:2016nnd, Atlas:2014}. The associated uncertainties are one of the dominant theoretical uncertainties in the recent $m_W$ determination by the ATLAS collaboration~\cite{Aaboud:2017svj}.

So far, mass effects from charm and bottom quarks in the initial state have been discussed extensively for inclusive heavy-quark induced cross sections, leading to the development of several variable-flavor number schemes in deep inelastic scattering and $pp$ collisions (see e.g.\ refs.~\cite{Aivazis:1993kh, Aivazis:1993pi, Thorne:1997ga, Kretzer:1998ju, Collins:1998rz, Forte:2010ta, Bonvini:2015pxa}).
On the other hand, analogous heavy-quark mass effects from initial-state radiation have received little attention so far in the context of resummed exclusive (differential) cross sections,
i.e.\ where the measurement of an additional (differential) observable restricts the QCD radiation into the soft-collinear regime requiring the resummation of the associated logarithms. While e.g.~for $m \ll q_T$ the mass effects in the resummed $q_T$ distribution are simply encoded by the matching between the parton distribution functions across a flavor threshold (e.g. matching four-flavor PDFs onto five-flavor PDFs including a $b$-quark PDF at the scale $m_b$, which happens much below the scale $q_T$), this description breaks down for $q_T \sim m$ or $q_T \ll m$.
A comprehensive treatment of these regimes in resummed predictions has been missing so far. This concerns in particular also parton-shower Monte-Carlo generators, which include massive quark effects primarily as kinematic effects and by using massive splitting functions. Since heavy-quark initiated corrections are one of the main differences between the $W$ and $Z$ boson spectra, this issue can play therefore an important role for $m_W$ measurements at the LHC.

\begin{figure}[t]
\centering
\hfill%
\subfigure[]{\includegraphics[scale=0.3]{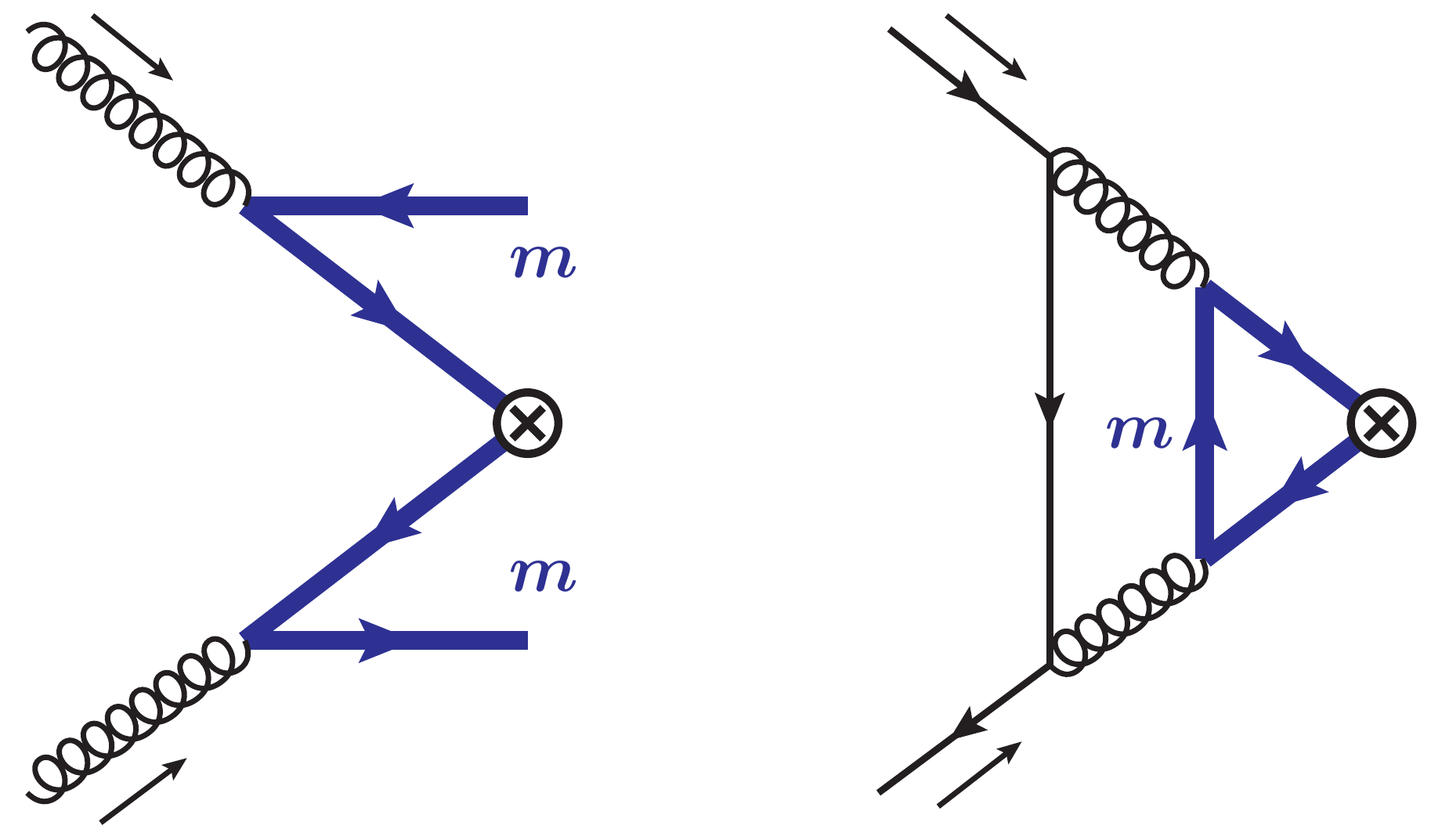}\label{fig:Primary}}%
\hfill\hfill%
\subfigure[]{\includegraphics[scale=0.3]{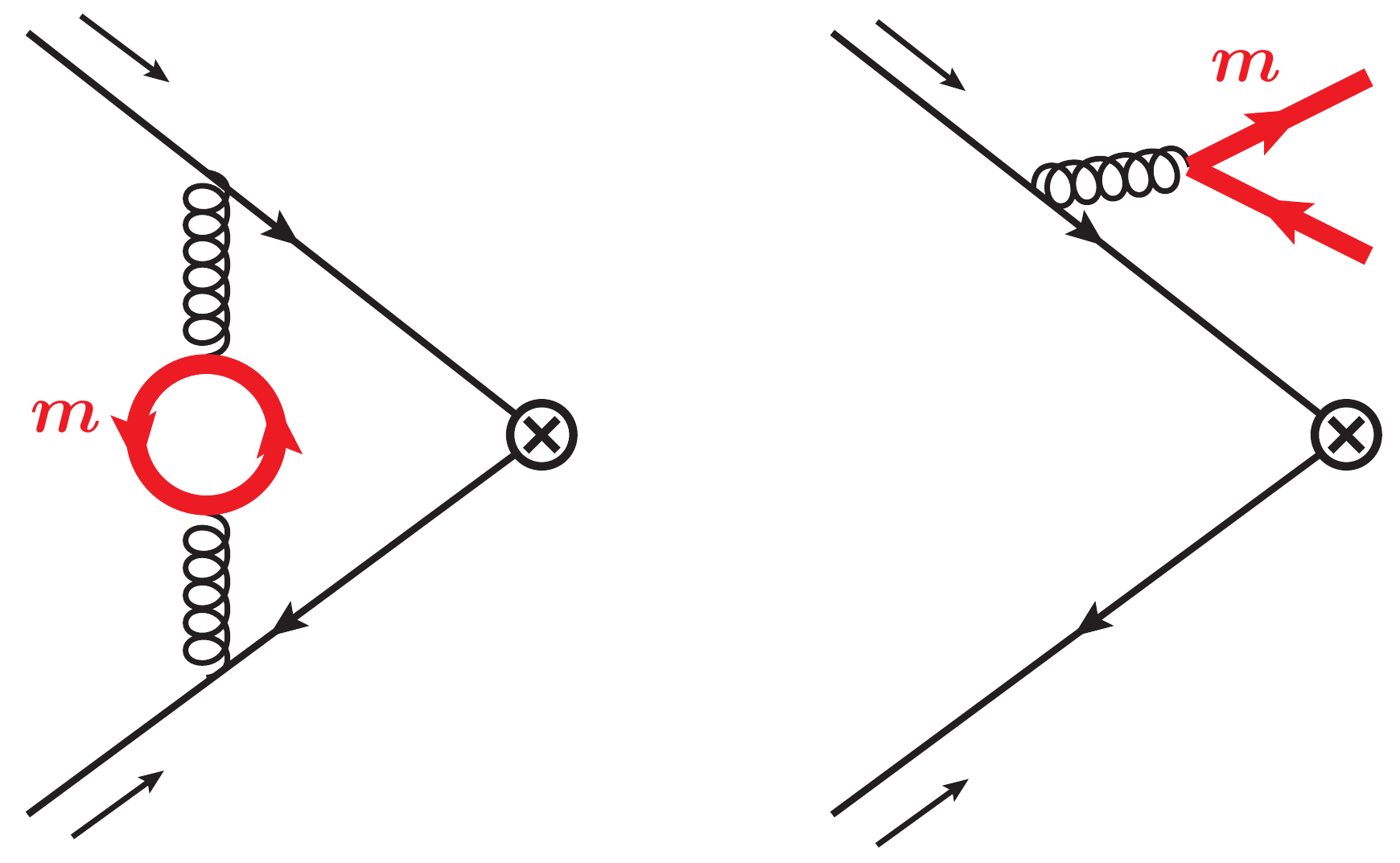}\label{fig:Secondary}}%
\hspace*{\fill}%
\caption{Primary (a) and secondary (b) heavy-quark mass effects for $Z$-boson production.}
\label{fig:PrimarySecondary}
\end{figure}

In general, one can distinguish two types of mass effects as illustrated in \fig{PrimarySecondary}, which have different characteristics: Contributions where the heavy-quark enters the hard interaction process are called {\it primary} mass effects. Contributions from a gluon splitting into a massive quark-antiquark pair with light quarks entering the hard interaction are called {\it secondary}.
For the $q_T$ spectrum, earlier treatments of the heavy-quark initiated primary contributions for $m \lesssim q_T$ have been given in refs.~\cite{Nadolsky:2002jr, Berge:2005rv, Belyaev:2005bs}, essentially combining the ACOT scheme with the standard CSS $q_T$ resummation. A complete setup also requires to account for secondary mass effects. Their systematic description for differential spectra in the various relevant hierarchies between mass and other physical scales has been established in the context of event shapes in $e^+ e^-$ collisions~\cite{Gritschacher:2013pha, Pietrulewicz:2014qza} and for threshold resummation in DIS~\cite{Hoang:2015iva}, see also \refcites{Butenschoen:2016lpz, Dehnadi:2016snl} for a recent utilization in the context of boosted heavy quark initiated jets. The application to differential spectra in $pp$ collisions will be part of the present paper.

We present a systematic effective-theory treatment of quark mass effects including both types of mass effects and all possible scale hierarchies using soft-collinear effective theory (SCET)~\cite{Bauer:2000ew, Bauer:2000yr, Bauer:2001ct, Bauer:2001yt}. We focus on the Drell-Yan process, $pp\to Z/\gamma^*\to \ell^+\ell^-$, and consider two types of observables that resolve additional QCD radiation and are used to constrain the process to the exclusive region, namely the transverse momentum $q_T$ of the gauge boson and beam thrust~\cite{Stewart:2009yx},
\begin{align}
q_T = |\vec{q}_T| = \abs{\vec{p}_{T\ell} + \vec{p}_{T\bar\ell}} = \Bigl\lvert \sum_i \vec{p}_{Ti} \Bigr\rvert
\,, \qquad
\Tau = \sum_i\min \{n_a \sdt p_i, n_b \sdt p_i\}
\,.\end{align}
Here, $p_i$ are all hadronic final-state momenta (i.e.\ excluding the color-singlet final state), and $n_{a,b}^\mu = (1, \pm \hat z)$ are lightlike vectors along the beam axes. Due to transverse momentum conservation $q_T$ measures the total transverse momentum of the final state hadronic radiation, while beam thrust measures the momentum projections of all hadronic particles onto the beam axis. The exclusive regime we are interested in corresponds to
$q_T \ll Q$ or $\Tau \ll Q$, where $Q = \sqrt{q^2}$ is the dilepton invariant mass.
These two observables restrict the allowed QCD radiation into the collinear and soft regime in different ways, leading to different effective-theory setups with distinct factorization and resummation properties, which are well-known in the massless limit up to high orders in the logarithmic counting (see e.g.~refs.~\cite{Collins:1981uk, Collins:1981va, Collins:1984kg, Balazs:1997xd, Balazs:2000wv, Bozzi:2010xn, deFlorian:2011xf, Becher:2010tm, Chiu:2011qc, Chiu:2012ir, GarciaEchevarria:2011rb} and \refcites{Stewart:2009yx, Stewart:2010pd, Berger:2010xi}). These two cases provide simple prototypical examples, which cover the essential features of the factorization with massive quarks that will also be relevant for including massive quark effects for other more complicated jet resolution variables whose factorization is known in the massless limit.
Throughout the paper we always consider the limit $\lqcd \ll q_T,\Tau$ allowing for a perturbative description of the physics at these kinematic scales. We then consider all relevant relative hierarchies between the heavy-quark mass $m$ and the kinematic scales set by the measurement of $q_T$ or $\Tau$, respectively.

In the second part of the paper, we explicitly compute all required ingredients for incorporating $m_b$ effects at NNLL$'$ order, which combines NNLL evolution with the full NNLO singular boundary conditions (hard, beam, and soft functions).
For $Z$-boson production at NNLL$'$, primary effects contribute via $\mathcal{O}(\alpha_s) \times \mathcal{O}(\alpha_s)$ heavy-quark initiated contributions, illustrated in \fig{Primary}. Secondary effects contribute as $\mathcal{O}(\alpha_s^2)$ corrections to light-quark initiated hard interactions, illustrated in \fig{Secondary}. Due to the strong CKM suppression primary $m_b$-effects do not play any significant role for $W$-production, which represents a key difference to $Z$-boson production. Primary $m_c$-effects enter $W$-production in the (sizeable) $cs$-channel, where they start already at NLL$'$ via $\mathcal{O}(\alpha_s)\times\ord{1}$ corrections. For this case, our explicit results for the regime $q_T \sim m_c$ allows for up to NNLL resummation. (Here, the resummation at NNLL$'$ would require the $\ord{\alpha_s^2}$ primary massive contributions.)

The paper is organized as follows: We first discuss in detail the effective field theory setup for the different parametric regimes for the case of $q_T$ in \sec{qT} and for $\Tau$ in \sec{Tau}. Here, we elaborate on the relevant mode setup in SCET, the resulting factorization formulae, and all-order relations between the factorization ingredients in the different regimes.
In \sec{results}, we give the $\mathcal{O}(\alpha_s)$ and $\mathcal{O}(\alpha_s^2)$ results for the various ingredients for NNLL$'$ resummation. We also verify the consistency of our results with the associated results in the massless limit.
Further details on all calculations are given in the appendices, where we also give the analytic results at fixed-order for the massive quark effects in the $q_T$ and $\Tau$ distributions in the singular limit $q_T, \Tau \ll Q$.
In \sec{nu_evolution}, we discuss the consequences of the secondary mass effects on the rapidity evolution, in particular for the $q_T$ distribution in the regime $q_T \sim m_b$.
As an outlook we provide in \sec{outlook} an estimate of the potential size of the bottom quark effects for low-$q_T$ Drell-Yan measurements. In \sec{conclusions} we conclude.

\section{\boldmath Factorization of quark mass effects for the $q_T$ spectrum}
\label{sec:qT}

\subsection{Factorization for massless quarks}
\label{subsec:fact_massless}

Before discussing the massive quark corrections, we first briefly summarize the EFT setup and factorization for massless quarks. The relevant modes for the measurement of $q_T$ in the limit $q_T \ll Q$ are $n_a$-collinear, $n_b$-collinear, and soft modes with the scaling
\begin{align} \label{eq:modes_massless}
n_a \text{-collinear:} 
&\quad
p_{n_a}^\mu \sim \Bigl(\frac{q_T^2}{Q},Q,q_T\Bigr)
\,, \nn \\
n_b \text{-collinear:} 
&\quad
p_{n_b}^\mu\sim \Bigl(Q,\frac{q_T^2}{Q},q_T\Bigr) 
\, , \nn\\
\text{soft:} 
&\quad
p_{s}^\mu\sim (q_T,q_T,q_T)
\,, 
\end{align}
which we have written in terms of light-cone coordinates along the beam axis,
\begin{align} \label{eq:lc}
  p^\mu = n_a \sdt p\,\frac{n_b^\mu}{2} + n_b \sdt p\,\frac{n_a^\mu}{2} + p_{\perp}^\mu \equiv  (n_a \sdt p,n_b \sdt p,p_{\perp}) \equiv  (p^+,p^-,p_\perp)
\,,\end{align}
with $\bar{n}_a \equiv n_b$. Besides these perturbative modes there are also nonperturbative collinear modes with the scaling $(\lqcd^2/Q,Q,\lqcd)$ and $(Q,\lqcd^2/Q,\lqcd)$, which describe the initial-state protons at the scale $\mu \sim \lqcd$, and which are unrelated to the specific jet resolution measurement. The typical invariant mass of the soft modes is parametrically the same as for the collinear modes, $p_{n_a}^2 \sim p_{n_b}^2 \sim p_s^2 \sim q_T^2$, which is the characteristic feature of a \SCETb theory. The soft and collinear modes are only separated in rapidity leading to the emergence of rapidity divergences and associated rapidity logarithms. The traditional approach for their resummation in QCD relies on the work by Collins, Soper, and Sterman~\cite{Collins:1981uk, Collins:1981va, Collins:1984kg}. In SCET the factorization and resummation were devised in refs.~\cite{Becher:2010tm, Chiu:2011qc, Chiu:2012ir, GarciaEchevarria:2011rb}.

Here we will use the rapidity renormalization approach of refs.~\cite{Chiu:2011qc, Chiu:2012ir}, where the rapidity divergences are regularized by a symmetric regulator and are renormalized by appropriate counterterms (by a $\MS$-type subtraction). The rapidity logarithms are then resummed by solving the associated rapidity renormalization group equations. Within this framework the factorized differential cross section with $n_f$ massless quarks reads%
\footnote{In principle there is also a corresponding contribution for a gluon initiated hard interaction. However, taking into account the decay of the electroweak boson into massless leptons this correction vanishes for onshell gluons and only contributes to the power suppressed terms of $\mathcal{O}(q_T/Q)$.}
\begin{align}\label{eq:factqT_ml}
\frac{\df \sigma}{\df q^2_T \,\df Q^2 \,\df Y}
&= \sum_{i,j \in \{q,\bar{q}\}}  H^{(n_f)}_{ij}(Q,\mu) \int\! \df^2 p_{Ta}\,\df^2 p_{Tb}\, \df^2 p_{Ts} \,
\delta(q_T^2-|\vec{p}_{Ta}+\vec{p}_{Tb}+\vec{p}_{Ts}|^2) \\
& \quad \times B^{(n_f)}_i\Bigl(\vec{p}_{Ta},x_a,\mu,\frac{\nu}{\omega_a}\Bigr)\, B^{(n_f)}_{j}\Bigl(\vec{p}_{Tb},x_b,\mu,\frac{\nu}{\omega_b}\Bigr) \, S^{(n_f)}(\vec{p}_{Ts},\mu,\nu) \, \Bigl[1+\mathcal{O}\Bigl(\frac{q_T}{Q}\Bigr)\Bigr] \nn
\, ,\end{align}
where
\begin{align}\label{eq:labelmomenta}
\omega_a = Q e^Y \, , \quad \omega_b = Q e^{-Y} \, , \quad x_{a,b} = \frac{\omega_{a,b}}{\Ecm}
\,,\end{align}
with $Y$ denoting the rapidity of the color-singlet state.

In \eq{factqT_ml}, the superscript $(n_f)$ on all functions indicates that the associated EFT operators and the strong coupling constant in these functions are renormalized with $n_f$ active quark flavors, which matters for the evolution already at LL. $H_{ij}$ denotes the process-dependent (but measurement-independent) hard function. It encodes the tree-level result and hard virtual corrections of the partonic process $ i j \to Z/W/\gamma^*$ at the scale $\mu \sim Q$. Following refs.~\cite{Collins:1981uw, Fleming:2006cd, Stewart:2009yx}, the renormalized transverse-momentum dependent (TMD) beam functions $B_{i}$, which are essentially equivalent to TMDPDFs, can be matched onto PDFs as
\begin{align}\label{eq:beamqT_ml}
B^{(n_f)}_i\Bigl(\vec{p}_{T},x,\mu,\frac{\nu}{\omega}\Bigr)
&= \sum_k \int_{x}^{1} \frac{\df z}{z} \, \mathcal{I}^{(n_f)}_{ik}\Bigl(\vec{p}_{T},z,\mu,\frac{\nu}{\omega}\Bigr) \,
f^{(n_f)}_k\Bigl(\frac{x}{z},\mu\Bigr) \biggl[1+\mathcal{O}\biggl(\frac{\lqcd^2}{|\vec{p}_T|^2}\biggr)\biggr]
\nn \\
&\equiv \sum_k \mathcal{I}^{(n_f)}_{ik}(\vec{p}_{T},x,\mu,\frac{\nu}{\omega}) \otimes_x f^{(n_f)}_k(x,\mu)
\, ,\end{align}
where the perturbative matching coefficients $\mathcal{I}_{ik}$ describe the collinear initial-state radiation at the invariant mass scale $\mu \sim q_T$ and rapidity scale $\nu \sim Q$, and the nonperturbative parton distribution functions (PDFs) are denoted by $f_k$.
In the following, we abbreviate the Mellin-type convolution in $x$ as in the second line above.
Finally, the soft function $S$ describes the wide-angle soft radiation at the invariant mass and rapidity scale $\mu \sim \nu \sim q_T$. The matching coefficients $\mathcal{I}_{ik}$ and the soft function are process-independent and have been computed to $\mathcal{O}(\alpha_s^2)$ in refs.~\cite{Catani:2012qa, Gehrmann:2012ze, Gehrmann:2014yya, Luebbert:2016itl} allowing for a full NNLL$'$ analysis of Drell-Yan for massless quarks. The three-loop noncusp rapidity anomalous dimension required for the resummation at N$^3$LL has recently become available~\cite{Li:2016axz,Li:2016ctv, Vladimirov:2016dll}.

In \eq{factqT_ml}, the logarithms of $q_T/Q$ are resummed by evaluating all functions at their characteristic renormalization scales and evolving them to common final scales $\mu$ and $\nu$ by solving the set of coupled evolution equations
\begin{align}\label{eq:RGE_ml}
\mu \frac{\df}{\df \mu} H^{(n_f)}_{ij}(Q,\mu)
&= \gamma_H^{(n_f)}(Q,\mu)\, H^{(n_f)}_{ij}(Q,\mu)
\, , \nn \\
\mu \frac{\df}{\df \mu} B^{(n_f)}_i\Bigl(\vec{p}_{T},x,\mu,\frac{\nu}{\omega}\Bigr)
& = \gamma_{B}^{(n_f)}\Bigl(\mu,\frac{\nu}{\omega}\Bigr) \, B^{(n_f)}_i\Bigl(\vec{p}_{T},x,\mu,\frac{\nu}{\omega}\Bigr)
\, ,\nn \\
\mu \frac{\df}{\df \mu} S^{(n_f)}(\vec{p}_{T},\mu,\nu)
& =  \gamma_{S}^{(n_f)}(\mu,\nu) \, S^{(n_f)}(\vec{p}_T,\mu,\nu)
\, ,\nn \\
\mu \frac{\df}{\df \mu} f_i^{(n_f)}(x,\mu)
& =  \sum_k \gamma_{f,ik}^{(n_f)}(x, \mu) \otimes_x f_k^{(n_f)}(x,\mu)
\, ,\nn \\
\nu \frac{\df}{\df \nu} B^{(n_f)}_i\Bigl(\vec{p}_{T},x,\mu,\frac{\nu}{\omega}\Bigr)
&= \int\! \df^2 k_T \, \gamma_{\nu,B}^{(n_f)}(\vec{p}_T-\vec{k}_T,\mu) \, B^{(n_f)}_i\Bigl(\vec{k}_T,x,\mu,\frac{\nu}{\omega}\Bigr)
\, ,\nn \\
\nu \frac{\df}{\df \nu} S^{(n_f)}(\vec{p}_{T},\mu,\nu)
& = \int \!\df^2 k_T \, \gamma_{\nu, S}^{(n_f)}(\vec{p}_T-\vec{k}_T,\mu) \, S^{(n_f)}(\vec{k}_T,\mu,\nu)
\, .\end{align}
Only the evolution of the PDF leads to flavor mixing. Consistency of RG running implies that
\begin{align}\label{eq:consistency_qT}
\gamma_H^{(n_f)}(Q,\mu) + \gamma_B^{(n_f)}\Bigl(\mu,\frac{\nu}{\omega_a}\Bigr)+ \gamma_B^{(n_f)}\Bigl(\mu,\frac{\nu}{\omega_b}\Bigr) + \gamma_S^{(n_f)}(\mu,\nu) &= 0
\, , \nn \\
2\gamma_{\nu,B}^{(n_f)}(\vec{p}_T,\mu) + \gamma_{\nu,S}^{(n_f)}(\vec{p}_T,\mu) &= 0
\,, \nn \\
\mu \frac{\df}{\df\mu} \gamma_{\nu,S}^{(n_f)}(\vec{p}_T,\mu) = \nu \frac{\df}{\df\nu} \gamma_{S}^{(n_f)}(\mu,\nu)\, \delta(\vec{p}_T)
&= - 4 \Gamma^{(n_f)}_\cusp[\alpha_s(\mu)]\, \delta(\vec{p}_T)
\, .\end{align}
Note that in practice, the evolution is usually performed in Fourier space, such that one actually
resums the conjugate logarithms $\ln(b \mu)$ where $b = \abs{\vec b_T} \sim 1/q_T$ is the Fourier-conjugate
variable to $q_T$. The $q_T$ spectrum is then obtained as the inverse Fourier transform of the resummed
$b$-spectrum. The exact solution and evolution directly in $q_T$ space, which directly resums the (distributional)
logarithms in $q_T$, has been recently discussed in~\cite{Ebert:2016gcn} (see also ref.~\cite{Monni:2016ktx}),
and turns out to be significantly more involved due to the intrinsic two-dimensional nature of $\vec q_T$.

\begin{figure}[t]
\subfigure[${q_T \ll m \sim Q} $]{\includegraphics[scale=0.25]{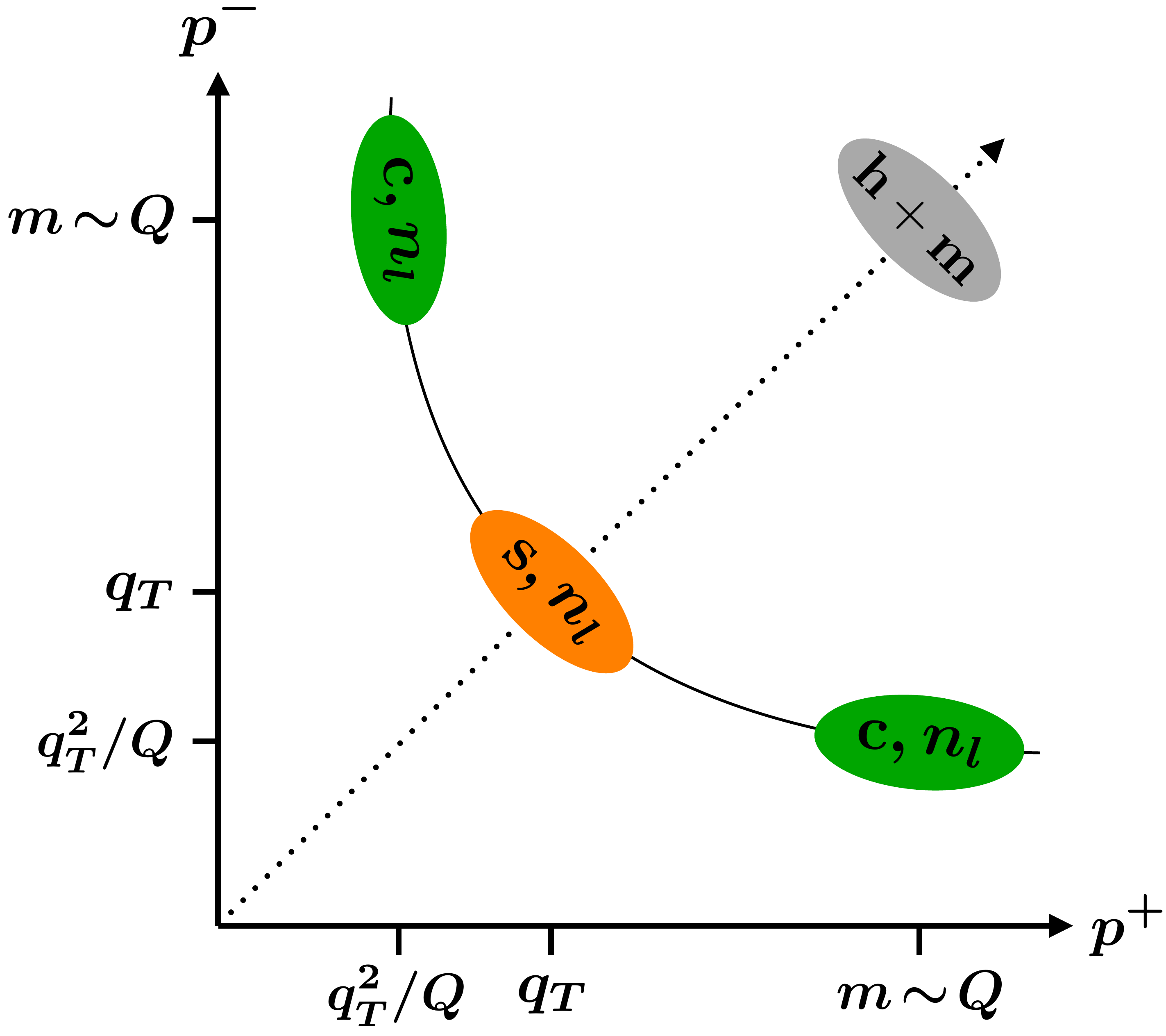}\label{fig:modes_pT_a}}%
\hfill%
\subfigure[${q_T \ll m \ll Q}$]{\includegraphics[scale=0.25]{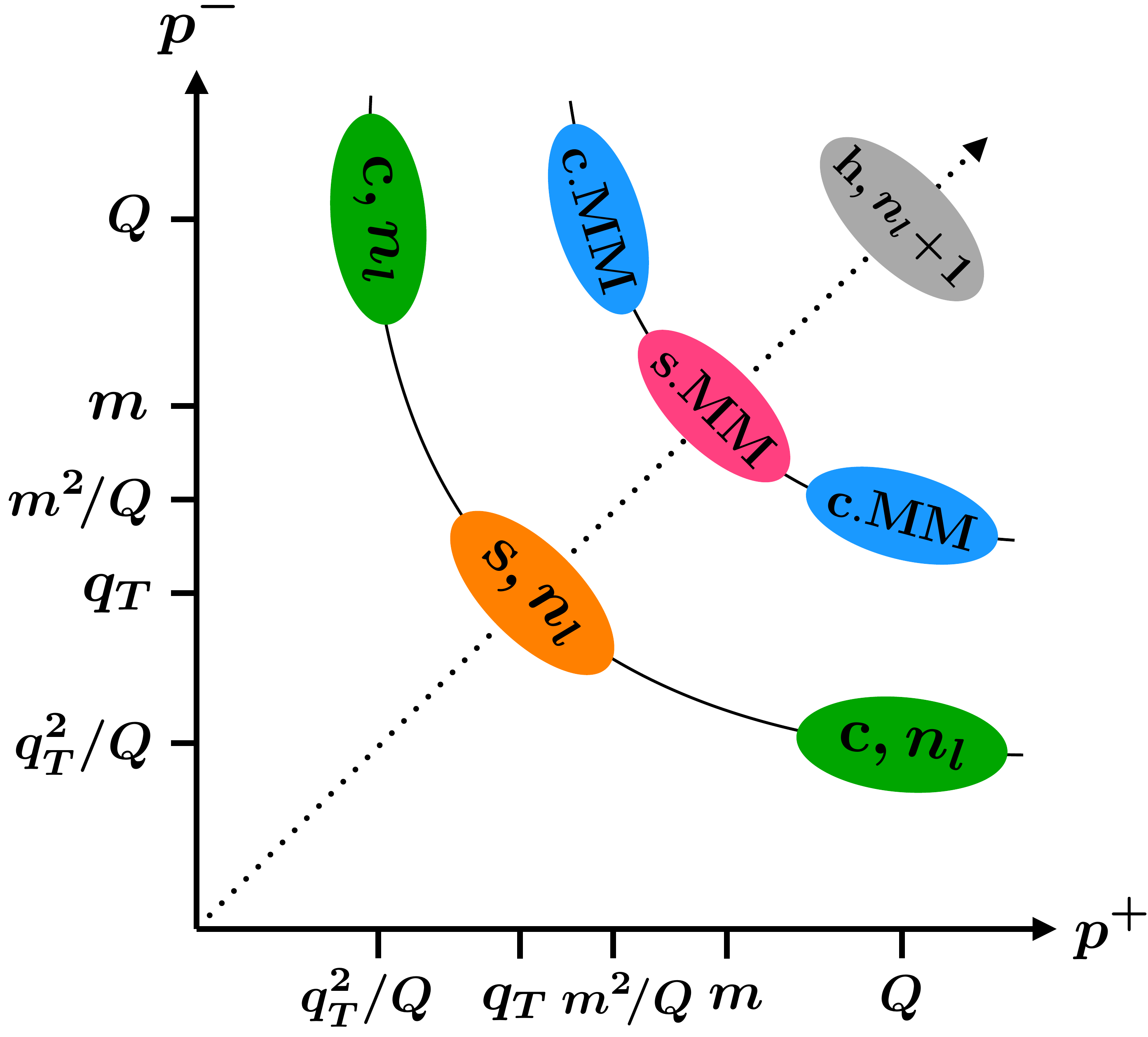}\label{fig:modes_pT_b}}%
\\
\subfigure[${q_T \sim m \ll Q}$]{\includegraphics[scale=0.25]{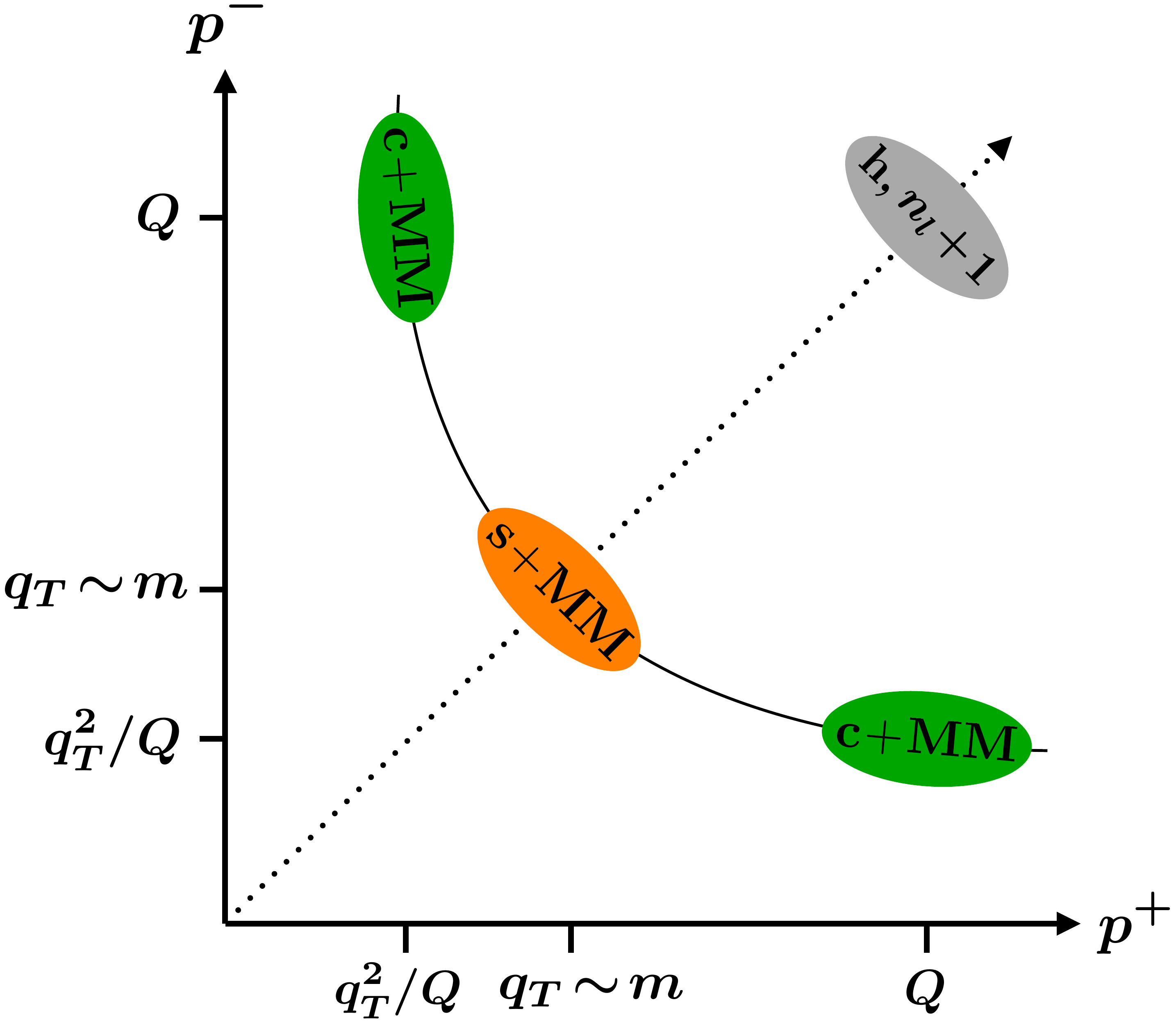}\label{fig:modes_pT_c}}%
\hfill%
\subfigure[${m \ll q_T \ll Q}$]{\includegraphics[scale=0.25]{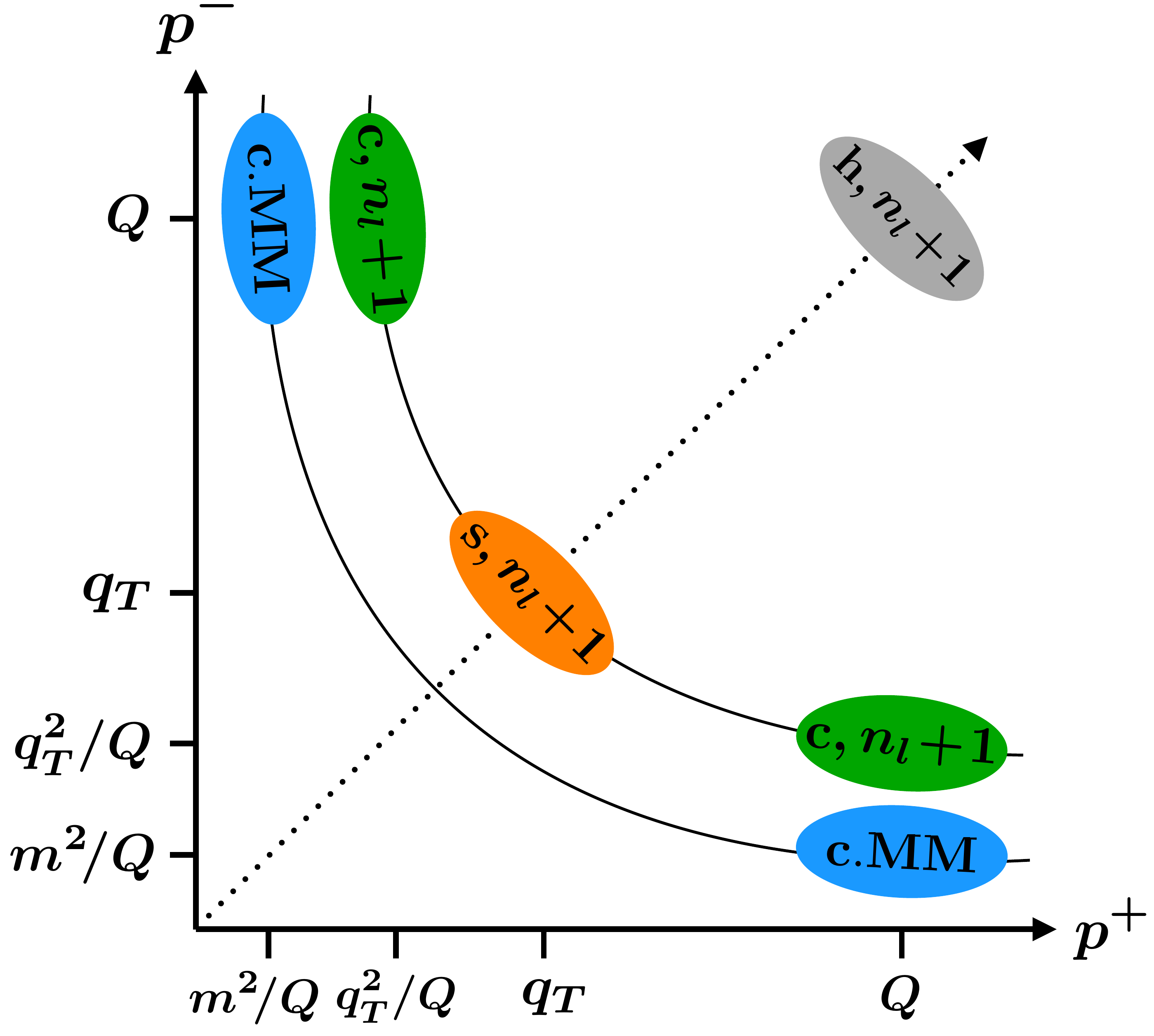}\label{fig:modes_pT_d}}%
\caption{Effective theory modes for the $q_T$ spectrum with massive quarks for $q_T \ll Q$ and $m \gg \Lambda_{\rm QCD}$.}
\label{fig:modes_pT}
\end{figure}

In the following subsections, we discuss how the mode and factorization setup changes when massive quark flavors are involved. These lead to the appearance of additional modes related to fluctuations around the mass shell as discussed extensively in refs.~\cite{Gritschacher:2013pha, Pietrulewicz:2014qza}. For the different hierarchies between the mass scale $m$ and the scales $Q$ and $q_T$ the relevant modes are illustrated in \fig{modes_pT}. In the first case, $q_T \ll m \sim Q$, the massive flavor is integrated out at the hard scale, which leads to the above massless case with $n_l$ massless flavors, as discussed in \subsec{fact_qT1}.
The second case, $q_T \ll m \ll Q$, where the quark mass is larger than the jet resolution variable, is analogous to the corresponding case for thrust in $e^+ e^- \to$ dijets in refs.~\cite{Gritschacher:2013pha, Pietrulewicz:2014qza} and DIS in the $x \to 1$ limit~\cite{Hoang:2015iva}. We refer to these papers for details and only summarize briefly the main features for this regime in \subsec{fact_qT2}.
Our main focus is on the hierarchies $q_T \sim m \ll Q$ and $m \ll q_T \ll Q$, which are important for bottom and charm quark mass effects at the LHC, and which are discussed in \subsecs{fact_qT3}{fact_qT4}.

\subsection{Quark mass effects for $m \sim Q$}
\label{subsec:fact_qT1}

If the quark mass represents a large scale $\sim Q$ (which concerns the top quark at the LHC), this quark flavor does not play a dynamic role in the low-energy effective theory and is integrated out at the hard scale in the matching from QCD to SCET. The relevant modes are shown in fig.~\ref{fig:modes_pT_a}. The massive quark only contributes via mass-dependent contributions to the hard function. This yields the factorization theorem
\begin{align}\label{eq:factqT_m1}
\frac{\df \sigma}{\df q^2_T \,\df Q^2 \,\df Y} &= \sum_{i,j \in \{q,\bar{q}\}}  H_{ij}(Q,m,\mu)\, \int \df^2 p_{Ta}\,\df^2 p_{Tb}\, \df^2 p_{Ts} \, \delta(q_T^2-|\vec{p}_{Ta}+\vec{p}_{Tb}+\vec{p}_{Ts}|^2)
\\ & \quad \times
B^{(n_l)}_i\Bigl(\vec{p}_{Ta},x_a,\mu,\frac{\nu}{\omega_a}\Bigr)\, B^{(n_l)}_{j}\Bigl(\vec{p}_{Tb},x_b,\mu,\frac{\nu}{\omega_b}\Bigr) \, S^{(n_l)}(\vec{p}_{Ts},\mu,\nu) \biggl[1+\mathcal{O}\Bigl(\frac{q_T^2}{m^2},\frac{q_T}{Q}\Bigr)\biggr]
\nn \,,\end{align}
which is essentially equivalent to the massless case in the previous subsection with $n_l$ massless flavors.
The hard function $H_{ij}(Q,m,\mu)$ can be evaluated either in the $(n_f = n_l)$ or the $(n_f = n_l+1)$ flavor scheme for $\alpha_s$, where $n_l$ is the number of light (massless) quark flavors. The associated massive quark corrections are directly related to the virtual contributions to the quark form factors, e.g. given at $\mathcal{O}(\alpha_s^2)$ by the virtual diagrams in \fig{Secondary}. In general both primary and secondary corrections contribute for initial (massless) quarks. Using the $(n_l)$ flavor scheme for $\alpha_s$ these vanish as $\mathcal{O}(Q^2/m^2)$ in the decoupling limit $m \gg Q$ for the conserved vector current. For the axial-vector current, contributing to $Z$-boson production, there are in addition also anomaly corrections starting at $\mathcal{O}(\alpha_s^2)$ from the massive quark triangle in \fig{Primary} that do not decouple.\footnote{Instead, for $m \gg Q$ the heavy quark can be integrated out around its mass scale and the axial current can be evolved between $m$ and $Q$ to resum the associated logarithms $\ln (m/Q)$.}
Since the massive quark does not appear as a dynamic flavor in the EFT below the hard scale $Q$, the entire RG evolution to sum the logarithms of $q_T$ is performed with $n_l$ massless flavors as in \eq{factqT_ml}.

\subsection{Quark mass effects for $q_T \ll m \ll Q$}
\label{subsec:fact_qT2}

Next, we consider the hierarchies where the quark mass is parametrically smaller than the hard scale, $m \ll Q$. These require a different factorization setup than $m \sim Q$ since fluctuations around the mass-shell are now parametrically separated from hard fluctuations, which would lead to large unresummed logarithms inside the hard function $H _{ij}(Q,m,\mu)$.
In this subsection, we start with the case where the transverse momentum is much smaller than the mass, $q_T \ll m \ll Q$, while
$q_T\sim m \ll Q$ and $m \ll q_T \ll Q$ are considered in the following subsections.

In a first step the QCD current is matched onto the SCET current with $n_l+1$ dynamic quark flavors at the scale $\mu \sim Q$. Since $m \ll Q$ this matching can be performed (at leading order in the expansion parameter $m/Q$) only with massless quarks, leading to the hard function with $n_l+1$ massless flavors, $H_{ij}^{(n_l+1)}$, with the strong coupling inside it renormalized with $n_l+1$ flavors. The matching is performed onto SCET containing $n_a$-collinear, $n_b$-collinear, and soft mass modes with the scaling
\begin{align} \label{eq:modes_mass}
n_a \text{-collinear MM:} 
&\quad
p_{m,n_a}^\mu \sim \Bigl(\frac{m^2}{Q},Q,m\Bigr)
\,, \nn \\
n_b \text{-collinear MM:} 
&\quad
p_{m,n_b}^\mu\sim\Bigl(Q,\frac{m^2}{Q},m\Bigr)
\, , \nn \\
\text{soft MM:} 
&\quad
p_{m,s}^\mu\sim (m,m,m)
\,, 
\end{align}
as illustrated in \fig{modes_pT_b}. These mass-shell fluctuations arise here purely from secondary virtual contributions.

In a second step at the scale $\mu \sim m$, the mass modes are integrated out and the SCET with $n_l$ massless and one massive flavor is matched onto SCET with $n_l$ massless flavors with the usual scaling as in the massless case in \eq{modes_massless}. Since the soft and collinear mass modes have the same invariant mass set by the quark mass and are only separated in rapidity, there are rapidity divergences in their (unrenormalized) collinear and soft contributions. Their renormalization and the resummation of the associated logarithms can be again handled using the rapidity RG approach in refs.~\cite{Chiu:2011qc,Chiu:2012ir}, which has been explicitly carried out in ref.~\cite{Hoang:2015vua}.%
\footnote{The matching in ref.~\cite{Hoang:2015vua} was performed with massive primary quarks yielding the matching functions denoted as $H_{m,n}$, $H_{m,\bar{n}}$ and $H_{m,s}$ there. However, this does not affect the structure of the rapidity logarithms arising from the secondary mass effects, which are independent of the primary quarks being massive or massless.}
In addition, all renormalized parameters like the strong coupling constant are  matched at the mass scale from $n_l+1$ to $n_l$ flavors taking into account that the massive flavor is removed as a dynamic degree of freedom.

After these steps, the factorization at the low scale $\sim q_T$ proceeds as in the massless case with all operator matrix elements depending on the $n_l$ massless flavors, which yields the factorization theorem
\begin{align}\label{eq:factqT_m2}
\frac{\df \sigma}{\df q^2_T \,\df Q^2 \,\df Y} &= \sum_{i,j \in \{q,\bar{q}\}}  H^{(n_l+1)}_{ij}(Q,\mu)\, H_{c}\Bigl(m,\mu,\frac{\nu}{\omega_a}\Bigr) H_{c}\Bigl(m,\mu,\frac{\nu}{\omega_b}\Bigr)  H_{s}(m,\mu,\nu)
\nn \\
& \quad \times
\int \df^2 p_{Ta}\,\df^2 p_{Tb}\, \df^2 p_{Ts} \, \delta(q_T^2-|\vec{p}_{Ta}+\vec{p}_{Tb}+\vec{p}_{Ts}|^2)\, B^{(n_l)}_i\Bigl(\vec{p}_{Ta},x_a,\mu,\frac{\nu}{\omega_a}\Bigr)
\nn \\
& \quad \times  B^{(n_l)}_{j}\Bigl(\vec{p}_{Tb},x_b,\mu,\frac{\nu}{\omega_b}\Bigr) \, S^{(n_l)}(\vec{p}_{Ts},\mu,\nu) \, \Bigl[1+\mathcal{O}\Bigl(\frac{q_T}{Q},\frac{q_T^2}{m^2},\frac{m^2}{Q^2}\Bigr)\Bigr]
\, .\end{align}
Here $H_{c}$ and $H_{s}$ denote the hard functions that arise from the matching at the mass scale $\mu \sim m$. Their natural rapidity scales are $\nu \sim Q$ for the collinear contributions and $\nu \sim m$ for the soft ones. They can be evaluated in either the $(n_l)$ or $(n_l+1)$ scheme for $\alpha_s$. We will give their expressions at $\mathcal{O}(\alpha_s^2)$ in \subsec{hard}. The resummation of all logarithms of ratios of $q_T$, $m$, and $Q$ is achieved by performing the evolution in $\mu$ and $\nu$ of all functions appearing in \eq{factqT_m2} from their natural scales.

\begin{figure}[t]
\subfigure[${q_T \ll m \ll Q}$]{\includegraphics[scale=0.25]{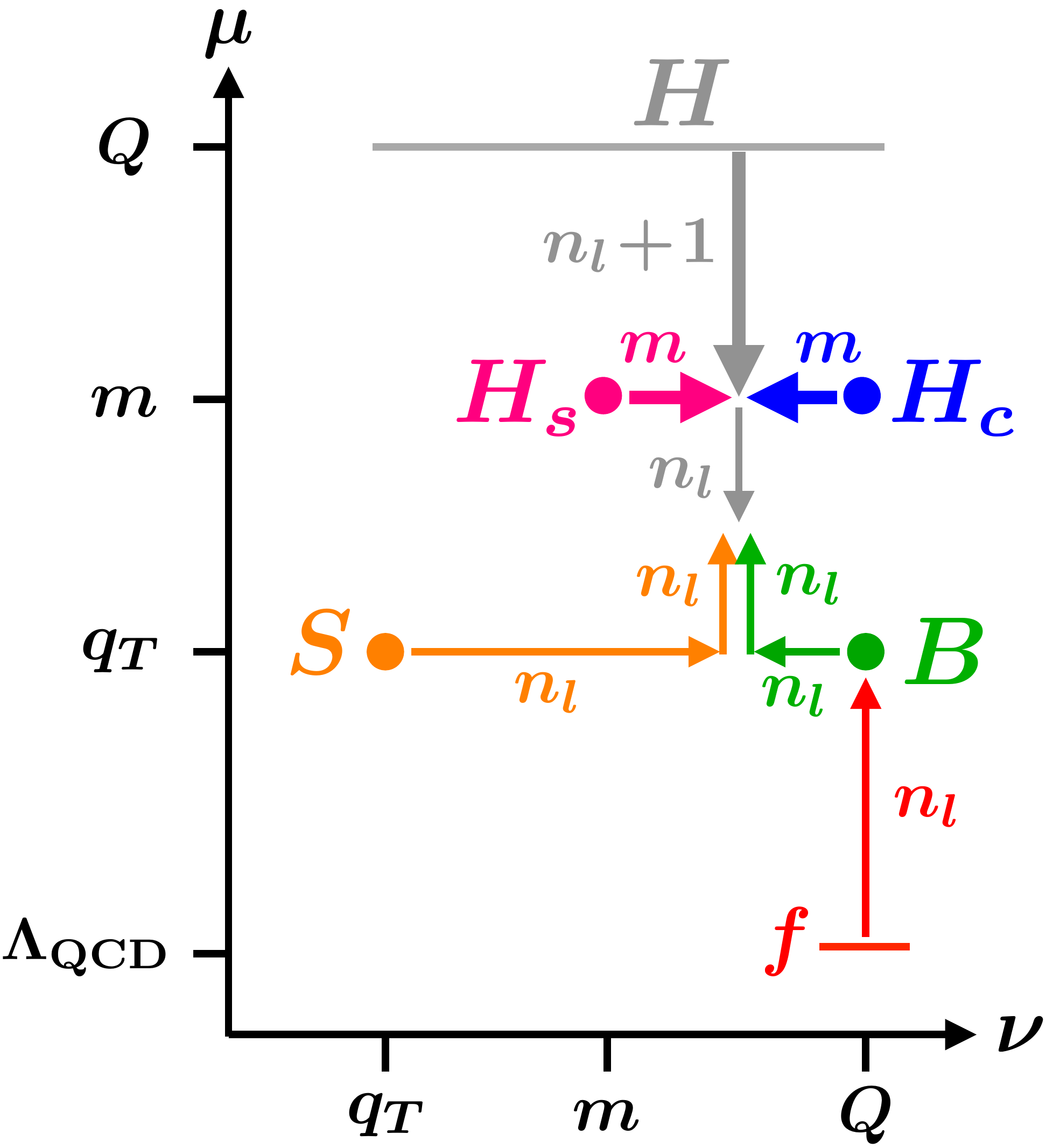}\label{fig:evolution_qT2}}%
\hfill%
\subfigure[${q_T \sim m \ll Q}$]{\includegraphics[scale=0.25]{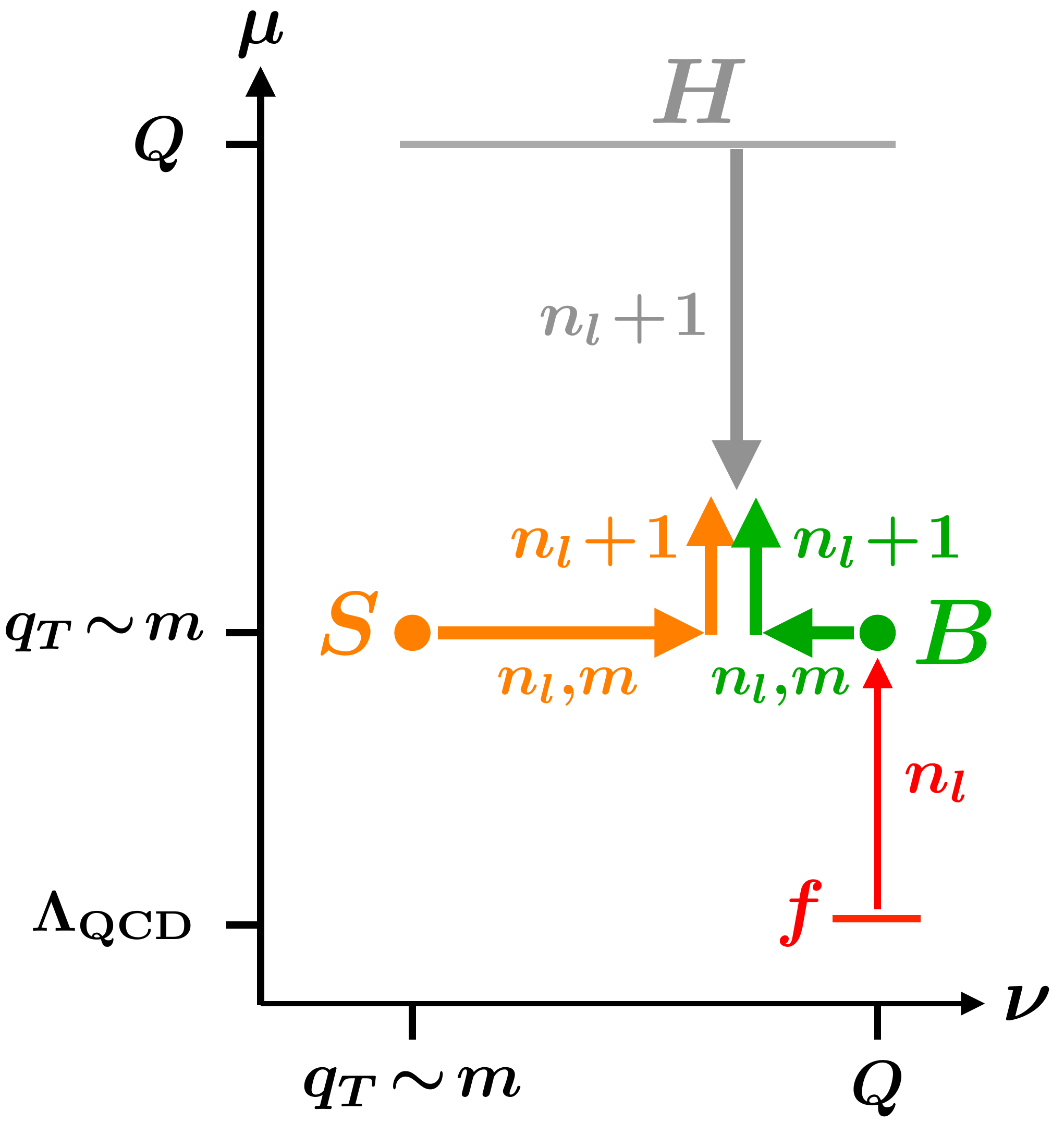}\label{fig:evolution_qT3}}%
\hfill%
\subfigure[${m \ll q_T \ll Q}$]{\includegraphics[scale=0.25]{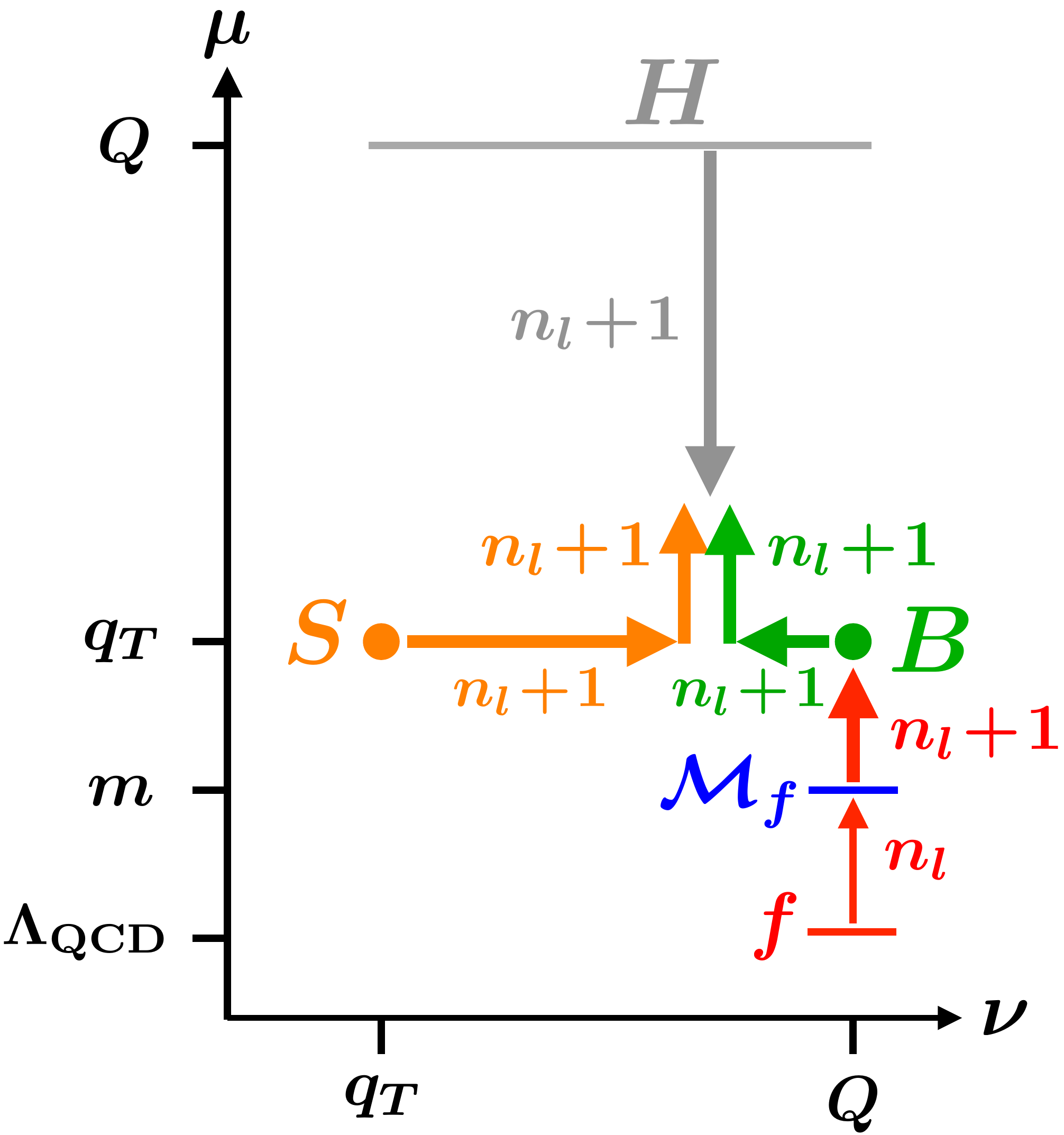}\label{fig:evolution_qT4}}%
\caption{Illustration of the renormalization group evolution for $q_T$ of the hard, beam, soft, and parton distribution functions in invariant mass and rapidity. The anomalous dimensions for each evolution step involve the displayed number of active quark flavors. The label $m$ indicates that the corresponding evolution is mass dependent.}
\label{fig:evolution_qT}
\end{figure}

In principle, the $\mu$ evolution can be performed by evolving all functions with their respective number of quark flavors without switching the flavor scheme, i.e.\ with $n_l+1$ flavors for $H$, $n_l$ flavors for $B$ and $S$ and an additional evolution for the collinear and soft matching functions $H_c$ and $H_s$.
The consistency of RG running for the factorization theorems in \eqs{factqT_m2}{factqT_m1}, and \eq{consistency_qT} with $n_l$ massless flavors, implies that the $\mu$-dependence of the mass-dependent hard functions $H_{c}$ and $H_{s}$ is precisely given by the difference between $n_l$ and $n_l+1$ active quark flavors in the evolution of the hard function $H_{ij}$,
\begin{align}\label{eq:gammaH_consistency}
\gamma_{H_c}\Bigl(m,\mu,\frac{\nu}{\omega_a}\Bigr) + \gamma_{H_c}\Bigl(m,\mu,\frac{\nu}{\omega_b}\Bigr) + \gamma_{H_s}(m,\mu,\nu) = \gamma^{(n_l)}_{H}(Q,\mu) -\gamma^{(n_l+1)}_H(Q,\mu)
\, ,\end{align}
where $\gamma^{(n_f)}_H$ is defined in \eq{RGE_ml}, and $\gamma_{H_c}$ and $\gamma_{H_s}$ are defined analogously.
At two loops this relation can be checked explicitly using the results in eqs.~(\ref{eq:gammaHs}),~(\ref{eq:gammaHn}) and~(\ref{eq:gammaH}).  As a result, the $\mu$ evolution for the hard functions can be conveniently implemented
as illustrated in \fig{evolution_qT2}, by carrying out the $\mu$ evolution with $n_l$ active quark flavors below the matching scale $\mu_m \sim m$ and with $n_l+1$ flavors above $\mu_m$, providing in this sense a ``variable-flavor number scheme''~\cite{Pietrulewicz:2014qza,Hoang:2015iva}.
(This effectively corresponds to using operator running for the hard scattering current, which is renormalized with $n_l+1$ flavors above the mass scale and with $n_l$ flavors below the mass scale.)
In addition there is also a rapidity evolution, which is carried out at $\mu_m = m$, i.e.\ at the border between the $(n_l +1)$ and $(n_l)$-flavor theories (see ref.~\cite{Hoang:2015vua}), which is governed by the mass-dependent rapidity anomalous dimensions for $H_s$ and $H_c$,
\begin{align}\label{eq:Hs_nuevolution}
\gamma_{\nu,H_s}(m,\mu) = -2\gamma_{\nu, H_c}(m,\mu) = \frac{\df}{\df \ln\nu} \ln H_s(m,\mu,\nu) \, .
\end{align}
  
\subsection{Quark mass effects for $q_T \sim m \ll Q$}
\label{subsec:fact_qT3}

If the $q_T$ scale is of the order of the quark mass, $q_T \sim m$, the massive quark becomes a dynamic degree of freedom, which contributes to the $q_T$ spectrum via real radiation effects. The mass modes in \eq{modes_mass} are now the same as the usual massless \SCETb modes for the $q_T$ measurement in \eq{modes_massless}, since their parametrically scaling coincides for $q_T\sim m$, as illustrated in fig.~\ref{fig:modes_pT_c}. In this case, there is only a single matching at the hard scale $\mu \sim Q$ from QCD onto SCET with these common soft and collinear modes. This hard matching gives again rise to the (mass-independent) hard function $H_{ij}^{(n_l+1)}$ for $n_l+1$ massless flavors. The SCET operator matrix elements at the scale $\mu \sim q_T$, i.e.~the beam and soft functions, now encode the effects of the massive quark. They are now renormalized with $n_l+1$ quark flavors and contain an explicit dependence on the quark mass. When integrating out the modes with the virtuality $q_T$ also the massive quark is integrated out and the collinear matching functions $\mathcal{I}_{ik}$ between the beam functions and the PDFs thus also contain the effect from changing from $n_l+1$ to $n_l$ flavors, i.e.\
\begin{align}\label{eq:Bm1_qT}
B_i^{(n_l+1)}\Bigl(\vec{p}_{T},m,x,\mu,\frac{\nu}{\omega}\Bigr)= \sum_{k \in \{q,\bar{q},g\}} \mathcal{I}_{ik}\Bigl(\vec{p}_{T},m,x,\mu,\frac{\nu}{\omega}\Bigr) \otimes_x f^{(n_l)}_k (x,\mu) \Bigl[1+\mathcal{O}\Bigl(\frac{\lqcd^2}{m^2},\frac{\lqcd^2}{q_T^2}\Bigr)\Bigr] \, .
\end{align}
Written out explicitly, the factorization theorem reads
\begin{align}\label{eq:factqT_m3}
\frac{\df \sigma}{\df q^2_T \,\df Q^2 \,\df Y} &= \sum_{i,j \in \{q,\bar{q},Q,\bar{Q}\}}  H^{(n_l+1)}_{ij}(Q,\mu)\, \int \df^2 p_{Ta}\,\df^2 p_{Tb}\, \df^2 p_{Ts} \, \delta(q_T^2-|\vec{p}_{Ta}+\vec{p}_{Tb}+\vec{p}_{Ts}|^2)
\nn \\ & \quad \times
\biggl[\sum_{k \in \{q,\bar{q},g\}} \mathcal{I}_{ik}\Bigl(\vec{p}_{Ta},m,x_a,\mu,\frac{\nu}{\omega_a}\Bigr) \otimes_x f^{(n_l)}_k (x_a,\mu)\biggr]
\nn \\ & \quad  \times
\biggl[\sum_{k \in \{q,\bar{q},g\}} \mathcal{I}_{jk}\Bigl(\vec{p}_{Tb},m,x_b,\mu,\frac{\nu}{\omega_b}\Bigr) \otimes_x f^{(n_l)}_k (x_b,\mu)\biggr]
\nn \\ & \quad \times
S(\vec{p}_{Ts},m,\mu,\nu)
\, \biggl[1+\mathcal{O}\Bigl(\frac{q_T}{Q},\frac{m^2}{Q^2},\frac{\lqcd^2}{m^2},\frac{\lqcd^2}{q_T^2}\Bigr)\biggr]
\,,\end{align}
where $i,j = Q, \bar Q$ denotes the massive quark flavor in the sum over flavors.
We stress that the renormalization of the bare soft and beam function with $n_l$ massless and one massive flavor is carried out in the $n_l+1$ flavor scheme for $\alpha_s$, while the strong coupling in the PDFs (which are defined in the lower theory with $n_l$ massless flavors) is renormalized with $n_l$ flavors. The \emph{renormalized} soft function and beam function coefficients $\mathcal{I}_{ik}$ can then be expressed in terms of either the $(n_l+1)$ or the $(n_l)$ flavor scheme for $\alpha_s$ without introducing large logarithms.

In this hierarchy quark mass effects enter in \eq{factqT_m3} at $\mathcal{O}(\alpha_s^2)$ in two ways:
There are secondary radiation effects appearing in the two-loop soft function $S^{(2)}$ and the flavor-diagonal beam function matching coefficients $\mathcal{I}^{(2)}_{qq}$. In addition, there are primary mass effects arising from a massive-quark initiated hard process. For $Z/\gamma^*$ production, this requires the production of the massive quarks via gluon splitting in both collinear sectors, which manifests itself in two one-loop collinear matching coefficients $\mathcal{I}^{(1)}_{Qg} \times \mathcal{I}^{(1)}_{\bar Qg}$.
For $W$-boson production, primary charm quark effects enter already at $\mathcal{O}(\alpha_s)$ from a single $\mathcal{I}^{(1)}_{Qg}$ with $Q = c$.

The resummation of logarithms $\ln(q_T/Q)$ and $\ln(m/Q)$ is again obtained by performing the RG evolution for
\eq{factqT_m3}, which is illustrated in \fig{evolution_qT3}.
While the evolution of the PDFs proceeds in $n_l$ flavors, the $\mu$-evolution for the hard, beam, and soft functions above the scale $m$ is now carried out purely with $n_l+1$ flavors. Consistency of RG running for \eq{Bm1_qT} implies that the matching coefficients $\mathcal{I}_{ik}$ satisfy
\begin{align}
\mu\frac{\df}{\df \mu} \,\mathcal{I}_{ik} \Bigl(\vec{p}_{T},m,z,\mu,\frac{\nu}{\omega}\Bigr) & = \Bigl[\gamma^{(n_l+1)}_{B_i}\times \mathcal{I}_{ik}\Bigr] \Bigl(\vec{p}_{T},m,z,\mu,\frac{\nu}{\omega}\Bigr) - \sum_{j \in q,\bar q,g} \Bigl[\mathcal{I}_{ij} \otimes \gamma^{(n_l)}_{f,jk}\Bigr] \Bigl(\vec{p}_{T},m,z,\mu,\frac{\nu}{\omega}\Bigr) \, .
\end{align}
Since the renormalization of the beam functions does not involve parton mixing, the one-loop primary mass contributions to $\cI_{Qg}^{(1)}$ cannot give rise to rapidity divergences and associated logarithms. On the other hand, the secondary mass effects change the rapidity evolution. In particular, the beam and soft $\nu$-anomalous dimensions become mass dependent%
\footnote{The fact that quark masses can affect the evolution was already pointed out in ref.~\cite{Collins:1984kg}.},
\begin{align}\label{eq:BSm_nuevolution}
\nu \frac{\df}{\df \nu} B_i^{(n_l+1)}\Bigl(\vec{p}_{T},m,\mu,\frac{\nu}{\omega}\Bigr)
& = \int\!\df^2 k_T \, \gamma^{(n_l+1)}_{\nu,B}(\vec{p}_T-\vec{k}_T,m,\mu) \, B_i^{(n_l+1)}\Bigl(\vec{k}_{T},m,\mu,\frac{\nu}{\omega}\Bigr)
\, , \nn \\
\nu \frac{\df}{\df \nu} S^{(n_l+1)}(\vec{p}_{T},m,\mu,\nu)
& = \int\! \df^2 k_T\, \gamma^{(n_l+1)}_{\nu,S}(\vec{p}_T-\vec{k}_T,m,\mu) \, S^{(n_l+1)}(\vec{k}_T,m,\mu,\nu)\, .
\end{align}
We discuss the implications of the mass dependence for the rapidity evolution in \sec{nu_evolution}.

\subsection{Quark mass effects for $m \ll q_T \ll Q$}
\label{subsec:fact_qT4}

If $q_T$ is much larger than the mass, the fluctuations around the mass-shell take place at a different scale than the jet resolution measurement. There are no relevant soft fluctuations scaling like $p_{m,s}^\mu \sim (m,m,m)$, since the measurement of $q_T$ is IR safe and is thus insensitive to the lower mass scale. (In other words, if we were to explicitly distinguish such soft mass modes their contribution would cancel as for an inclusive observable since they are not constrained by the $q_T$-measurement.) This means that the soft modes are described by a soft function with $n_l+1$ massless flavors at the scale $\mu \sim q_T$. Due to the collinear sensitivity of the initial-state radiation there are still relevant collinear mass modes scaling like $p_{m,n_a}^\mu \sim (m^2/Q,Q,m)$ and $p_{m,n_b}^\mu \sim (Q,m^2/Q,m)$, as illustrated in \fig{modes_pT_d}. Thus there are collinear modes in SCET at different invariant mass scales, which can be disentangled by a multistage matching. First, the beam functions are matched onto
the PDFs with $n_l$ massless and one massive flavor. Since this matching takes place at the scale $\mu_B \sim q_T \gg m$ this gives just the matching coefficients $\mathcal{I}_{ik}$ for $n_l+1$ massless flavors,
\begin{align}\label{eq:beam_matching1}
B_i^{(n_l+1)}\Bigl(\vec{p}_{T},m,x,\mu,\frac{\nu}{\omega}\Bigr)= \sum_{k \in \{q,\bar{q},Q,\bar{Q},g\}} \mathcal{I}^{(n_l+1)}_{ik}\Bigl(\vec{p}_{T},x,\mu,\frac{\nu}{\omega}\Bigr) \otimes_x f^{(n_l+1)}_k (x,m,\mu) \Bigl[1+\mathcal{O}\Bigl(\frac{m^2}{q_T^2}\Bigr)\Bigr] \, .
\end{align}
In a second step, at the mass scale $\mu_m \sim m$, the PDFs including the massive quark effects are matched onto PDFs with $n_l$ massless quarks, and with $\alpha_s$ in the $(n_l)$ flavor scheme,
\begin{align}\label{eq:PDF_matching_function}
f^{(n_l+1)}_i (x,m,\mu)= \sum_{k \in \{q,\bar{q},g\}} \mathcal{M}_{ik}(x,m,\mu) \otimes_x f^{(n_l)}_k (x,\mu) \biggl[1+\mathcal{O}\Bigl(\frac{\lqcd^2}{m^2}\Bigr)\biggr]
\,.\end{align}
The PDF matching functions $\mathcal{M}_{ik}$ can be expressed in either the $(n_l)$ or the $(n_l+1)$ flavor scheme for $\alpha_s$.

In total, the factorization theorem reads
\begin{align}\label{eq:factqT_m4}
\frac{\df \sigma}{\df q^2_T \,\df Q^2 \,\df Y} &= \sum_{i,j \in \{q,\bar{q},Q,\bar{Q}\}}  H^{(n_l+1)}_{ij}(Q,\mu)\, \int \df^2 p_{Ta}\,\df^2 p_{Tb}\, \df^2 p_{Ts} \, \delta(q_T^2-|\vec{p}_{Ta}+\vec{p}_{Tb}+\vec{p}_{Ts}|^2)
\nn \\
& \quad \times
\biggl[\sum_{k \in \{q,\bar{q},Q,\bar{Q},g\}}
\sum_{l \in \{q,\bar{q},g\}} \mathcal{I}^{(n_l+1)}_{ik}\Bigl(\vec{p}_{Ta},x_a,\mu,\frac{\nu}{\omega_a}\Bigr) \otimes_x \mathcal{M}_{kl} (x_a,m,\mu) \otimes_x f^{(n_l)}_l (x_a,\mu)\biggr]
\nn \\
& \quad \times \biggl[\sum_{k \in \{q,\bar{q},Q,\bar{Q},g\}}
 \sum_{l\in \{q,\bar{q},g\}} \mathcal{I}^{(n_l+1)}_{jk}\Bigl(\vec{p}_{Tb},x_b,\mu,\frac{\nu}{\omega_b}\Bigr)\otimes_x \mathcal{M}_{kl} (x_b,m,\mu) \otimes_x f^{(n_l)}_l (x_b,\mu)\biggr]
\nn \\
& \quad \times S^{(n_l+1)}(\vec{p}_{Ts},\mu,\nu) \biggl[1+\mathcal{O}\Bigl(\frac{q_T}{Q},\frac{m^2}{q_T^2},\frac{\lqcd^2}{m^2}\Bigr)\biggr]
\, .\end{align}
As in \subsec{fact_qT3}, massive quark corrections can arise at $\mathcal{O}(\alpha_s^2)$ either via primary mass effects involving the product of two one-loop PDF matching corrections $\mathcal{M}^{(1)}_{Qg}$ (for $Z/\gamma^*$) generating a massive quark-antiquark pair that initiates the hard interaction, or via secondary mass effects involving one two-loop contribution $\mathcal{M}^{(2)}_{qq}$. Note that also the running of the light quark and gluon PDFs above $\mu_m$ generates an effective massive quark PDF via evolution factors $U^{(n_l+1)}_{f,Qk}(\mu_B,\mu_m)$, which for large hierarchies $m \ll q_T$ can give $\mathcal{O}(1)$ contributions.

The evolution of the hard, beam, and soft functions can be performed purely with $n_l+1$ massless flavors, while the PDFs are evolved with $n_l$ flavor below $\mu_m \sim m$ and with $n_l+1$ flavors above $\mu_m \sim m $, see \fig{evolution_qT4}, as usual in any variable-flavor number scheme. The $\mu_m$ dependence is canceled order-by-order by the matching factors $\mathcal{M}_{ij}$,
\begin{align}
\mu\frac{\df}{\df \mu} \,\mathcal{M}_{ik} (m,z,\mu)
&= \sum_{j \in \{q,\bar q, Q,\bar Q, g\}}\Bigl[ \gamma^{(n_l+1)}_{f,ij} \otimes_z \mathcal{M}_{jk}\Bigr]  (m,z,\mu) -  \sum_{j \in \{q,\bar q, g\}} \Bigl[\mathcal{M}_{ij} \otimes_z \gamma^{(n_l)}_{f,jk}\Bigr] (m,z,\mu) \, .
\end{align}
The absence of soft mass modes in this regime implies there is no rapidity evolution at the mass scale, while the rapidity evolution between beam and soft functions is the same as for $n_l+1$ massless flavors.

In this regime, the mass dependence is thus fully contained in the collinear sectors. Within each collinear sector, the EFT setup is completely analogous to that of the heavy-quark induced inclusive cross section discussed in detail in \refcite{Bonvini:2015pxa}, with the beam  functions here playing the role of the inclusive cross section there and the $q_T$ scale here playing the role of the hard scale there.

\subsection{Relations between hierarchies}
\label{subsec:qT_hierarchies}

\begin{figure}
\centering
\includegraphics[scale=0.25]{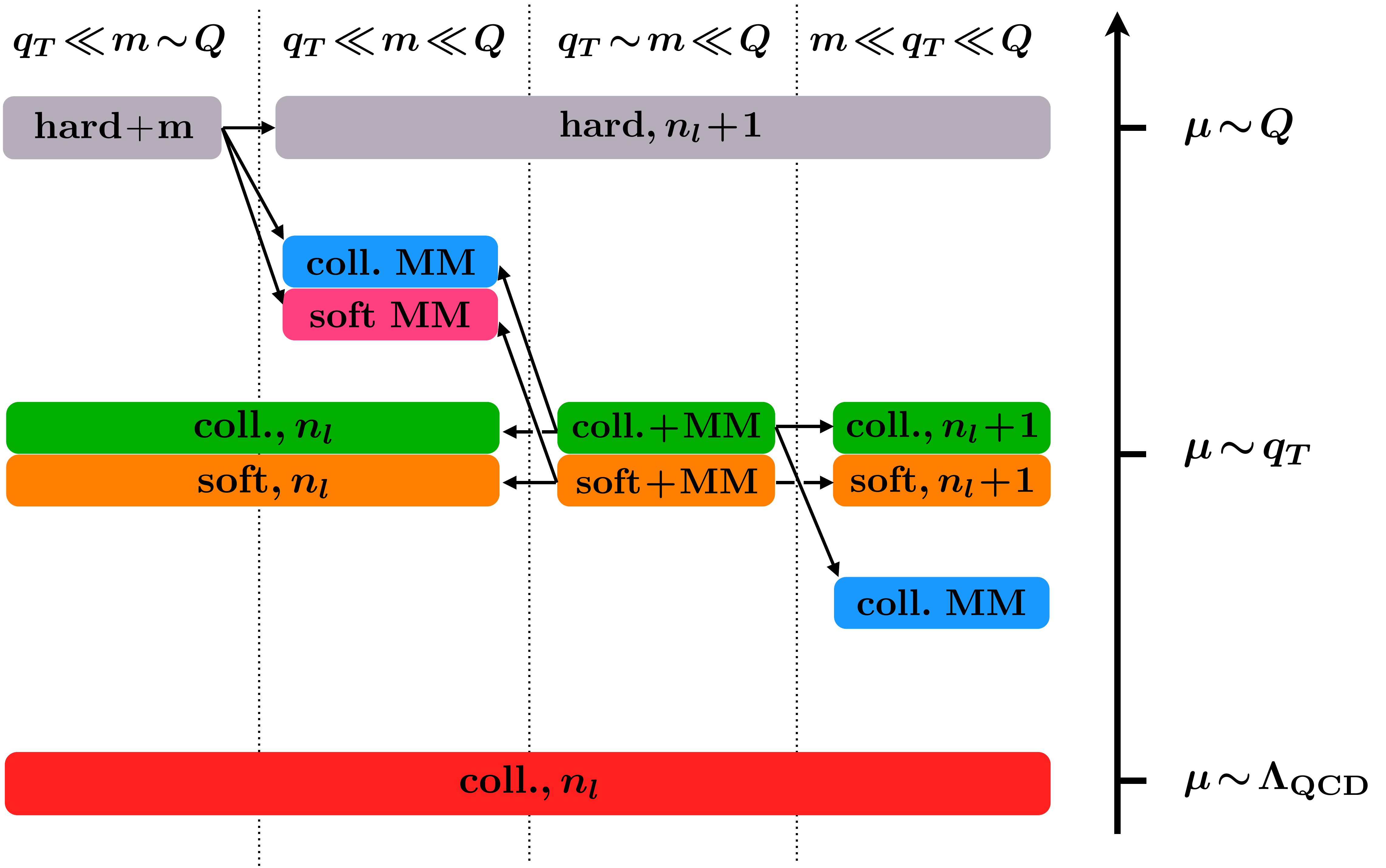}
\caption{Relevant modes for the $q_T$ spectrum with $q_T \ll Q$ for different hierarchies between the quark mass $m$ and the scales $q_T$ and $Q$. The arrows indicate the relations between the modes and their associated contributions.}
\label{fig:scales_pT}
\end{figure}

After discussing all hierarchies separately, we now show how the ingredients in each of the associated factorization theorems are related to each other. This will also make it obvious how the mass-dependent fixed-order corrections that are kept in one hierarchy but are dropped in another can be combined with the resummation of logarithms to obtain a systematic inclusion of the mass effects over the whole $q_T$ spectrum. The relations between the modes and their contributions between the different regimes are summarized in \fig{scales_pT}.

The hard functions appearing in the hierarchy $q_T \ll m \ll Q$ in \eq{factqT_m2} are related to the hard function for $q_T \ll m \sim Q$ in \subsec{fact_qT1} as follows%
\footnote{Here and in the following it is implied that at a specific fixed order the functions on both sides have to be expanded in the same renormalization scheme for $\alpha_s$. Since $H^{(n_l+1)}_{ij}$ is generically written with $n_l+1$ (massless) flavors, the $(n_l+1)$-flavor scheme is most convenient here and also to extract the $\ord{m^2/Q^2}$ power corrections on the right-hand side of \eq{consistency_hard} from its fixed-order expansion.}
\begin{align}\label{eq:consistency_hard}
H_{ij} (Q,m,\mu) = H^{(n_l+1)}_{ij}(Q,\mu)\, H_{c}\Bigl(m,\mu,\frac{\nu}{\omega_a}\Bigr) H_{c}\Bigl(m,\mu,\frac{\nu}{\omega_b}\Bigr)  H_{s}(m,\mu,\nu)
\biggl[1 + \mathcal{O}\Bigl(\frac{m^2}{Q^2}\Bigr)\biggr]
\, .\end{align}
In the product of functions on the right-hand side, which appear in \eq{factqT_m2}, the logarithms $\ln(m/Q)$ can be resummed to all orders. One can construct a smooth description of the cross section for $q_T \ll m$ that resums these logarithms and also includes the associated mass-dependent $\ord{m^2/Q^2}$ power corrections by simply adding the latter to the hard function $H^{(n_l+1)}(Q,\mu)$ at the scale $\mu \sim Q$.

The fixed-order contributions to the operator matrix elements appearing in the hierarchy $q_T \ll m$ are encoded in the ones for $q_T \sim m$. The mass-dependent beam function matching coefficients for $q_T\sim m$ are related to those for $q_T \ll m$ and the collinear mass-mode function $H_c$ by
\begin{align}\label{eq:consistencyqT_beam1}
\mathcal{I}_{ik}\Bigl(\vec{p}_T, x,m,\mu,\frac{\nu}{\omega}\Bigr)
=  H_{c}\Bigl(m,\mu,\frac{\nu}{\omega}\Bigr)\,\mathcal{I}_{ik}^{(n_l)} \Bigl(\vec{p}_T, x,\mu,\frac{\nu}{\omega}\Bigr)
\biggl[1+\mathcal{O}\Bigl(\frac{p_T^2}{m^2}\Bigr)\biggr]
\, .\end{align}
Similarly, the mass-dependent soft function for $q_T\sim m$ is related to the one for $q_T \ll m$ and the soft mass-mode function $H_s$ by
\begin{align}\label{eq:consistencyqT_soft1}
S(\vec{p}_T, m,\mu,\nu)
=  H_{s}(m,\mu,\nu)\,S^{(n_l)} (\vec{p}_T, \mu,\nu) \biggl[1+\mathcal{O}\Bigl(\frac{p_T^2}{m^2}\Bigr)\biggr]
\, .\end{align}
In the products on the right-hand sides, which appear in \eq{factqT_m2}, logarithms $\ln(q_T/m)$ are resummed to all orders in the limit $q_T \ll m$. One can include the associated $\ord{q_T^2/m^2}$ power corrections that are important for $q_T\sim m$, by obtaining them from the fixed-order expansions of \eqs{consistencyqT_beam1}{consistencyqT_soft1} and adding them to the $(n_l)$-flavor beam function coefficients and soft function at the scale $\mu \sim q_T$.

Finally, the fixed-order contributions for the operator matrix elements appearing in the hierarchy $m \ll q_T$ are also encoded in the corresponding ones for $q_T \sim m$. Hence, the mass-dependent beam function matching coefficients are related to those for $m \ll q_T$ and the PDF matching functions by
\begin{align}\label{eq:consistencyqT_beam2}
\mathcal{I}_{ik} \Bigl(\vec{p}_T, m,x,\mu,\frac{\nu}{\omega}\Bigr)
=  \sum_{j=q,\bar{q},g}\mathcal{I}_{ij}^{(n_l+1)} \Bigl(\vec{p}_T, x,\mu,\frac{\nu}{\omega}\Bigr) \otimes_x \mathcal{M}_{jk}(m,x,\mu)
\biggl[1+\mathcal{O}\Bigl(\frac{m^2}{p_T^2}\Bigr)\biggr]
\, .\end{align}
Similarly, the mass-dependent and massless soft function are related by
\begin{align}\label{eq:consistencyqT_soft2}
S (\vec{p}_T, m,\mu,\nu) =  S^{(n_l+1)} (\vec{p}_T, \mu,\nu)  \biggl[1+\mathcal{O}\Bigl(\frac{m^2}{p_T^2}\Bigr)\biggr]
\,,\end{align}
since there are no relevant soft IR fluctuations below the mass scale.
In the functions on the right-hand sides, which appear in \eq{factqT_m4}, logarithms $\ln(m/q_T)$ can be resummed to all orders in the limit $m \ll q_T$. This can be combined with the associated $\ord{m^2/q_T^2}$ power corrections relevant for $q_T \sim m$, by obtaining them from the fixed-order expansions of \eqs{consistencyqT_beam2}{consistencyqT_soft2} and adding them to the $(n_l+1)$-flavor beam function matching coefficients and soft function at the scale $\mu \sim q_T$.

By including the various power corrections, one combines the factorization theorems in the different hierarchies
and obtains a theoretical description that is valid across the whole $q_T$ spectrum and includes the resummation
of logarithms in all relevant limits. This can be considered a variable-flavor scheme for the resummed $q_T$ spectrum.
(In addition one should of course also include the usual $q_T/Q$ nonsingular corrections to reproduce the full
fixed-order result for $q_T\sim Q$.)

We stress that different specific ways of how to incorporate the various power corrections are formally equivalent
as long as the correct fixed-order expansion and the correct resummation is reproduced in each
limit. Any differences then amount to resummation effects at power-suppressed level and are thus beyond
the formal (leading-power) resummation accuracy.

A particular scheme (``S-ACOT'') to merge the $m \ll q_T$ and $q_T\sim m$ regimes was discussed in ref.~\cite{Nadolsky:2002jr} for the primary massive quark corrections.
In practice, for the numerical study of $b$-quark mass effects at low $q_T \ll m \ll Q$ the off-diagonal evolution factor $U_{f,bg}$ and thus the effective $b$-quark PDF at the scale $q_T$ are still quite small, so that one may effectively count $f_b(\mu_B) \sim \mathcal{O}(\alpha_s)$. In particular, this counting facilitates the seamless combination with the nonsingular corrections for $m\sim q_T$ encoded in the beam function matching coefficients in \eq{factqT_m3}.
This was discussed in ref.~\cite{Bonvini:2015pxa} in the context of the inclusive $b\bar{b} H$ production cross section, and the analogous discussion applies here as well.
In \refcites{Pietrulewicz:2014qza,Hoang:2015iva}, the power corrections were included implicitly in the construction of the variable-flavor number schemes for thrust in $e^+e^-$ and DIS in the endpoint region by applying different renormalization schemes for the massive quark contributions to the EFT operators above and below the mass scale.

\section{Factorization of mass effects for beam thrust}
\label{sec:Tau}

\subsection{Factorization for massless quarks}
\label{subsec:fact_massless_Tau}

For the measurement of beam thrust with $\Tau \ll Q$ the relevant EFT modes are $n_a$-collinear, $n_b$--collinear and usoft modes with the scaling
\begin{align}
n_a \text{-collinear:} 
&\quad
p_{n_a}^\mu \sim (\Tau,Q,\sqrt{Q\Tau})
\,, \nn \\
n_b \text{-collinear:} 
&\quad
p_{n_b}^\mu\sim (Q,\Tau,\sqrt{Q\Tau})
\, , \nn\\
\text{usoft:}
&\quad
p_{us}^\mu\sim (\Tau,\Tau,\Tau)
\,. 
\end{align}
The usoft and collinear modes are now separated in invariant mass, $p_{us}^2 \sim \Tau^2 \ll p_{n_a}^2 \sim p_{n_b}^2 \sim Q\Tau$, which is the characteristic feature of a \SCETa theory. In this case, there are no rapidity logarithms and the renormalization and evolution is solely in invariant mass. The resulting factorization formula reads~\cite{Stewart:2009yx}
\begin{align}\label{eq:factTau_ml}
\frac{\df \sigma}{\df Q^2 \,\df Y \,\df \Tau} &=\sum_{i,j \in \{q,\bar{q}\}}  H^{(n_f)}_{ij}(Q,\mu)  \int \df t_a \,\df t_b\,  B^{(n_f)}_i(t_a,x_a,\mu) \, B^{(n_f)}_{j}(t_b,x_b,\mu) 
\nn \\
& \quad  \times S^{(n_f)}\Bigl(\Tau- \frac{t_a}{\omega_a}- \frac{t_b}{\omega_b},\mu\Bigr)\,
\biggl[1+\mathcal{O}\Bigl(\frac{\Tau}{Q}\Bigr)\biggr]
\,.\end{align}
This as well as the expressions including mass effects in the subsequent subsections are valid for the primary hard scattering, and do not account for spectator forward (multiparton) scattering effects, since the Glauber Lagrangian of ref.~\cite{Rothstein:2016bsq} has been neglected. (There are also corrections from perturbative Glauber effects starting at ${\cal O}(\alpha_s^4)$~\cite{Gaunt:2014ska, Zeng:2015iba}, which are well beyond the order we are interested in, but can be calculated and included using the Glauber operator framework of ref.~\cite{Rothstein:2016bsq}.) This is sufficient for our purposes of discussing the mass effects in a prototypical \SCETa scenario. Our results are also directly relevant to include massive quark effects in the Geneva Monte-Carlo program~\cite{Alioli:2012fc, Alioli:2015toa}, which employs $\Tau$ as the jet resolution variable for the primary interaction and where multiparton effects are included~\cite{Alioli:2016wqt} via the combination with Pythia8 and its MPI model~\cite{Sjostrand:1987su, Sjostrand:2004pf, Sjostrand:2004ef}.

The hard function $H_{ij}$ in \eq{factTau_ml} is measurement independent and the same as in \eq{factqT_ml}. The beam and soft functions depend on the measurement and are different from those in \eq{factqT_ml}. The virtuality-dependent beam functions $B_i$ can be factorized into perturbative matching coefficients $\mathcal{I}_{ik}$ at the scale $\mu \sim t \sim \sqrt{Q\Tau}$ and the standard nonperturbative PDFs~\cite{Stewart:2009yx, Stewart:2010qs}
\begin{align}\label{eq:beamTau_ml}
B^{(n_f)}_i(t,x,\mu) = \sum_k \mathcal{I}^{(n_f)}_{ik}(t,x,\mu) \otimes_x f^{(n_f)}_k(x,\mu) \biggl[1+\mathcal{O}\biggl(\frac{\lqcd^2}{t}\biggr)\biggr]
\,.\end{align}
The matching coefficients $\mathcal{I}_{ik}$ have been calculated to $\mathcal{O}(\alpha_s^2)$~\cite{Gaunt:2014xga, Gaunt:2014cfa}. The soft function at the scale $\mu \sim \Tau$ is equivalent to the thrust soft function~\cite{Kang:2015moa}, which is known to $\mathcal{O}(\alpha_s^2)$~\cite{Kelley:2011ng, Monni:2011gb}. The noncusp anomalous dimensions required at N$^3$LL
are available from existing results~\cite{Stewart:2010qs}.

The resummation of logarithms $\ln(\Tau/Q)$ is performed by evaluating all functions at their characteristic scales and evolving them to a common final scale $\mu$ using the solutions of the RGEs
\begin{align}\label{eq:gammaBS_Tau}
\mu \frac{\df}{\df \mu} B^{(n_f)}_i(t,x,\mu) & =\int \df t' \, \gamma_{B}^{(n_f)}(t-t',\mu) \, B^{(n_f)}_i(t',x,\mu) \, ,\nn \\
\mu \frac{\df}{\df \mu} S^{(n_f)}(\ell,\mu) & =  \int \df \ell' \, \gamma_{S}^{(n_f)}(\ell-\ell',\mu) \, S^{(n_f)}(\ell',\mu)\, .
\end{align}
In contrast to \eq{factqT_ml}, there is no rapidity evolution in \SCETa for massless quarks. Consistency of the RG evolution implies that
\begin{align}\label{eq:consistency_Tau}
\omega_a\gamma_{B}^{(n_f)}(\omega_a \ell ,\mu) + \omega_b\gamma_{B}^{(n_f)}(\omega_b \ell ,\mu) + \gamma_{S}^{(n_f)}(\ell ,\mu)= \gamma_H^{(n_f)} (Q,\mu) \,\delta(\ell)\, .
\end{align}

\begin{figure}[t]
\subfigure[${\Tau \ll m \sim\sqrt{Q\Tau}}$]{\includegraphics[scale=0.25]{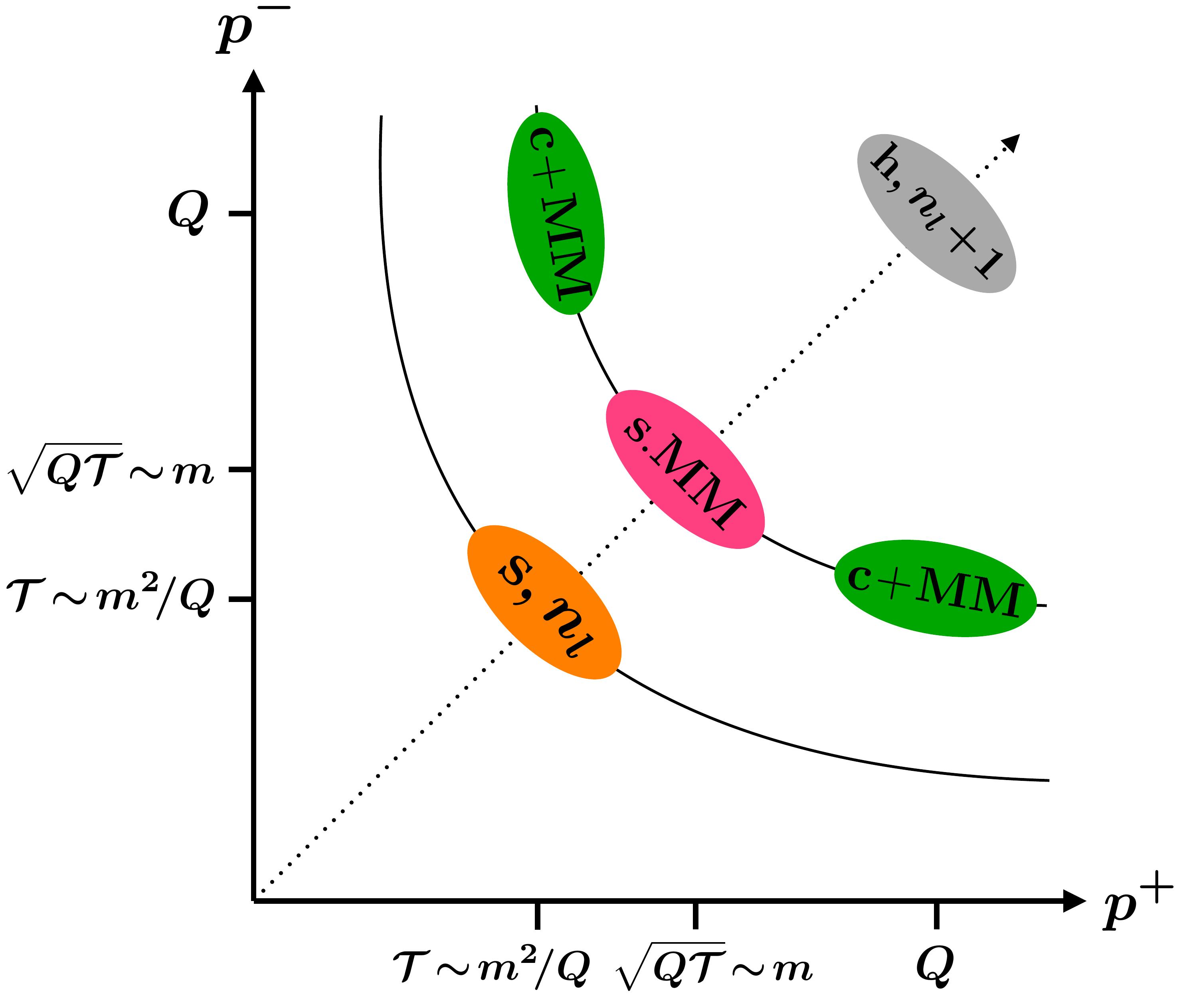}\label{fig:modes_Tau1}}%
\hfill%
\subfigure[${\Tau \ll m \ll \sqrt{Q\Tau}}$]{\includegraphics[scale=0.25]{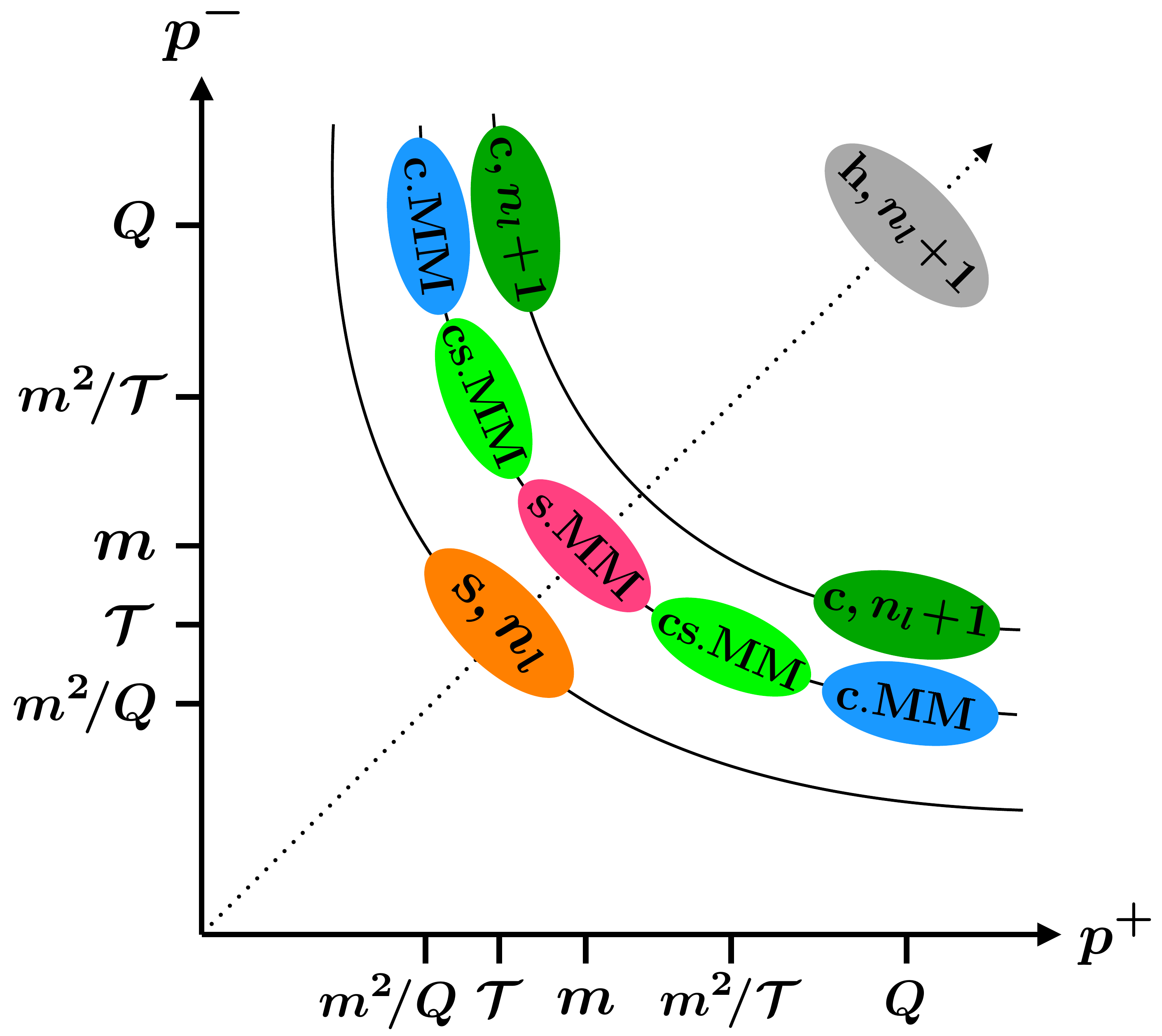}\label{fig:modes_Tau2}}%
\\
\subfigure[${\Tau \sim m \ll \sqrt{Q\Tau}}$]{\includegraphics[scale=0.25]{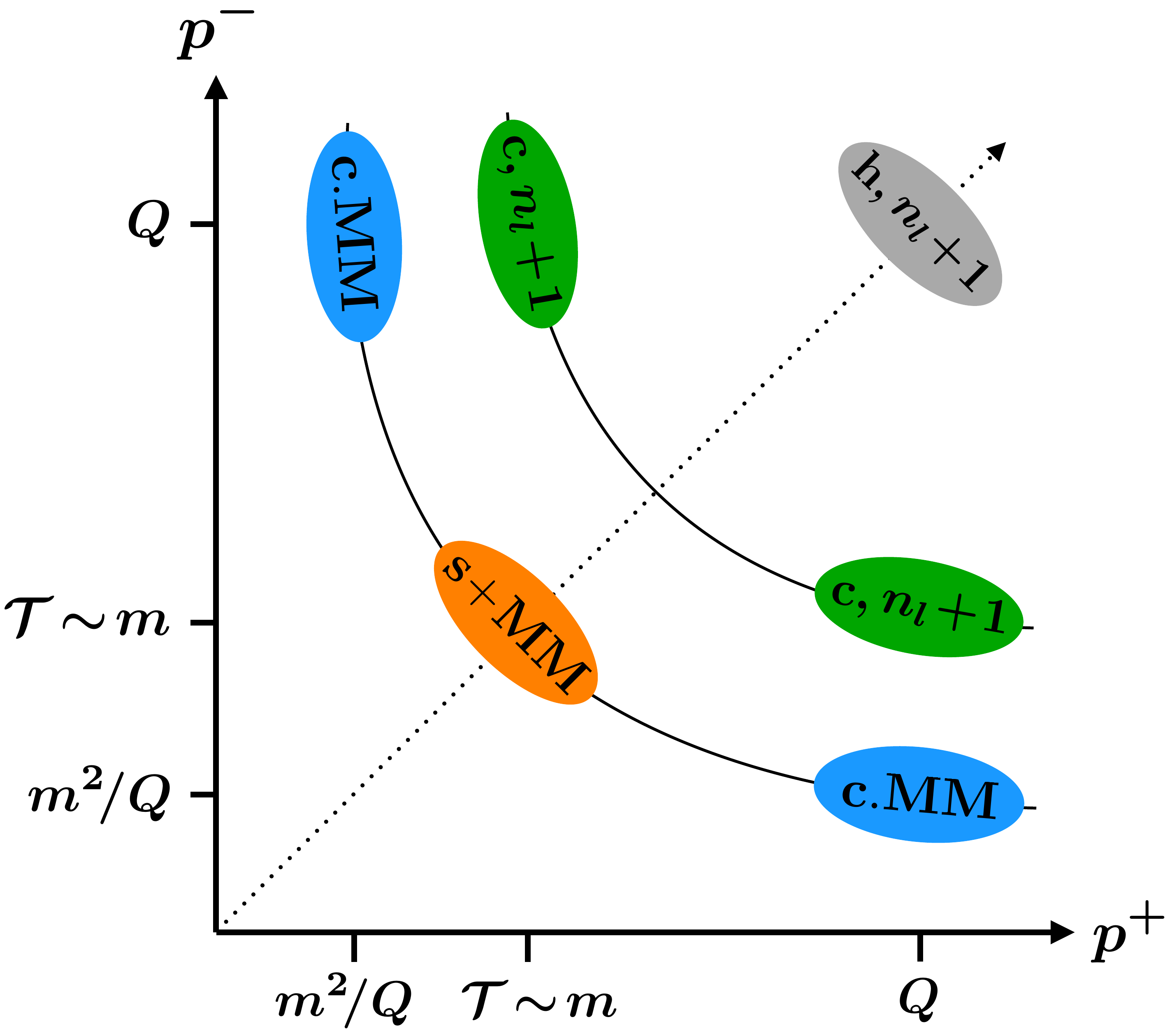}\label{fig:modes_Tau3}}%
\hfill%
\subfigure[${m \ll \Tau \ll \sqrt{Q\Tau}}$]{\includegraphics[scale=0.25]{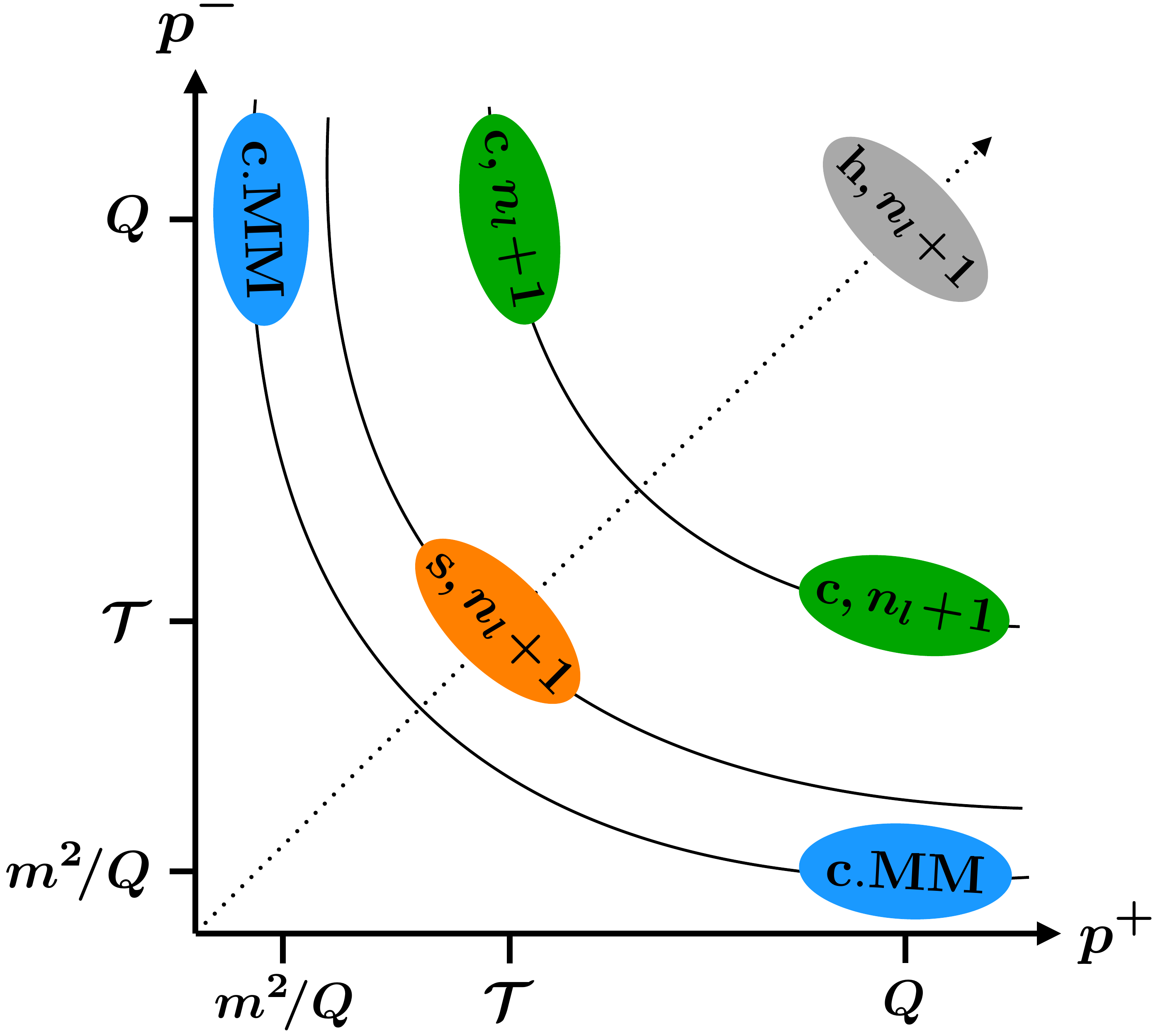}\label{fig:modes_Tau4}}%
\caption{Effective theory modes for the beam thrust spectrum with massive quarks for $m^2/Q \lesssim \Tau \ll Q$ and $m \gg \Lambda_{\rm QCD}$.}
\label{fig:modes_Tau}
\end{figure}

For beam thrust the number of possible scale hierarchies with a massive quark is larger due to the fact that the (massless) collinear and soft modes have different invariant mass scales. The discussion for the hierarchies with $\sqrt{Q\Tau} \ll m$ where the massive quark cannot be produced via real emissions, is completely identical to $q_T \ll m$, since the quark mass effects in these cases are independent of the low-energy measurement.
For $m\sim Q$, all mass effects are encoded by using the mass-dependent hard function from \subsec{fact_qT1} in \eq{factTau_ml} together with $n_f = n_l$ everywhere else.
Similarly, the case $\sqrt{Q\Tau} \ll m \ll Q$ is described by using \eq{factTau_ml} with $n_f = n_l$, and replacing the hard function by the product of massless $(n_l+1)$-flavor hard function and the soft and collinear mass-mode functions $H_s$ and $H_c$, as for the case $q_T \ll m \ll Q$ in \subsec{fact_qT2}.
We therefore proceed directly to the hierarchies $m \lesssim \sqrt{Q \Tau}$, where the massive quark can be produced in collinear and/or soft real radiation. The four possible hierarchies and the relevant EFT modes in the $p^+ p^-$-plane are illustrated in \fig{modes_Tau}, and are discussed in the following subsections.

\subsection{Quark mass effects for $\sqrt{Q \Tau}\sim m \ll Q$}
\label{subsec:fact_Tau1}

For $\sqrt{Q \Tau}\sim m \ll Q$ massive quarks can be produced via collinear initial-state radiation, but not via soft real radiation. After the hard matching, carried out with $n_l+1$ massless quark flavors as discussed in \subsec{fact_qT2}, the degrees of freedom in the EFT are collinear and soft modes with the scaling
\begin{align}
n_a \text{-collinear + MM:} 
&\quad
p_{n_a}^\mu \sim (\Tau,Q,\sqrt{Q\Tau})  \sim \Bigl(\frac{m^2}{Q},Q,m\Bigr)
\,, \nn \\
n_b \text{-collinear + MM:} 
&\quad
p_{n_b}^\mu\sim (Q,\Tau,\sqrt{Q\Tau})  \sim \Bigl(Q,\frac{m^2}{Q},m\Bigr)
\, , \nn\\
\text{soft MM:} 
&\quad
p_{s}^\mu\sim (m,m,m)
\, , \nn\\
\text{usoft:}
&\quad
p_{us}^\mu\sim (\Tau,\Tau,\Tau)
\,,\end{align}
as illustrated in \fig{modes_Tau1}.
While the usual usoft modes live at a lower virtuality scale than the collinear modes,
the soft mass-modes are separated from the collinear modes only in rapidity, leading to
a mix of \SCETa and \SCETb features. In particular, there will be mass-related rapidity
divergences.

At the scale $\mu \sim \sqrt{Q\Tau} \sim m$ this theory with $n_l+1$ dynamical quark flavors is matched onto a theory with $n_l$ flavors integrating out also fluctuations related to initial-state collinear radiation of massless particles. The matching in the collinear sectors leads to mass-dependent beam function coefficients $\mathcal{I}_{ik}$,
\begin{align}\label{eq:Bm1_Tau}
B_i^{(n_l+1)}\Bigl(t,m,x,\mu,\frac{\nu}{\omega}\Bigr)
= \sum_{k \in \{q,\bar{q},g\}} \mathcal{I}_{ik}\Bigl(t,m,x,\mu,\frac{\nu}{\omega}\Bigr) \otimes_x f^{(n_l)}_k (x,\mu)
\biggl[1+\mathcal{O}\Bigl(\frac{\lqcd^2}{m^2},\frac{\lqcd^2}{t}\Bigr)\biggr]
\,,\end{align}
analogous to \eq{Bm1_qT}.
The dependence on the rapidity scale $\nu$ here arises due to virtual secondary massive quark corrections
and is the same as for the collinear mass-mode function $H_c$ in \eq{factqT_m2}, i.e.,
\begin{align}\label{eq:BTau_nuevolution}
\nu\frac{\df}{\df \nu} B_i^{(n_l+1)}\Bigl(t,m,x,\mu,\frac{\nu}{\omega}\Bigr)
= \gamma_{\nu, H_c}(m,\mu)\, B_i^{(n_l+1)}\Bigl(t,m,x,\mu,\frac{\nu}{\omega}\Bigr)\, .
\end{align}
In the soft sector the soft mass modes are integrated out, leaving only the usoft modes. This gives exactly the soft mass-mode function $H_s$ in \eq{factqT_m2}, which encodes the effects of virtual secondary massive quark radiation. As usual, also the strong coupling constant has to be matched from $n_l+1$ to $n_l$ flavors. The remaining contributions at the lower scales, the soft function and the PDFs, are given in terms of $n_l$ massless flavors and in the $(n_l)$-scheme for $\alpha_s$.
The resulting factorized cross section reads
\begin{align}\label{eq:factTau_m1}
\frac{\df \sigma}{\df Q^2 \,\df Y \,\df \Tau} &=\sum_{i,j \in \{q,\bar{q},Q,\bar{Q}\}}  H^{(n_l+1)}_{ij}(Q,\mu)  \,H_{s}(m,\mu,\nu)
\int\! \df t_a \, \df t_b \,
\nn \\ & \quad \times
\biggl[\sum_{k \in \{q,\bar{q},g\}} \mathcal{I}_{ik}\Bigl(t_a,m,x_a,\mu,\frac{\nu}{\omega_a}\Bigr) \otimes_x f^{(n_l)}_k (x_a,\mu)\biggr]
\nn \\ & \quad  \times
\biggl[\sum_{k \in \{q,\bar{q},g\}} \mathcal{I}_{jk}\Bigl(t_b,m,x_b,\mu,\frac{\nu}{\omega_a}\Bigr) \otimes_x f^{(n_l)}_k (x_b,\mu)\biggr]
\nn \\ & \quad  \times
S^{(n_l)}\Bigl(\Tau- \frac{t_a}{\omega_a}- \frac{t_b}{\omega_b},\mu\Bigr)\,
\biggl[1+\mathcal{O}\Bigl(\frac{\Tau}{Q},\frac{m^2}{Q^2},\frac{\Tau^2}{m^2},\frac{\lqcd}{\Tau}\Bigr)\biggr]
\,.\end{align}

\begin{figure}[t]
\subfigure[${\Tau \ll m \sim\sqrt{Q \Tau}}$]{\includegraphics[scale=0.25]{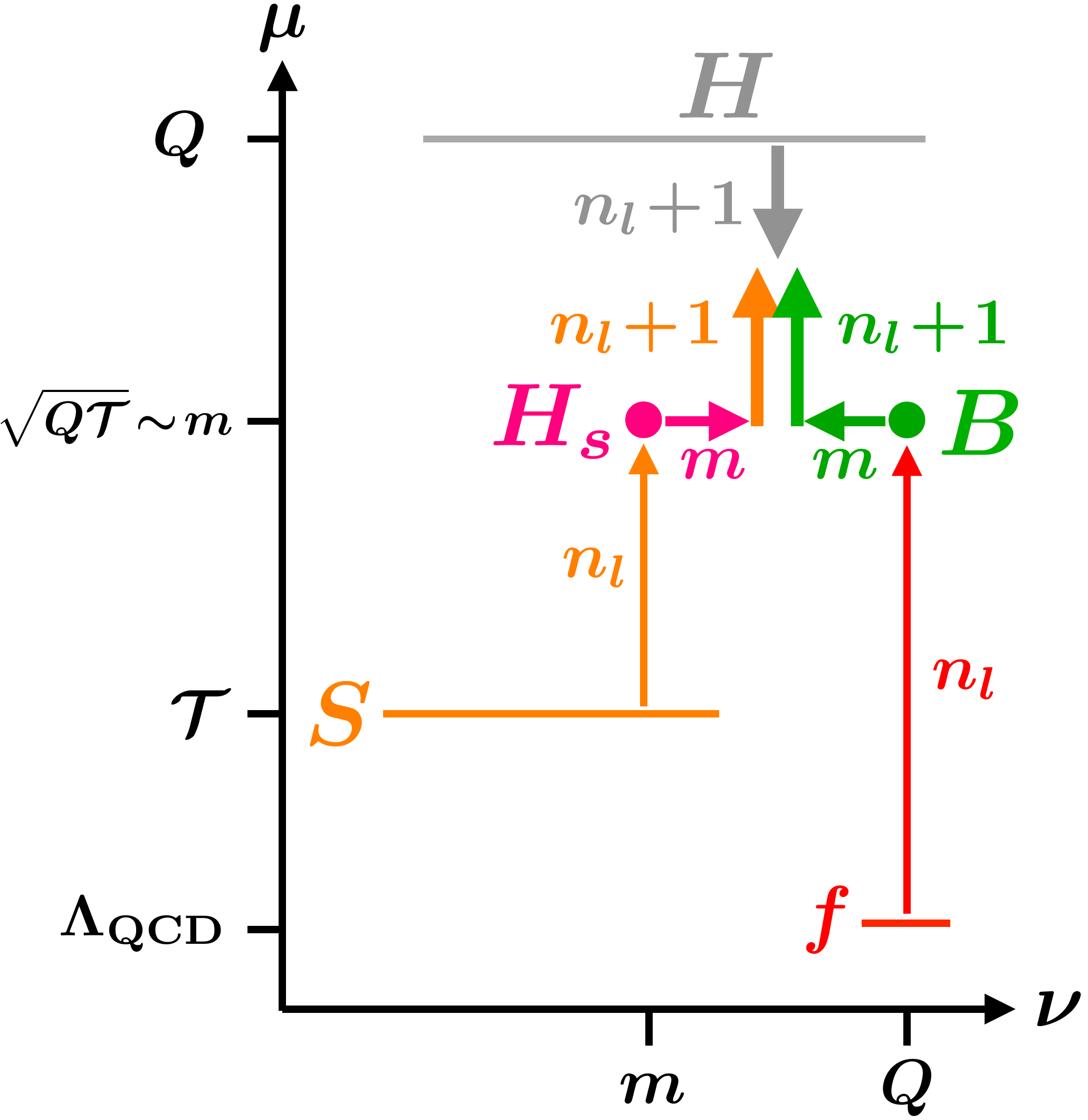}\label{fig:evolution_Tau1}}%
\hfill%
\subfigure[${\Tau \ll m \ll\sqrt{Q \Tau}}$]{\includegraphics[scale=0.25]{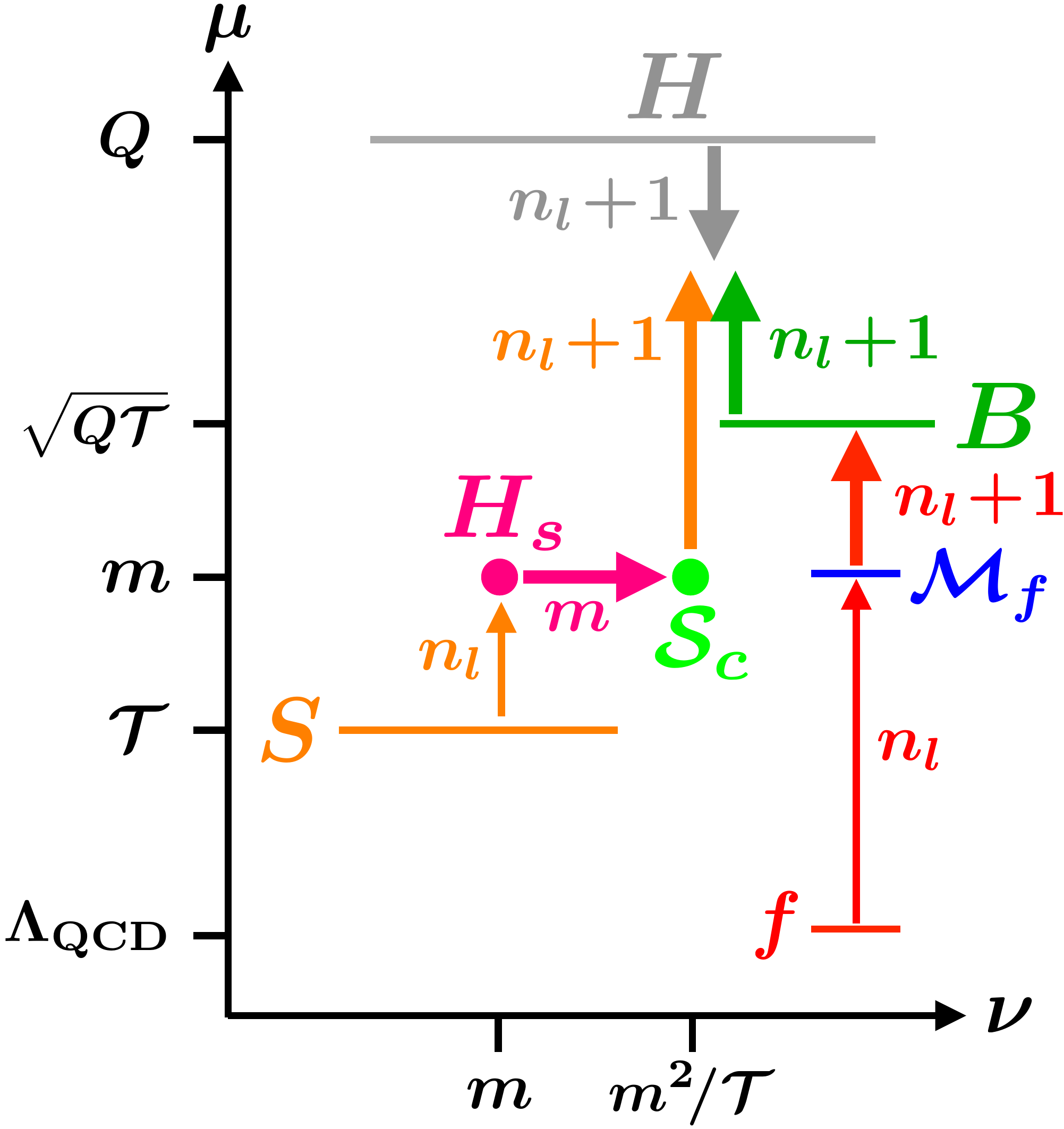}\label{fig:evolution_Tau2}}%
\hfill%
\subfigure[${\Tau \sim m \ll \sqrt{Q \Tau}}$]{\includegraphics[scale=0.25]{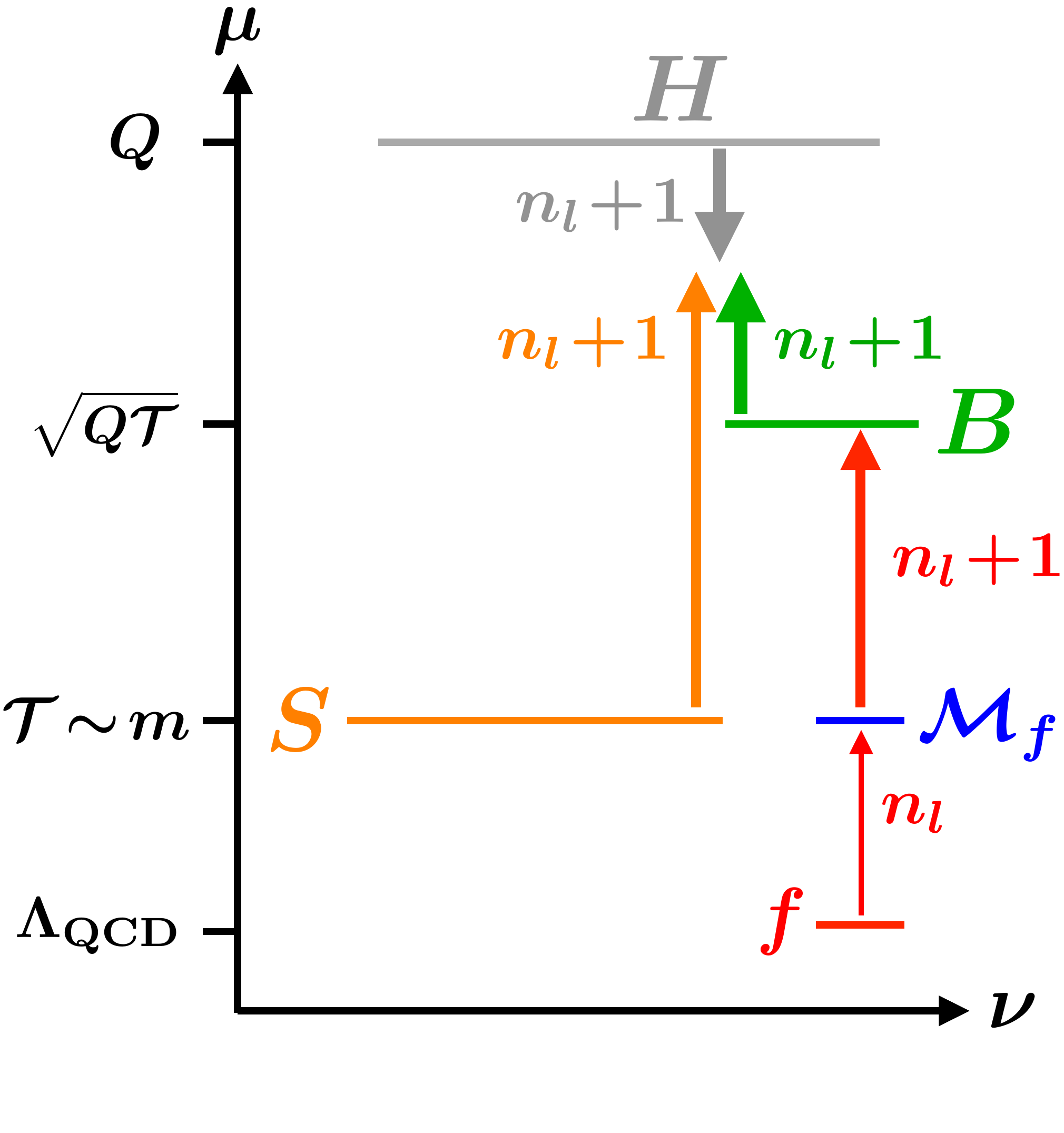}\label{fig:evolution_Tau3}}%
\caption{Illustration of the renormalization group evolution for beam thrust of the hard, beam, soft, and parton distribution function in invariant mass and rapidity. The anomalous dimensions for each evolution step involve the displayed number of active quark flavors. The label $m$ indicates that the corresponding evolution is mass dependent.}
\label{fig:evolution_Tau}
\end{figure}

The resummation of logarithms in \eq{factTau_m1} is obtained by evolving all functions from their natural scales, as illustrated
in \fig{evolution_Tau1}.
The mass-dependent $\nu$ evolution, which resums the rapidity logarithms $\ln(Q/m)$, is identical to the one for the hard functions $H_c$ and $H_s$ in \subsec{fact_qT2}. The $\mu$ evolution can be conveniently carried out by evolving the hard, beam, and soft functions with $n_l+1$ active flavors above the mass scale and with $n_l$ active flavors below the mass scale, which automatically takes into account the $\mu$ dependence of $H_S$. To see this, the consistency of RG running for \eq{factTau_m1} together with the consistency relation for $n_l+1$ massless quarks in \eq{consistency_Tau} implies
\begin{align}\label{eq:consistency_Tau1}
&\omega_a\gamma_{B,m}^{(n_l+1)}\Bigl(\omega_a \ell,m,\mu,\frac{\nu}{\omega_a}\Bigr)+\omega_b\gamma_{B,m}^{(n_l+1)}\Bigl(\omega_b\ell,m,\mu,\frac{\nu}{\omega_b}\Bigr)+\gamma_{S}^{(n_l)}(\ell,\mu)+\gamma_{H_s,\mu}(m,\mu,\nu)\,\delta(\ell)
\nn \\ &\qquad
=\omega_a \gamma_{B}^{(n_l+1)}(\omega_a \ell,\mu)+ \omega_b \gamma_{B}^{(n_l+1)}(\omega_b \ell,\mu)+\gamma_{S}^{(n_l+1)}(\ell,\mu)  \, ,
\end{align}
where $\gamma_{S}^{(n_l)}$, $\gamma_{S}^{(n_l+1)}$, $\gamma_{B}^{(n_l+1)}$ are the anomalous dimensions for the soft and beam functions with $n_l$ and $n_l+1$ massless flavors as defined in \eq{gammaBS_Tau}, and $\gamma_{B,m}^{(n_l+1)}(t,m,\mu,\nu/\omega)$ is the anomalous dimension of the mass-dependent beam function,
\begin{align}\label{eq:BTau_muevolution}
\mu\frac{\df}{\df \mu} B_i^{(n_l+1)}\Bigl(t,m,x,\mu,\frac{\nu}{\omega}\Bigr) = \int \df t' \,\gamma_{B,m}^{(n_l+1)}\Bigl(t-t',m,\mu,\frac{\nu}{\omega}\Bigr)\, B_i^{(n_l+1)}\Bigl(t',m,x,\mu,\frac{\nu}{\omega}\Bigr)\, .
\end{align}
The consistency relation in \eq{consistency_Tau1} can be confirmed explicitly at two loops with the expressions in eqs.~(\ref{eq:gammaB_thrust}), (\ref{eq:gamma_soft}), (\ref{eq:gammaHs}), and (\ref{eq:gammaBm_mu}). Note that this relation does {\it not} imply that $\gamma_{B,m}^{(n_l+1)}(t,m,\mu,\nu/\omega)$ and $\gamma_{B}^{(n_l+1)}(t,\mu)$ are the same, which is indeed not the case for the  massive quark corrections as we will see explicitly in \sec{beamfct_results}. The reason is that the presence of the quark mass leads to a \SCETb-type theory, in which the required rapidity regularization redistributes the $\mu$ anomalous dimension between soft and collinear corrections with individually regularization scheme dependent pieces. Only their sum, as given on the left-hand side of \eq{consistency_Tau1}, is independent of the regularization scheme and yields the combined running for beam and soft functions with $n_l+1$ massless flavors above $\mu_m \sim m$, as on the right-hand side of \eq{consistency_Tau1}.
 
\subsection{Quark mass effects for $\Tau \ll m \ll \sqrt{Q \Tau}$}
\label{subsec:fact_Tau2}

When the beam scale becomes larger than the mass scale, but the soft scale is still larger than the mass, which happens for $m^2/Q \ll \Tau \ll m$, the beam function matching coefficients $\mathcal{I}_{ik}$ encode only fluctuations related to initial-state collinear radiation with $n_l+1$ massless quarks. The EFT below $\sqrt{Q\Tau}$ contains the usual collinear and soft mass modes scaling as $p_{m,n_a}^\mu \sim (m^2/Q,Q,m)$, $p_{m,n_b}^\mu \sim (Q,m^2/Q,m)$, and $p_{m,s}^\mu \sim (m,m,m)$, which do not contribute to the beam thrust measurement. However, besides these there are also additional modes with fluctuations around the mass scale which can have a dynamic impact on the $\Tau$ spectrum in this hierarchy, as illustrated in \fig{modes_Tau2}. Their scaling is precisely determined by this condition and the on-shell constraint, yielding the scaling
\begin{align}\label{eq:csoft_scaling}
n_a \text{-csoft MM:} 
&\quad
p_{cs,n_a}^\mu \sim \Bigl(\Tau,\frac{m^2}{\Tau},m\Bigr) 
\,, \nn \\
n_b \text{-csoft MM:} 
&\quad
p_{cs,n_b}^\mu\sim \Bigl(\frac{m^2}{\Tau},\Tau,m\Bigr) 
\,.\end{align}
We refer to these intermediate modes as collinear-soft (csoft), since they are simultaneously boosted (by a factor $m/\Tau$) but are softer than the standard collinear modes, thus coupling to the latter via Wilson lines and leading to a \SCETp theory~\cite{Bauer:2011uc}. This type of intermediate \SCETp modes have appeared in various contexts~\cite{Bauer:2011uc, Procura:2014cba, Larkoski:2015zka, Pietrulewicz:2016nwo}.
The setup here is similar to the case of double-differential distributions with a simultaneous $q_T$ and beam thrust measurement discussed in \refcite{Procura:2014cba}. Also there, several hierarchies are possible ranging from a \SCETb regime for $q_T \sim \Tau$ to a \SCETa regime for $q_T \sim \sqrt{Q\Tau}$ with a \SCETp regime in between. The csoft modes in their \SCETp regime are separated from the collinear modes only in rapidity. In our case here, the csoft mass modes are separated in invariant mass from the standard \SCETa soft and collinear modes and in rapidity from their \SCETb-type soft mass-mode cousins.

The matching in the collinear sector can be performed in two steps as in \eqs{beam_matching1}{PDF_matching_function}. After integrating out all of the mass modes, the PDF and the soft function are still given in a $(n_l)$-flavor theory. Thus the factorization formula reads
\begin{align}\label{eq:factTau_m2}
\frac{\df \sigma}{\df Q^2 \,\df Y \,\df \Tau}
&=\sum_{i,j \in \{q,\bar{q},Q,\bar{Q}\}}  H^{(n_l+1)}_{ij}(Q,\mu)  \,H_{s}(m,\mu,\nu) \int\!\df k_a \, \df k_b \,\mathcal{S}_c(k_a,m,\mu,\nu) \,\mathcal{S}_c(k_b,m,\mu,\nu)
\nn \\ & \quad \times
\int\!\df t_a \,
\biggl[\sum_{k \in \{q,\bar{q},Q,\bar{Q},g\}} \sum_{l \in \{q,\bar{q},g\}} \mathcal{I}^{(n_l+1)}_{ik}(t_a,x_a,\mu) \otimes_x \mathcal{M}_{kl} (x_a,m,\mu)\otimes_x f^{(n_l)}_k (x_a,\mu)\biggr]
\nn \\ & \quad \times
\int\! \df t_b \,
\biggl[\sum_{k \in \{q,\bar{q},Q,\bar{Q},g\}} \sum_{l \in \{q,\bar{q},g\}} \mathcal{I}^{(n_l+1)}_{jk}(t_b,x_b,\mu) \otimes_x \mathcal{M}_{kl} (x_b,m,\mu)\otimes_x f^{(n_l)}_n (x_b,\mu)\biggr]
\nn \\ & \quad  \times
S^{(n_l)}\Bigl(\Tau- \frac{t_a}{\omega_a}- \frac{t_b}{\omega_b}-k_a-k_b,\mu\Bigr) \Bigl[1+\mathcal{O}\Bigl(\frac{\Tau}{Q},\frac{m^2}{Q \Tau},\frac{\Tau^2}{m^2},\frac{\lqcd}{\Tau}\Bigr)\Bigr]
\,.\end{align}
The functions $\mathcal{S}_c$ here are the csoft matching functions encoding the interactions of the collinear-soft radiation at the invariant mass scale $\mu \sim m$ and the rapidity scale $\nu \sim m^2/\Tau$. The $\mathcal{M}_{ij}$ correspond to the well-known PDF matching correction incorporating the effect of the collinear mass modes, as in \eq{factqT_m4}. The virtual soft massive quark corrections are still described by the function $H_s$ at the rapidity scale $\nu \sim m$ as in \eq{factTau_m1}.

The RG evolution for \eq{factTau_m2} is illustrated in \fig{evolution_Tau2}.
The csoft function satisfies the same rapidity RGE as the collinear mass-mode function $H_c$ in \eq{factqT_m2} and the massive beam functions in \eq{BTau_nuevolution}, i.e.,
\begin{align}\label{eq:Csoft_nuevolution}
\nu\frac{\df}{\df \nu} \mathcal{S}_c(k,m,\mu,\nu) = \gamma_{\nu, H_c}(m,\mu) \, \mathcal{S}_c(k,m,\mu,\nu) \, .
\end{align}
The only difference with respect to the rapidity evolution in \eq{factTau_m1} is that it now happens between $H_s$ and $\mathcal{S}_c$ with $\nu_{\mathcal{S}_c} \sim m^2/\Tau$ rather than between $H_s$ and the beam functions with $\nu_B \sim Q$, such that now the (smaller) rapidity logarithms $\ln(m/\Tau)$ are resummed. The $\mu$ evolution can be performed with $n_l+1$ flavors for the hard function $H_{ij}$, the beam and soft function above the mass scale and with $n_l$ flavors below. This automatically accounts for the $\mu$ dependence of $\mathcal{S}_c$ and $H_s$ above $\mu_m \sim m$, which precisely gives the difference between the evolution of the soft function with $n_l+1$ and $n_l$ flavors, as implied by the consistency of RG running for \eq{factTau_m2} and the relation in \eq{consistency_Tau} with $n_l+1$ massless quarks,
\begin{align}\label{eq:consistency_Tau2}
 \gamma_{S}^{(n_l)}(\ell,\mu)+2\gamma_{\mathcal{S}_c}(\ell,m,\mu,\nu)+\delta(\ell)\,\gamma_{H_s}(m,\mu,\nu) =\gamma_{S}^{(n_l+1)}(\ell,\mu)\, ,
\end{align}
where 
\begin{align}\label{eq:Csoft_muevolution}
\mu\frac{\df}{\df \mu} \mathcal{S}_c(k,m,\mu,\nu) = \int \df k'\, \gamma_{\mathcal{S}_c}(k-k',m,\mu,\nu) \, \mathcal{S}_c(k,m,\mu,\nu) \, .
\end{align}
At two loops, the consistency relation \eq{consistency_Tau2} can be explicitly confirmed with the expressions in eqs.~(\ref{eq:gamma_soft}), (\ref{eq:gamma_csoft}), and (\ref{eq:gammaHs}).

\subsection{Quark mass effects for $\Tau \sim m$ and $m \ll \Tau$}
\label{subsec:fact_Tau3}

For $\Tau\sim m$ the csoft and soft mass modes in the previous section merge with the usual usoft modes,
\begin{align}
\text{usoft:}
&\quad
p_{s}^\mu\sim (\Tau,\Tau,\Tau) \sim \Bigl(\Tau,\frac{m^2}{\Tau},m\Bigr) \sim (m,m,m)
\, .\end{align}
In this hierarchy massive quarks can be also produced in soft real radiation leading to a soft function at the scale $\mu \sim \Tau$ that depends on the quark mass.
In addition, there are the usual collinear modes as well as the collinear mass modes,
as illustrated in \fig{modes_Tau3}. Since we still have $m \ll \sqrt{Q\Tau}$, the matching
in the collinear sectors is the same as in the previous subsection.
The factorization formula reads
\begin{align}\label{eq:factTau_m3}
\frac{\df \sigma}{\df Q^2 \,\df Y \,\df \Tau} &=\sum_{i,j \in \{q,\bar{q},Q,\bar{Q}\}}  H^{(n_l+1)}_{ij}(Q,\mu)
\int\! \df t_a \, \df t_b
\nn \\ & \quad \times
\biggl[\sum_{k \in \{q,\bar{q},Q,\bar{Q},g\}} \sum_{l \in \{q,\bar{q},g\}} \mathcal{I}^{(n_l+1)}_{ik}(t_a,x_a,\mu) \otimes_x \mathcal{M}_{kl} (x_a,m,\mu)\otimes_x f^{(n_l)}_k (x_a,\mu)\biggr]
\nn \\ & \quad \times
\biggl[\sum_{k \in \{q,\bar{q},Q,\bar{Q},g\}} \sum_{l \in \{q,\bar{q},g\}} \mathcal{I}^{(n_l+1)}_{jk}(t_b,x_b,\mu) \otimes_x \mathcal{M}_{kl} (x_b,m,\mu)\otimes_x f^{(n_l)}_k (x_b,\mu)\biggr]
\nn \\ & \quad  \times
S\Bigl(\Tau- \frac{t_a}{\omega_a}- \frac{t_b}{\omega_b},m,\mu\Bigr)
\biggl[1+\mathcal{O}\Bigl(\frac{\Tau}{Q},\frac{m^2}{Q \Tau},\frac{\lqcd}{\Tau},\frac{\lqcd^2}{m^2}\Bigr)\biggr]
\,.\end{align}
Now all rapidity divergences cancel within the soft function and do not leave behind any potentially large rapidity logarithms.
The RG evolution for this case is illustrated in \fig{evolution_Tau3}.

Finally, for $m \ll \Tau$ the mass dependence in the IR insensitive soft function vanishes, if expressed in terms of the $(n_l+1)$-flavor scheme for $\alpha_s$. Otherwise, \eq{factTau_m3} remains unchanged, such that now the only dependence on the mass scale arises in the PDF matching corrections $\mathcal{M}_{ij}$. The hard, beam, and soft functions can now be always evolved with $n_l+1$ massless flavors and only the evolution of the PDF changes, when crossing the flavor threshold.

\subsection{Relations between hierarchies}
\label{subsec:hierarchies_Tau}

\begin{figure}
\centering
\includegraphics[width=\textwidth]{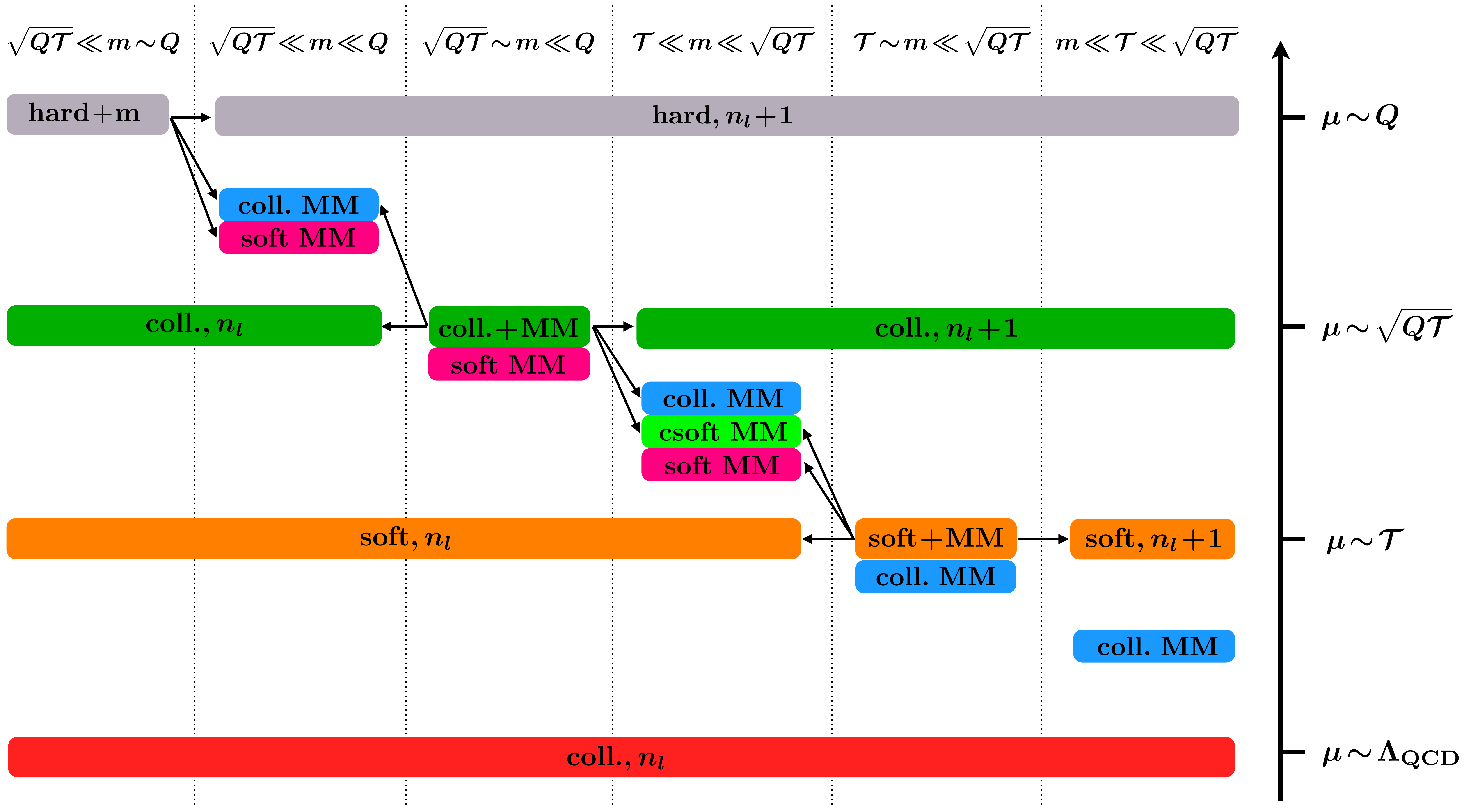}
\caption{Relevant modes for the beam thrust spectrum with $\Tau \ll Q$ for different hierarchies between the quark mass $m$ and the scales $\Tau$, $\sqrt{Q\Tau}$ and $Q$. The arrows indicate the relations between the modes and their associated contributions.}
\label{fig:Scales_Tau}
\end{figure}

We now discuss how the ingredients appearing in the different factorization formulae are related to each other. The relations between the modes and their contributions are illustrated in \fig{Scales_Tau} for the different possible hierarchies.
As in \subsec{qT_hierarchies}, these relations show how one can combine the resummation of logarithms relevant in one regime with the power-suppressed fixed-order content that becomes important in the neighboring regimes, enabling a systematic inclusion of mass corrections across the entire $\Tau$ spectrum.

Similar to \eq{consistencyqT_beam1}, the mass-dependent beam function coefficients appearing for $\sqrt{Q\Tau}\sim m$ (incorporating massive quark fluctuation as discussed in \subsec{fact_Tau1}) are related to those for $\sqrt{Q\Tau} \ll m$ with $n_l$ massless quarks and the collinear mass-mode function $H_c$ by
\begin{align}\label{eq:consistencyTau_beam1}
\mathcal{I}_{ik} \Bigl(t, m, x, \mu,\frac{\nu}{\omega}\Bigr) =  H_{c}\Bigl(m,\mu,\frac{\nu}{\omega}\Bigr)\,\mathcal{I}_{ik}^{(n_l)} (t, x,\mu) \Bigl[1+\mathcal{O}\Bigl(\frac{t}{m^2}\Bigr)\Bigr]
\, .\end{align}
At the same time, the mass-dependent beam function also encodes information about the fixed-order content for $\Tau \ll m \ll \sqrt{Q\Tau}$. Comparing \eqs{factTau_m1}{factTau_m2}, they are related to those with $n_l+1$ massless flavors, the PDF matching functions, and the csoft function $\mathcal{S}_c$ by
\begin{align}\label{eq:consistencyTau_beam2}
\mathcal{I}_{ik}\Bigl(t, m, x,\mu,\frac{\nu}{\omega}\Bigr)
&= \!\!\sum_{j=\{q,\bar{q},Q,\bar Q,g\}} \int\! \df \ell\,
\mathcal{I}_{ij}^{(n_l+1)} (t-Q\ell, x,\mu) \otimes_x \mathcal{M}_{jk}(x,m,\mu) \, \mathcal{S}_c(\ell, m,\mu,\nu)
\nn \\& \quad \times
\biggl[1+\mathcal{O}\Bigl(\frac{m^2}{Q\Tau}\Bigr)\biggr]
\, .\end{align}

The mass-dependent soft function for $\Tau \sim m$ in \eq{factTau_m3} contains massive quark fluctuations that for $\Tau \ll m$ get split into the massless soft function with $n_l$ flavors, the soft mass mode function $H_s$, and the csoft function $\mathcal{S}_c$ in \eq{factTau_m2} as
\begin{align}\label{eq:consistencyTau_soft1}
S(\ell, m,\mu) = H_s(m,\mu,\nu)\int\! \df \ell'\, S^{(n_l)} (\ell-\ell', \mu)\, \mathcal{S}_c (\ell',m, \mu,\nu) \Bigl[1+\mathcal{O}\Bigl(\frac{\ell^2}{m^2}\Bigr)\Bigr]
\, .\end{align}
Finally, as already mentioned below \eq{factTau_m3}, the soft function approaches its massless limit for $m \ll \Tau$,
\begin{align}\label{eq:consistencyTau_soft2}
S(\ell, m,\mu) =S^{(n_l+1)} (\ell, \mu)  \Bigl[1+\mathcal{O}\Bigl(\frac{m^2}{\ell^2}\Bigr)\Bigr]
\, .\end{align}

\subsection{Relation to previous literature}

Here, we briefly comment on the connection of the factorization setup presented here for beam thrust to the closely related \SCETa setup in \refcites{Gritschacher:2013pha,Pietrulewicz:2014qza} for thrust in $e^+ e^-$-collisions (or similarly also for DIS with $x \to 1$ \cite{Hoang:2015iva}). Besides the fact that the jet functions appearing for thrust in $e^+e^-$ are replaced by virtuality-dependent beam functions for beam thrust in $pp$ collisions, there are also some differences in the description of the different regimes. While we have discussed each possible hierarchy in a strict EFT sense identifying a single operator matrix element or matching function with each EFT mode, refs.~\cite{Gritschacher:2013pha,Pietrulewicz:2014qza} already set up their factorization theorems in a way that they apply for neighboring hierarchies (e.g.~$\Tau \ll m\sim \sqrt{Q\Tau}$ and $\Tau\ll m \ll \sqrt{Q\Tau}$). Using appropriate renormalization conditions, the mass dependent corrections to the jet and soft functions were assigned such that they directly give the massless results in the small mass limit and decouple in the infinite mass limit. In addition, the factorization theorems contained mass mode matching functions for hard, jet, and soft function, whenever the evolution of one of the matrix elements crossed the mass scale. In our setup this essentially amounts to a specific practical choice how to incorporate the power corrections in eqs.~(\ref{eq:consistencyTau_beam1})--(\ref{eq:consistencyTau_soft2}). Although the final outcome is thus essentially the same once the correct rapidity scales are chosen in the mass mode matching functions, it is perhaps more transparent conceptionally to first distinguish all hierarchies with the associated modes as we do here, and separately discuss the possible ways to add the nonsingular corrections later. In particular, for the hierarchy $\Tau\ll m \ll \sqrt{Q\Tau}$, this leads us to identify the csoft modes as a relevant degree of freedom with a corresponding function evaluated naturally at the rapidity scale $\nu \sim m^2/\Tau$. In contrast, refs.~\cite{Gritschacher:2013pha,Pietrulewicz:2014qza} the corresponding corrections appeared inside the mass mode matching functions as soft-bin contributions that had to be evaluated at this rapidity scale to minimize large rapidity logarithms.

\section{Results for massive quark corrections}
\label{sec:results}

In this section we present our results for the contributions from primary massive quarks at $\mathcal{O}(\alpha_s)$ and from secondary massive quarks at $\mathcal{O}(\alpha_s^2)$ to all components of the various factorization theorems discussed in \secs{qT}{Tau}, providing all required ingredients for the Drell-Yan spectrum at NNLL$'$. The results in this section are only given for a single massive quark flavor and with the  rapidity divergences regularized by the symmetric Wilson line regulator introduced in refs.~\cite{Chiu:2011qc, Chiu:2012ir}. The actual computations of the primary and secondary massive quark corrections to the beam and soft functions are carried out in some detail in \app{calculations}.
In sec.~\ref{sec:massless_and_decoupling}, we show explicitly that the results satisfy the small and large mass limits, and illustrate the numerical size of the mass-dependent corrections for the case of $b$ quarks.

The fixed-order results for the mass-dependent corrections can be expanded either in terms of the $(n_l)$-flavor or $(n_l+1)$-flavor scheme for $\alpha_s$. For definiteness we expand in this section any function $F(m)$ using $\alpha_s^{(n_l+1)}$,
\begin{align}\label{eq:alphas_expansion}
 F(m)=\sum_{n=0}^{\infty}\biggl(\frac{\alpha_s^{(n_l+1)}(\mu)}{4\pi}\biggr)^n\,F^{(n)}(m)\;.
\end{align}
The different two-loop contributions to $F^{(2)}(m)$ are written as
\begin{align}
 F^{(2)}(m)=T_Fn_l\,F^{(2,l)}+T_FF^{(2,h)}(m)+\,.\,.\,.\;,
\end{align}
where $F^{(2,h)}$ contains all mass dependent two-loop corrections and $F^{(2,l)}$ the associated contributions for massless flavors. The expansion of $F$ in terms of $\alpha_s^{(n_l)}$ can be easily obtained by using the matching relation for $\alpha_s$,
\begin{align}
\alpha_s^{(n_l+1)}(\mu) = \alpha_s^{(n_l)}(\mu)\biggl[1- \frac{\alpha_s^{(n_l)}(\mu) T_F}{4\pi}\, \frac{4}{3}L_m +\mathcal{O}(\alpha_s^2)\biggr] \, ,
\end{align}
where here and in the following we abbreviate
\begin{align}
L_m \equiv \ln\frac{m^2}{\mu^2} \, .
\end{align}

\subsection{Hard matching functions}
\label{subsec:hard}

All hard matching functions, i.e.\ the hard function $H$ at the scale $Q$ and the mass mode matching functions $H_c$ and $H_s$ at the scale $m\ll Q$, are insensitive to the measurement performed at a lower scale and are therefore the same for $q_T$ and beam thrust $\Tau$. Since the QCD and SCET currents are the same as for $e^+e^-\to 2 \;\text{jets}$, the results can be read off from the corresponding ones in refs.~\cite{Pietrulewicz:2014qza,Hoang:2015vua}.

\subsubsection{Massive quark corrections to the hard function}

The secondary massive quark corrections to the hard function in \eq{factqT_m1} read  
\begin{align}\label{eq:hard_massive}
H^{(2,h)}(Q,m,\mu)& = H^{(0)}(Q) \biggl\{C_F h_{\rm virt}\Bigl(\frac{m^2}{Q^2}\Bigr) +\frac{4}{3} L_m\, H^{(1)}(Q,\mu) \biggr\} \, ,
\end{align}
where $H^{(0)}$ denotes the tree-level normalization and $H^{(1)}$ the massless one-loop contribution given in \eq{hard_massless}. The function $h_{\rm virt}$ contains the $\mathcal{O}(\alpha_s^2C_F T_F)$ virtual massive quark bubble correction in full QCD shown in \fig{PrimarySecondary}. It has been calculated in \refcites{Kniehl1990, Hoang:1995fr} and is given by
\begin{align}\label{eq:f_QCD}
h_{\rm virt}(x) & = \Bigl(16 x^2-\frac{8}{3}\Bigr) \biggl[-4\Li_3\Bigl(\frac{r-1}{r+1}\Bigr)-\frac{1}{3}\ln^3\frac{r-1}{r+1} +\frac{2\pi^2}{3}\ln\frac{r-1}{r+1} + 4\zeta_3\biggr]
\nn \\ & \quad
+ r\Bigl(\frac{184}{9}x+\frac{76}{9} \Bigr)
\biggl[4\Li_2\Bigl(\frac{r-1}{r+1}\Bigr) +\ln^2\frac{r-1}{r+1} - \frac{2\pi^2}{3}\biggr]+\Bigl(\frac{880}{9}x+\frac{1060}{27}\Bigr)\ln x
\nn \\ &\quad
+ \frac{1904}{9}x+\frac{6710}{81}
\,,\end{align}
with $r= \sqrt{1+4x}$. For $m\to\infty$ the massive quark decouples such that $h_{\rm virt}(x) \to 0$ for $x \to \infty$.

For $Z$-boson production there is an additional primary massive quark contribution to the axial vector current, namely the massive quark triangle correction in \fig{PrimarySecondary}, which we denote by $\Delta h_{\rm axial}$ with the same prefactor as for $h_{\rm virt}$ using the narrow width approximation for notational simplicity. It has been computed in refs.~\cite{Kniehl:1989qu,Gonsalves:1991qn,Bernreuther:2005rw} and is given by
\begin{align}\label{eq:h_axial}
\Delta h_{\rm axial}(Q,m,\mu) &= \frac{8 a_q a_Q}{v_q^2+a_q^2} \biggl[3 \ln\frac{Q^2}{\mu^2} - 9 +
\frac{\pi^2}{3}+\theta(Q^2-4m^2)\,G_1\Bigl(\frac{m^2}{Q^2}\Bigr)
\nn \\ & \qquad\qquad
+\theta(4m^2-Q^2)\,G_2 \Bigl(\frac{m^2}{Q^2}\Bigr)\biggr]
\, ,\end{align}
where the vector and axial vector couplings for up- and down-type quarks are proportional to $v_u=1-8/3 \sin^2\theta_W$, $v_d=-1+4/3 \sin^2\theta_W$, $a_u=1$, $a_d=-1$. The functions $G_1$ and $G_2$ are given in eqs.~(2.8) and~(2.9) of ref.~\cite{Gonsalves:1991qn}. In the small mass limit $m\ll Q$ the function $G_1(m^2/Q^2)$ vanishes, such that $\Delta h_{\rm axial}$ gives the same result as for a massless flavor in the loop,
\begin{align}
\Delta h_{\rm axial}(Q,m,\mu) =  \frac{8 a_q a_Q}{v_q^2+a_q^2} \biggl[3 \ln\frac{Q^2}{\mu^2} -9 + \frac{\pi^2}{3} +\mathcal{O}\Bigl(\frac{m^2}{Q^2}\Bigr)\biggr]
\,.\end{align}
For a massless isospin partner this correction is thus canceled within the $SU(2)_L$ doublet, while for different masses (as for $m_b \ll m_t$) there is a ($\mu$-independent) remainder. Note that for $Q \ll m$ the function $\Delta h_{\rm axial}$ gives a nonvanishing contribution
\begin{align}
\Delta h_{\rm axial}(Q,m,\mu) = \frac{8 a_q a_Q}{v_q^2+a_q^2} \biggl[3 \ln\frac{m^2}{\mu^2} + \frac{3}{2} +\mathcal{O}\Bigl(\frac{Q^2}{m^2}\Bigr)\biggr]
\, .\end{align}
In this case one would integrate out the heavy quark at the scale $\mu_m \sim m$ and evolve the axial current to $\mu_H \sim Q$ to resum logarithms $\ln(m^2/Q^2)$.

\subsubsection{Soft and collinear mass-mode matching functions}

The contributions to the mass-mode matching functions originate only from secondary radiation. The soft mass-mode function $H_s$ appearing in eqs.~\eqref{eq:factqT_m2}, \eqref{eq:factTau_m1}, and \eqref{eq:factTau_m2} has been computed at two loops with the symmetric $\eta$-regulator in \refcite{Hoang:2015vua}. It is given by
\begin{align}\label{eq:Hs2}
 H_s(m,\mu,\nu)&=1+\frac{\alpha_s^2 C_F T_F}{16\pi^2}\biggl[-\Bigl(\frac{16}{3}L_m^2+\frac{160}{9}L_m+\frac{448}{27}\Bigr)\ln\frac{\nu}{\mu}+\frac{8}{9}L_m^3+\frac{40}{9}L_m^2 \nn \\
 & \quad  +\Big(\frac{448}{27}-\frac{4\pi^2}{9}\Big)L_m+\frac{656}{27}-\frac{10\pi^2}{27}-\frac{56\zeta_3}{9}\biggr]
 +\mathcal{O}(\alpha_s^3)
\,.\end{align}
Since there are no $\ord{\alpha_s}$ corrections, the flavor scheme for $\alpha_s$ does not affect the results at $\mathcal{O}(\alpha_s^2)$. Its anomalous dimensions are
\begin{align}\label{eq:gammaHs}
\gamma_{H_s}(m,\mu,\nu)
&=\frac{\alpha_s^2 C_F T_F}{16\pi^2} \biggl[\Bigl(\frac{64}{3}L_m+\frac{320}{9}\Bigr)\ln\frac{\nu}{\mu}-\frac{448}{27}+\frac{8\pi^2}{9}\biggr] +\mathcal{O}(\alpha_s^3)
\,, \nn \\
\gamma_{\nu, H_s}(m,\mu)
&=\frac{\alpha_s^2 C_F T_F}{16\pi^2} \biggl[-\frac{16}{3}L_m^2-\frac{160}{9}L_m-\frac{448}{27}\biggr]
+ \mathcal{O}(\alpha_s^3)
\, .\end{align}
The rapidity anomalous dimension is even known at $\mathcal{O}(\alpha_s^3)$, see ref.~\cite{Hoang:2015iva}.

The result for the collinear mass-mode function $H_c$ in \eq{factqT_m2} can be inferred at $\mathcal{O}(\alpha_s^2)$ from the computations in refs.~\cite{Pietrulewicz:2014qza,Hoang:2015vua} and reads
\begin{align}\label{eq:Hn2}
H_c\Bigl(m,\mu,\frac{\nu}{\omega}\Bigr)
&= 1 + \frac{\alpha_s^2 C_F T_F}{16\pi^2} \biggl[\left(\frac{8}{3}L_m^2+\frac{80}{9}L_m+\frac{224}{27}\right)\ln\frac{\nu}{\omega} \nn \\
 & \quad +2L_m^2+\Bigl(\frac{2}{3}+\frac{8\pi^2}{9}\Bigr)L_m+\frac{73}{18}+\frac{20\pi^2}{27}-\frac{8\zeta_3}{3}\biggr]
 + \mathcal{O}(\alpha_s^3)
\,.\end{align}
Its anomalous dimensions are
\begin{align}\label{eq:gammaHn}
\gamma_{H_c}\Bigl(m,\mu,\frac{\nu}{\omega}\Bigr)
&= \frac{\alpha_s^2 C_F T_F}{16\pi^2} \biggl[-\Bigl(\frac{32}{3}L_m+\frac{160}{9}\Bigr)\ln\frac{\nu}{\omega}-8L_m-\frac{4}{3}-\frac{16\pi^2}{9}\biggr] + \mathcal{O}(\alpha_s^3)
\,, \nn \\
\gamma_{\nu, H_c}(m,\mu)
&= \frac{\alpha_s^2 C_F T_F}{16\pi^2} \biggl(\frac{8}{3}L_m^2+\frac{80}{9}L_m+\frac{224}{27}\biggr)
+ \mathcal{O}(\alpha_s^3)
\,.\end{align}

One can easily verify that the relation in \eq{consistency_hard} between the massive hard function in \eq{hard_massive}, the hard function contribution for a massless flavor in \eq{hard_massless}, and the two mass-mode functions in \eqs{Hs2}{Hn2} is satisfied,
\begin{align}
 H^{(2,h)}(Q,m,\mu)& =H^{(2,l)}(Q,\mu)+H_c^{(2)}\Bigl(m,\mu,\frac{\nu}{\omega_a}\Bigr)+H_c^{(2)}\Bigl(m,\mu,\frac{\nu}{\omega_b}\Bigr)+H_s^{(2)}(m,\mu,\nu) +\mathcal{O}\Bigl(\frac{m^2}{Q^2}\Bigr)
.\end{align}

\subsection{Beam functions}
\label{sec:beamfct_results}

Here we give our results for the massive quark beam function coefficient $\mathcal{I}_{Qg}$ at $\mathcal{O}(\alpha_s)$ and the secondary massive quark corrections to the light-quark coefficients $\mathcal{I}_{qq}$ at $\mathcal{O}(\alpha_s^2)$, which appear in \eqs{factqT_m3}{factTau_m1} for the $q_T$ and beam thrust measurement. We also give the massive quark contributions to the beam function anomalous dimensions. We also give the well-known results for the corresponding PDF matching coefficients $\mathcal{\mathcal{M}}_{Qg}$ at $\mathcal{O}(\alpha_s)$ and $\mathcal{\mathcal{M}}_{qq}$ at $\mathcal{O}(\alpha_s^2)$ appearing in eqs.~\eqref{eq:factqT_m4}, \eqref{eq:factTau_m2} and \eqref{eq:factTau_m3}.

\subsubsection{TMD beam function coefficients}

\begin{figure}
\center
\includegraphics[scale=0.6]{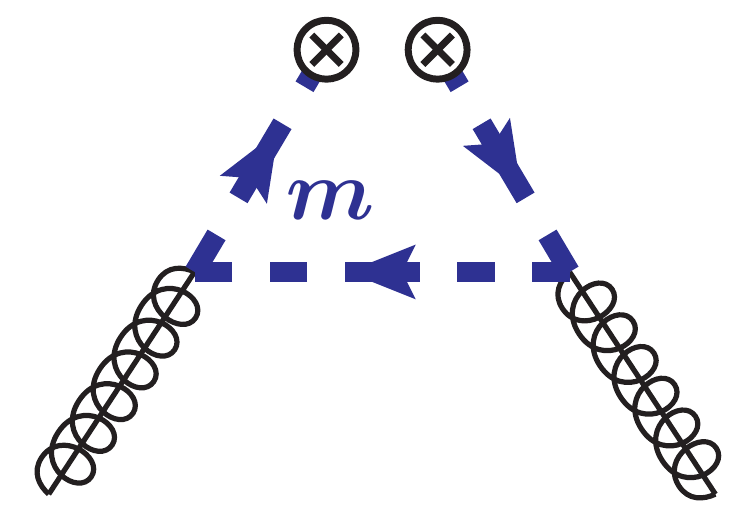}
\caption{Feynman diagram for the massive quark beam function at one loop.}
\label{fig:beam_primary}
\end{figure}

The matching coefficient $\mathcal{I}_{Qg}$ generating a massive beam function from a gluon splitting is calculated at $\mathcal{O}(\alpha_s)$ in \sec{Calc_primarymassive} and corresponds to the diagram shown in fig.~\ref{fig:beam_primary}. The result reads ($p_T^2=|\vec{p}_T|^2$)
\begin{align} \label{eq:TMD_beam_coefficient}
\mathcal{I}_{Qg}(\vec{p}_T,m,z)
= \mathcal{I}_{\bar{Q}g}(\vec{p}_T,m,z)
= \frac{\alpha_sT_F}{4\pi^2}\,\theta(z)\,\theta(1-z)\,\frac{2}{p_T^2+m^2} \biggl[ P_{qg}(z)+\frac{2m^2z(1-z)}{p_T^2+m^2}\biggr]
+ \mathcal{O}(\alpha_s^2)
\,,\end{align}
with the splitting function
\begin{align}
P_{qg}(z) =z^2+(1-z)^2 \, .
\end{align}
This result is equivalent to the Fourier transform of the mass-dependent matching functions $\mathcal{C}_{h/G}$ in ref.~\cite{Nadolsky:2002jr}. After performing an appropriate crossing it also agrees with the massive final-state splitting functions~\cite{Catani:2000ef, Kang:2016ofv} or fragmenting jet function~\cite{Bauer:2013bza}.

\begin{figure}
\hfill\subfigure[]{\includegraphics[scale=0.5]{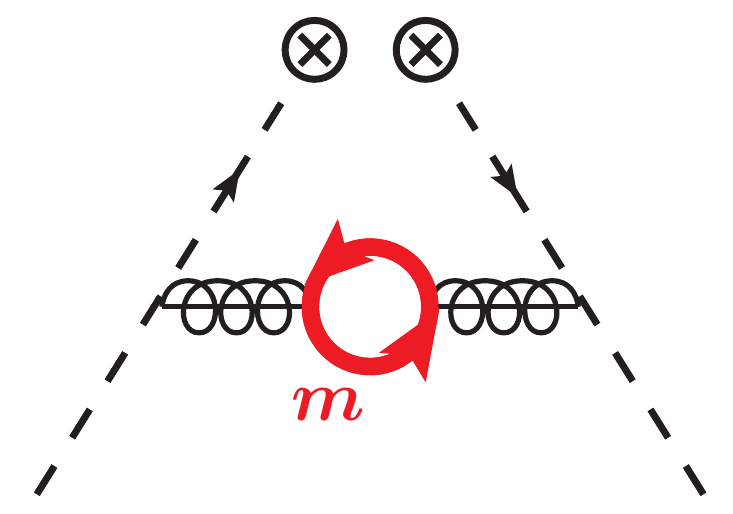}\label{fig:beam_secondary_a}}%
\hfill\subfigure[]{\includegraphics[scale=0.5]{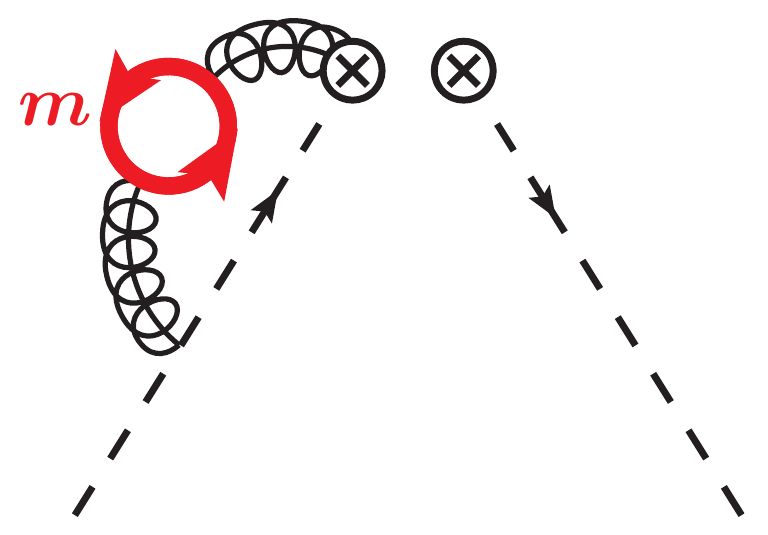}\label{fig:beam_secondary_b}}%
\hfill\subfigure[]{\includegraphics[scale=0.5]{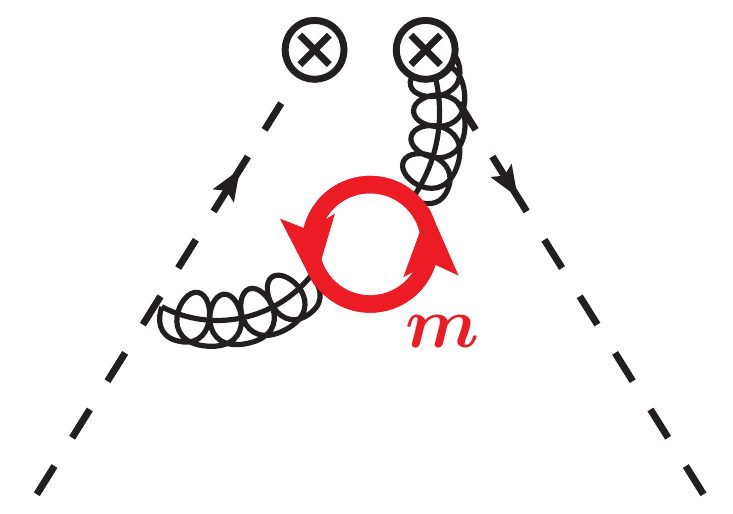}\label{fig:beam_secondary_c}}%
\hspace*{\fill}
\caption{Secondary massive quark corrections for the light-quark beam function at $\ord{\alpha_s^2}$. In addition, also the wave-function renormalization correction and the mirror diagrams for (b) and (c) have to be included.}
\label{fig:beam_secondary}
\end{figure}

The contributions from secondary massive quarks to the matching coefficient $\mathcal{I}_{qq}$ are computed in \sec{Calc_secondarymassive} at $\mathcal{O}(\alpha_s^2)$. The corresponding diagrams are shown in fig.~\ref{fig:beam_secondary}. The result is given by
\begin{align}\label{eq:TMD_Iqq2_massive}
&\mathcal{I}_{qq}^{(2,h)}\Bigl(\vec{p}_T,m,z,\mu,\frac{\nu}{\omega}\Bigr)
\nn \\ & \quad
=\theta(z) C_F
\biggl\{\delta^{(2)}(\vec{p}_T)\,\delta(1-z)
\biggl[\Bigl(\frac{8}{3}L_m^2+\frac{80}{9}L_m+\frac{224}{27}\Bigr)\ln\frac{\nu}{\omega}+2L_m^2+\Bigl(\frac{2}{3}+\frac{8\pi^2}{9}\Bigr)L_m
\nn \\ & \qquad\quad
+\frac{73}{18} +\frac{20\pi^2}{27}-\frac{8\zeta_3}{3}\biggr]
\nn \\ & \qquad
+\frac{16}{9\pi p_T^2}
\Bigl[\mathcal{L}_0(1-z)-\delta(1-z)\ln\frac{\nu}{\omega}\Bigr]
\Bigl[-5+12\hat{m}^2+3c(1-2\hat{m}^2)\ln\frac{c+1}{c-1} \Bigr]
\nn \\ & \qquad
+\frac{16}{9\pi p_T^2} \theta(1-z) \biggl[\frac{3}{2d(1-z)} \bigl[(1+z^2)(1+2\hat{m}^2z) + 4\hat{m}^4z^2(-5+6z-5z^2)\bigr] \ln\frac{d+1}{d-1}
\nn \\ & \qquad\quad
-\frac{3c(1-2\hat{m}^2)}{1-z}\ln\frac{c+1}{c-1} + 1+4z+3\hat{m}^2(-4+z-5z^2) \biggr]
\biggr\}
\nn \\ & \qquad
+\frac{4}{3}L_m \, \mathcal{I}_{qq}^{(1)}\Bigl(\vec{p}_T,z,\mu,\frac{\nu}{\omega}\Bigr)
\,,\end{align}
where
\begin{align}\label{eq:mhat}
\hat{m} \equiv \frac{m}{p_T} \, ,\qquad c=\sqrt{1+4\hat{m}^2} \, , \qquad d=\sqrt{1+4\hat{m}^2z}
\,,\end{align}
and the one-loop term $\mathcal{I}_{qq}^{(1)}$ is given in eq.~\eqref{eq:TMD_Iqq1}.
Here $\mathcal{L}_n(1-z)$ denotes the standard plus distribution as defined in appendix~\ref{sec:distributions}.

In the $(n_l+1)$-flavor scheme for $\alpha_s$ there is also a correction from a virtual massive quark loop to the flavor-nondiagonal matching coefficient $\mathcal{I}^{(2)}_{qg}$. This contribution is trivial, since it factorizes into a vacuum polarization correction corresponding to the matching of $\alpha_s$ between the $(n_l)$ and $(n_l+1)$-flavor schemes, and the one-loop contribution, such that
\begin{align}\label{eq:Iqg2TF2}
\mathcal{I}_{qg}^{(2,h)}(\vec{p}_T,m,z,\mu)= \frac{4}{3} L_m \, \mathcal{I}_{qg}^{(1)}(\vec{p}_T,z,\mu) \, ,
\end{align}
with $\mathcal{I}_{qg}^{(1)}$ given in \eq{TMD_Iqg1}. In the $(n_l)$-flavor scheme for $\alpha_s$ the $\mathcal{I}_{qg}^{(2,h)}$ contribution vanishes.

The contributions from a massive flavor to the beam function anomalous dimensions are
\begin{align}\label{eq:anomdim_TMD}
\gamma_{B}^{(2,h)}\Bigl(\frac{\nu}{\omega}\Bigr)
&=C_F\,\biggl(-\frac{160}{9}\ln\frac{\nu}{\omega}-\frac{4}{3}-\frac{16\pi^2}{9}\biggr)
\,, \nn \\
\gamma_{\nu, B}^{(2,h)}(\vec{p}_T,m,\mu)
&= C_F\,\biggl\{-\frac{16}{3}L_m \mathcal{L}_0(\vec{p}_T,\mu)+\delta^{(2)}(\vec{p}_T)\Bigl(\frac{8}{3}L_m^2+\frac{80}{9}L_m+\frac{224}{27}\Bigr)
\nn \\ & \quad
+\frac{16}{9\pi p_T^2} \biggl[5-12\hat{m}^2-3c(1-2\hat{m}^2)\ln\frac{c+1}{c-1} \biggr] \biggr\}
\,.\end{align}
The $\mathcal{L}_0(\vec{p}_T,\mu)$ distribution is defined in appendix~\ref{sec:distributions}.
The $\mu$ anomalous dimension here is the same as for a massless quark flavor, $\gamma_{B}^{(2,h)}=\gamma_{B}^{(2,l)}$ [see \eq{gammaB}]. The rapidity anomalous dimension is explicitly mass dependent and only reproduces the result for a massless flavor in the limit $m \ll p_T$.

\subsubsection{Virtuality-dependent beam function coefficients}

The massive quark-gluon virtuality beam function matching coefficient at $\mathcal{O}(\alpha_s)$ shown in fig.~\ref{fig:beam_primary} is given by
\begin{align} \label{eq:virtuality_beam_coefficient}
\mathcal{I}_{Qg}(t,m,z)
= \frac{\alpha_sT_F}{4\pi}\,\theta(t)\,\theta(z)\,\theta\biggl[\frac{t(1-z)}{z}-m^2\biggr]\,\frac{2}{t}\biggl[P_{qg}(z)+\frac{2m^2z^2}{t}\biggr]
+ \mathcal{O}(\alpha_s^2)
\,.\end{align}

The contributions from secondary massive quarks to the light-quark coefficient at $\mathcal{O}(\alpha_s^2)$ as shown in fig.~\ref{fig:beam_secondary} are given by
\begin{align}\label{eq:Virt_Iqq2_massive}
&\mathcal{I}_{qq}^{(2,h)}\Bigl(t,m,z,\mu,\frac{\nu}{\omega}\Bigr)
\nn \\ & \quad
=\theta(z) C_F
\biggl\{ \delta(t)\,\delta(1-z)\biggl[\Bigl(\frac{8}{3}L_m^2+\frac{80}{9}L_m+\frac{224}{27}\Bigr)\ln\frac{\nu}{\omega} +2L_m^2+\Bigl(\frac{2}{3}+\frac{8\pi^2}{9}\Bigr)L_m
\nn \\ &\qquad\quad
+\frac{73}{18}+\frac{20\pi^2}{27}-\frac{8\zeta_3}{3}  \biggr]
\nn \\ &\qquad
+ \theta\biggl(t-\frac{4m^2z}{1-z}\biggr)\frac{8}{9t(1-z)}\biggl[-\frac{3}{u} \Bigl[ (1+z^2)(1 -2\hat{m}_t^2 z) -4 \hat{m}_t^4 z^2(2-3z+5z^2)\Bigr]\ln\frac{u-v}{u+v}
\nn \\ & \qquad\quad
-2v\Bigl[4-3z+4z^2 +\frac{z(11-21z+29z^2-15z^3)}{1-z}\,\hat{m}_t^2\Bigr]\biggr]\biggr\}
+\frac{4}{3}L_m\,\mathcal{I}_{qq}^{(1)}(t,z,\mu)
\,,\end{align}
with 
\begin{align}\label{eq:u_and_v}
\hat{m}_t= \frac{m}{\sqrt{t}} \, \qquad u=\sqrt{1-4\hat{m}_t^2 z} \, , \qquad  v=\sqrt{1-\frac{4\hat{m}_t^2z}{1-z}}
\,,\end{align}
and the one-loop term $\mathcal{I}_{qq}^{(1)}$ is given in eq.~\eqref{eq:Virt_Iqq1}.

In the $(n_l+1)$-flavor scheme for $\alpha_s$ there is also the analogous contribution to \eq{Iqg2TF2} to the flavor-nondiagonal coefficient
\begin{align}\label{eq:Iqg2TF2_Tau}
\mathcal{I}_{qg}^{(2,h)}(t,m,z,\mu)= \frac{4}{3} L_m \, \mathcal{I}_{qg}^{(1)}(t,z,\mu) \, ,
\end{align}
with $\mathcal{I}_{qg}^{(1)}$ given in \eq{Virt_Iqq1}. In the $(n_l)$-flavor scheme for $\alpha_s$ the
$\mathcal{I}_{qg}^{(2,h)}$ contribution vanishes.

The contribution from the massive flavor to the $\mu$ anomalous dimension at $\mathcal{O}(\alpha_s^2)$ is given by
\begin{align}\label{eq:gammaBm_mu}
\gamma^{(2,h)}_{B,m}\Bigl(t,\frac{\nu}{\omega}\Bigr)
&=C_F\,\delta(t)\biggl(-\frac{160}{9}\ln\frac{\nu}{\omega}-\frac{4}{3}-\frac{16\pi^2}{9}\biggr)\,.
\end{align}
We emphasize that the massive quark contribution to the $\mu$ anomalous dimension is not the same as for a massless flavor, but is in fact the same as for the TMD beam function in \eq{anomdim_TMD}. This is required by consistency with the large mass limit $Q\Tau,q_T \ll m$, where the massive flavor can only contribute to the (local) running of the common current operators, which are independent of the measurement. Only in combination with the soft mass-mode function $H_s$ and the soft function, the combined $\mu$ evolution above the mass scale is the same as for $n_l+1$ massless flavors as discussed in \eq{consistency_Tau1}.

The secondary massive quarks introduce rapidity divergences and associated logarithms also in the virtuality-dependent beam function. The $\nu$ anomalous dimension induced by the secondary massive effects is the same as for the collinear mass-mode function, see \eq{BTau_nuevolution}, given in \eq{gammaHn}.

\subsubsection{PDF matching coefficients}

The matching coefficients relating the PDFs in the $(n_l+1)$ and the $(n_l)$-flavor scheme are all known at two loops \cite{Buza:1996wv} and partially beyond (see e.g.~\refcites{Ablinger:2014nga,Ablinger:2014vwa,Ablinger:2014lka} and references therein). The matching coefficient for a primary massive quark originating from an initial-state gluon at $\mathcal{O}(\alpha_s)$ is
\begin{align}\label{eq:quark_gluon_PDF_matching}
\mathcal{M}_{Qg}(m,z,\mu)=-\frac{\alpha_s T_F}{4\pi}\,\theta(1-z)\theta(z)\,2 P_{qg}(z) L_m +\mathcal{O}(\alpha_s^2)\,.
\end{align}
The matching coefficient coming from secondary massive quark corrections to the light-quark PDFs reads up to $\mathcal{O}(\alpha_s^2)$
\begin{align}\label{eq:Mqq2}
\mathcal{M}_{qq}(m,z,\mu)
&=  1+ \frac{\alpha_s^2 C_F T_F}{16\pi^2} \,\theta(z) \biggl\{\mathcal{L}_0(1-z) \biggl(\frac{8}{3}L_m^2 +\frac{80}{9}L_m + \frac{224}{27}\biggr)
\nn \\ &  \qquad
+ \delta(1-z)\biggl[2 L_m^2 +\left(\frac{2}{3}+\frac{8\pi^2}{9}\right)L_m +\frac{73}{18} + \frac{20 \pi^2}{27} - \frac{8\zeta_3}{3} \biggr]
\nn \\ & \qquad
+ \theta(1-z)\ \biggl[-\frac{4}{3}L_m^2 (1+z)+L_m\biggl(\frac{8}{9}-\frac{88}{9}z + \frac{8}{3}\frac{1+z^2}{1-z}\,\ln z\biggr)+ \frac{2}{3}\frac{1+z^2}{1-z}\,\ln^2 z
\nn \\ &  \qquad
+ \frac{\ln z}{1-z}\biggl(\frac{44}{9}-\frac{16}{3}z+\frac{44}{9}z^2\biggr) +\frac{44}{27}- \frac{268}{27} z \biggr]\biggr\} +\mathcal{O}(\alpha_s^3)
\,.\end{align}
The matching coefficient between the gluon PDF in the $(n_l)$ and $(n_l+1)$-flavor schemes at $\mathcal{O}(\alpha_s)$, which is also required for Drell-Yan at $\mathcal{O}(\alpha_s^2)$, is equivalent to the matching relation for $\alpha_s$
\begin{align}
\mathcal{M}_{gg}(m,z,\mu) =\delta(1-z) +\frac{\alpha_s T_F}{4\pi} \, \delta(1-z) \,\frac{4}{3} L_m +\mathcal{O}(\alpha_s^2)
\,.\end{align}
Note that taking into account the nondiagonal evolution of the PDFs the known $\mathcal{O}(\alpha_s^2)$ corrections for all matching factors $\mathcal{M}_{ij}$ become relevant at NNLL$^\prime$.

\subsection{Soft and collinear-soft functions}

Here we give all massive quark corrections at $\mathcal{O}(\alpha_s^2)$ to the soft and csoft functions. They arise exclusively from secondary radiation. Note that the soft functions satisfy Casimir scaling at this order and can be thus applied also to color-singlet production in gluon-fusion by replacing an overall $C_F \to C_A$.

\subsubsection{TMD soft function}

 \begin{figure}
\hfill\subfigure[]{\includegraphics[scale=0.5]{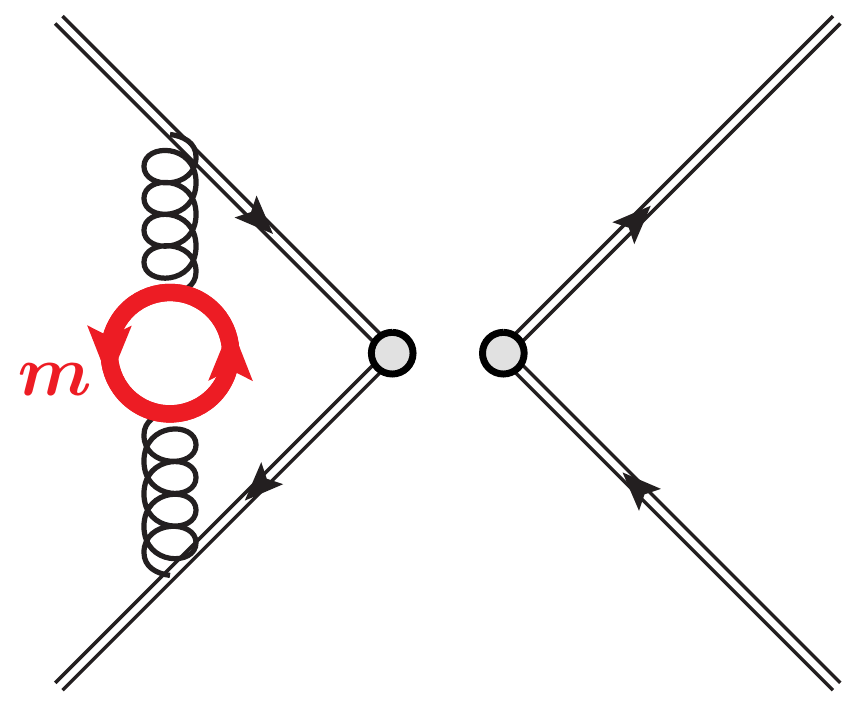}\label{fig:soft_secondary_a}}%
\hfill\subfigure[]{\includegraphics[scale=0.5]{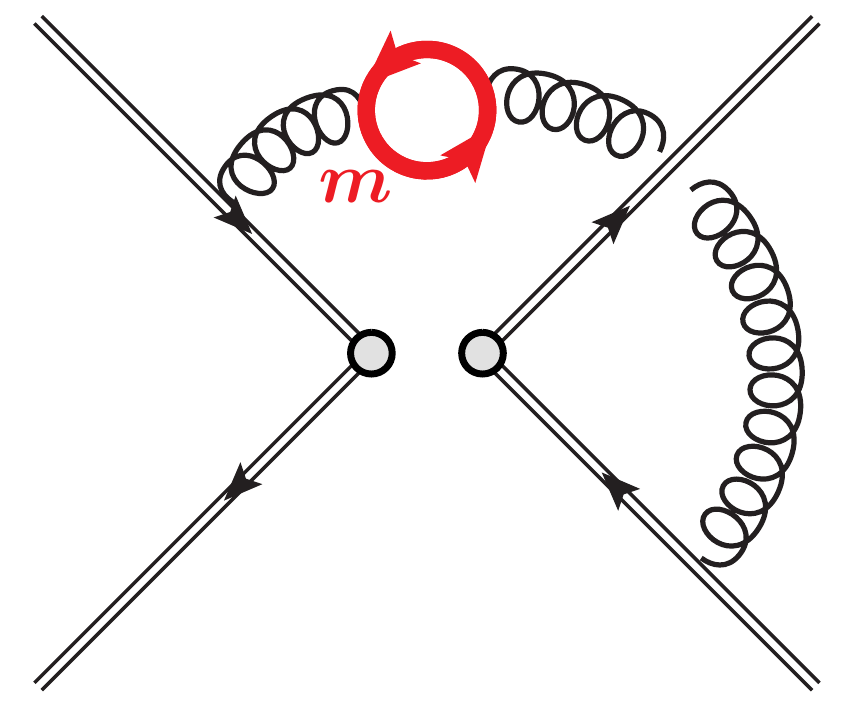}\label{fig:soft_secondary_b}}%
\hspace*{\fill}
\caption{Corrections from secondary massive quarks to the (c)soft function. Also the mirror diagrams need to be included.}
\label{fig:soft_secondary}
\end{figure}

The contributions from secondary massive quarks to the TMD soft function, which appears in \eq{factqT_m3} for $q_T\sim m$, are calculated in \app{Calc_softfct} at $\mathcal{O}(\alpha_s^2)$ and correspond to the diagrams shown in fig.~\ref{fig:soft_secondary}.
The result reads
\begin{align} \label{eq:pT_soft_massive_ren}
S^{(2,h)}(\vec{p}_T,m,\mu,\nu)
&=C_F\,\biggl\{ \delta^{(2)}(\vec{p}_T)\biggl[\Bigl(-\frac{16}{3}L_m^2-\frac{160}{9}L_m-\frac{448}{27}\Bigr)\ln\frac{\nu}{\mu} +\frac{8}{9}L_m^3+\frac{40}{9}L_m^2
\nn \\ & \qquad
+ \Bigl(\frac{448}{27}-\frac{4\pi^2}{9}\Bigr)L_m +\frac{656}{27}-\frac{10\pi^2}{27}-\frac{56\zeta_3}{9}\biggr]
\nn\\& \quad
+\frac{16}{9\pi p_T^2}\biggl[2\Bigl[-5+12\hat{m}^2+3c(1-2\hat{m}^2)\ln\frac{c+1}{c-1} \Bigr]\ln\frac{\nu}{m}
\nn \\ & \qquad
+3c(1-2\hat{m}^2)\biggl[\Li_2\biggl(\frac{(c-1)^2}{(c+1)^2}\biggr)+2\ln\frac{c+1}{c-1} \ln\frac{\hat{m}(c+1)}{2c} -
\frac{\pi^2}{6}\biggr]
\nn \\ & \qquad
+c(5-16\hat{m}^2)\ln\frac{c+1}{c-1} + 8\hat{m}^2 \biggr]\biggr\}
+\frac{4}{3}L_m\,S^{(1)}(\vec{p}_T,\mu,\nu)
\,,\end{align}
where $\hat{m}=m/p_T$ and $c=\sqrt{1+4\hat{m}^2}$ as in \eq{mhat} and the one-loop soft function $S^{(1)}$ given in \eq{TMD_soft_massless_1}.

The massive quark contributions to the anomalous dimensions of the soft function are
\begin{align}\label{eq:gammaS_m}
\gamma_{S}^{(2,h)}(\vec{p}_T,\mu,\nu)
&= C_F\,\biggl(\frac{320}{9}\ln\frac{\nu}{\mu}-\frac{448}{27}+\frac{8\pi^2}{9}\biggr)
\,, \nn \\
\gamma_{\nu,S}^{(2,h)}(\vec{p}_T,m,\mu)
&= C_F\,\biggl\{
\frac{32}{3}L_m\mathcal{L}_0(\vec{p}_T,\mu)
+ \delta^{(2)}(\vec{p}_T)\biggl(-\frac{16}{3}L_m^2-\frac{160}{9}L_m-\frac{448}{27}\biggr)
\nn \\ & \quad
+\frac{32}{9\pi p_T^2} \biggl[-5+12\hat{m}^2+3c(1-2\hat{m}^2)\ln\frac{c+1}{c-1} \biggr]\biggr\}
\,.\end{align}
The $\mu$ anomalous dimension here is the same as for an additional massless flavor, $\gamma_{S}^{(2,h)}=\gamma_{S}^{(2,l)}$  [see \eq{SpT}]. The rapidity anomalous dimension is explicitly mass dependent and only reduces to the result for a massless flavor in the limit $m \ll p_T$.

\subsubsection{Csoft function for beam thrust}

The csoft function is a matching coefficient between an eikonal matrix element in the $n_l+1$ and $n_l$ flavor theories appearing for the hierarchy $\Tau\ll m \ll \sqrt{Q\Tau}$ in \eq{factTau_m2}. The relevant diagrams at $\mathcal{O}(\alpha_s^2)$ are shown in fig.~\ref{fig:soft_secondary} and are calculated in \sec{csoft}. The result is given by
\begin{align}\label{eq:csoft}
\mathcal{S}_c(\ell,m,\mu,\nu)
&= \delta(\ell) + \frac{\alpha_s^2 C_F T_F}{16\pi^2}\biggl\{
\frac{\nu}{\mu^2}\mathcal{L}_0\Bigl(\frac{\ell\, \nu}{\mu^2}\Bigr) \Bigl( \frac{8}{3}L_m^2+\frac{80}{9}L_m+\frac{224}{27}\Bigr)
\\\nn & \quad
+ \delta(\ell)\biggl[-\frac{8}{9}L_m^3-\frac{40}{9}L_m^2+\Bigl(-\frac{448}{27}+\frac{4\pi^2}{9}\Bigr)L_m-\frac{656}{27}+\frac{10\pi^2}{27} +\frac{56\zeta_3}{9} \biggr]
\biggr\}
+\mathcal{O}(\alpha_s^3)
.\end{align}
We can see that with the scale choices $\mu \sim m$ and $\nu \sim \mu^2/\ell \sim m^2/\Tau$ all large logarithms (including the implicit one inside the plus distribution) are minimized. The $\mu$ anomalous dimensions of the csoft matching function is given by
\begin{align}\label{eq:gamma_csoft}
\gamma_{\mathcal{S}_c}(\ell,m,\mu,\nu)
&=\frac{\alpha_s^2 C_F T_F}{16\pi^2}\biggl[
- \frac{\nu}{\mu^2}\mathcal{L}_0\Bigl(\frac{\ell\, \nu}{\mu^2}\Bigr) \Bigl( \frac{32}{3}L_m+\frac{160}{9} \Bigr)
+ \delta(\ell) \Bigl(\frac{448}{27}-\frac{8\pi^2}{9}\Bigr) \biggr]
+ \mathcal{O}(\alpha_s^3)
\,.\end{align}
The $\nu$ anomalous dimension is the same as for the collinear mass mode function in \eq{gammaHn}, $\gamma_{\nu, \mathcal{S}_c} =\gamma_{\nu, H_c}$.

\subsubsection{(Beam) thrust soft function}

The secondary massive quark corrections to the (beam) thrust soft function at $\mathcal{O}(\alpha_s^2)$ were calculated in \refcite{Gritschacher:2013tza} and are given by
\begin{align}\label{eq:soft_tau}
S^{(2,h)}(\ell,m,\mu) &=C_F\,\biggl\{
\frac{1}{\mu}\mathcal{L}_0\Bigl(\frac{\ell}{\mu}\Bigr) \biggl(\frac{16}{3}L_m^2+\frac{160}{9}L_m+\frac{448}{27}\biggr)
\nn \\ & \quad
+ \delta(\ell)\biggl[-\frac{8}{9}L_m^3 -\frac{40}{9}L_m^2+\Bigl(-\frac{448}{27}+\frac{4\pi^2}{9}\Bigr)L_m-\frac{656}{27} +\frac{10\pi^2}{27}+\frac{56}{9}\zeta_3 \biggr]
\nn \\  & \quad
+\theta(\ell-2m)\frac{1}{\ell}\biggl[\frac{64}{3} \,\Li_2\Bigl(\frac{w-1}{w+1}\Bigr)
+\frac{16}{3}\ln^2\frac{1-w}{1+w} - \frac{64}{3}\ln\frac{1-w}{1+w} \ln \hat{m}_{\ell}
\nn \\ & \qquad
-\frac{160}{9} \ln\frac{1-w}{1+w} - w\Bigl(\frac{896}{27}+\frac{256}{27}\hat{m}_\ell^2\Bigr) + \frac{16\pi^2}{9}\biggr]
\biggr\}
\nn \\ & \quad
+ \Delta S_{\tau}(\ell,m) +\frac{4}{3}L_m\,S^{(1)}(\ell,\mu)
\,,\end{align}
where
\begin{align}
\hat{m}_\ell \equiv \frac{m}{\ell} \, , \qquad w = \sqrt{1-4\hat{m}_\ell^2}\, ,
\label{eq:wdef}
\end{align}
and the one-loop soft function $S^{(1)}$ is given in \eq{soft_massless_1}.
The term $\Delta S_{\tau}(\ell,m)$ contains the correction from two real final-state emissions entering two opposite hemispheres, which vanishes both for $\ell \ll m$ and $m\ll \ell$ and is currently only known numerically. The integral expression for this numerically small contribution is given in eq.~(61) of \refcite{Gritschacher:2013tza}, and a precise parametrization can be found in ref.~\cite{Pietrulewicz:2014qza}.

The massive quark contribution to the anomalous dimension is the same as for a massless flavor,
$\gamma_S^{(2,h)}(\ell, \mu) = \gamma_S^{(2,l)}(\ell, \mu)$, given in \eq{gamma_soft}.

\subsection{Small and large mass limits}\label{sec:massless_and_decoupling}

In secs.~\ref{subsec:qT_hierarchies} and \ref{subsec:hierarchies_Tau} we explained how the ingredients in the factorization theorems for different hierarchies are related to each other. Here we verify these relations for the beam and soft functions up to $\mathcal{O}(\alpha_s^2)$. We also scrutinize the numerical impact of the power corrections for these functions.
We focus in particular on the $\mathcal{O}(m^2/q_T^2)$ corrections the $q_T$ spectrum for $b$ quarks, which are contained in the factorization theorem \eq{factqT_m3} for $q_T \sim m$ but not in the massless limit for $m\ll q_T$ in \eq{factqT_m4}, as these are phenomenologically important hierarchies for $b$-quark mass effects at the LHC.

For the numerical results we use the MMHT2014 NNLO PDFs~\cite{Harland-Lang:2014zoa} and evaluate the contributions for $\mu=m_b=4.8 $ GeV, $\omega=m_Z$, and $E_{\rm cm}=13$ TeV. The main  qualitative features of the results do not depend on these specific input parameters.

\subsubsection{Limiting behavior for $q_T$}

We first consider the primary mass effects at one loop, which are encoded in the TMD beam function matching coefficient $\mathcal{I}_{Qg}^{(1)}$ in \eq{TMD_beam_coefficient}. In the limit $p_T \ll m$ the primary massive quarks decouple, which is manifest in the result,
\begin{align}
\mathcal{I}_{Qg}^{(1)}(\vec{p}_T,m,z)=\mathcal{O}\Bigl(\frac{p_T^2}{m^2}\Bigr)
\,.\end{align}
On the other hand, in the opposite limit $m\ll p_T$ it becomes
\begin{align}\label{eq:IQg_ml}
&\mathcal{I}_{Qg}^{(1)}(\vec{p}_T,m,z)
\nn \\ & \quad
=T_F\,\theta(1-z)\theta(z)\,\Bigl\{2P_{qg}(z)\mathcal{L}_0(\vec{p}_T,\mu) +\delta^{(2)}(\vec{p}_T)\Bigl[-2P_{qg}(z)L_m+4z(1-z)\Bigr]\!+\mathcal{O}\Bigl(\frac{m^2}{p_T^2}\Bigr)\biggr\}
\nn \\ & \quad
=\mathcal{I}_{qg}^{(1)}(\vec{p}_T,z,\mu)+\delta^{(2)}(\vec{p}_T)\mathcal{M}_{Qg}^{(1)}(m,z,\mu)+\mathcal{O}\Bigl(\frac{m^2}{p_T^2}\Bigr)
\,,\end{align}
confirming that the relation in \eq{consistencyqT_beam2} is satisfied at $\mathcal{O}(\alpha_s)$.
The massless one-loop matching coefficient $\mathcal{I}_{qg}^{(1)}$ can be found in \eq{TMD_Iqg1} and the PDF matching coefficient $\mathcal{M}_{Qg}^{(1)}$ in \eq{quark_gluon_PDF_matching}. To account for the correct distributive structure in $\vec p_T$ that emerges in the massless limit, one can integrate the expressions with massive quarks and identify the distributions at the cumulant level.

\begin{figure}
\includegraphics[height=5.3cm]{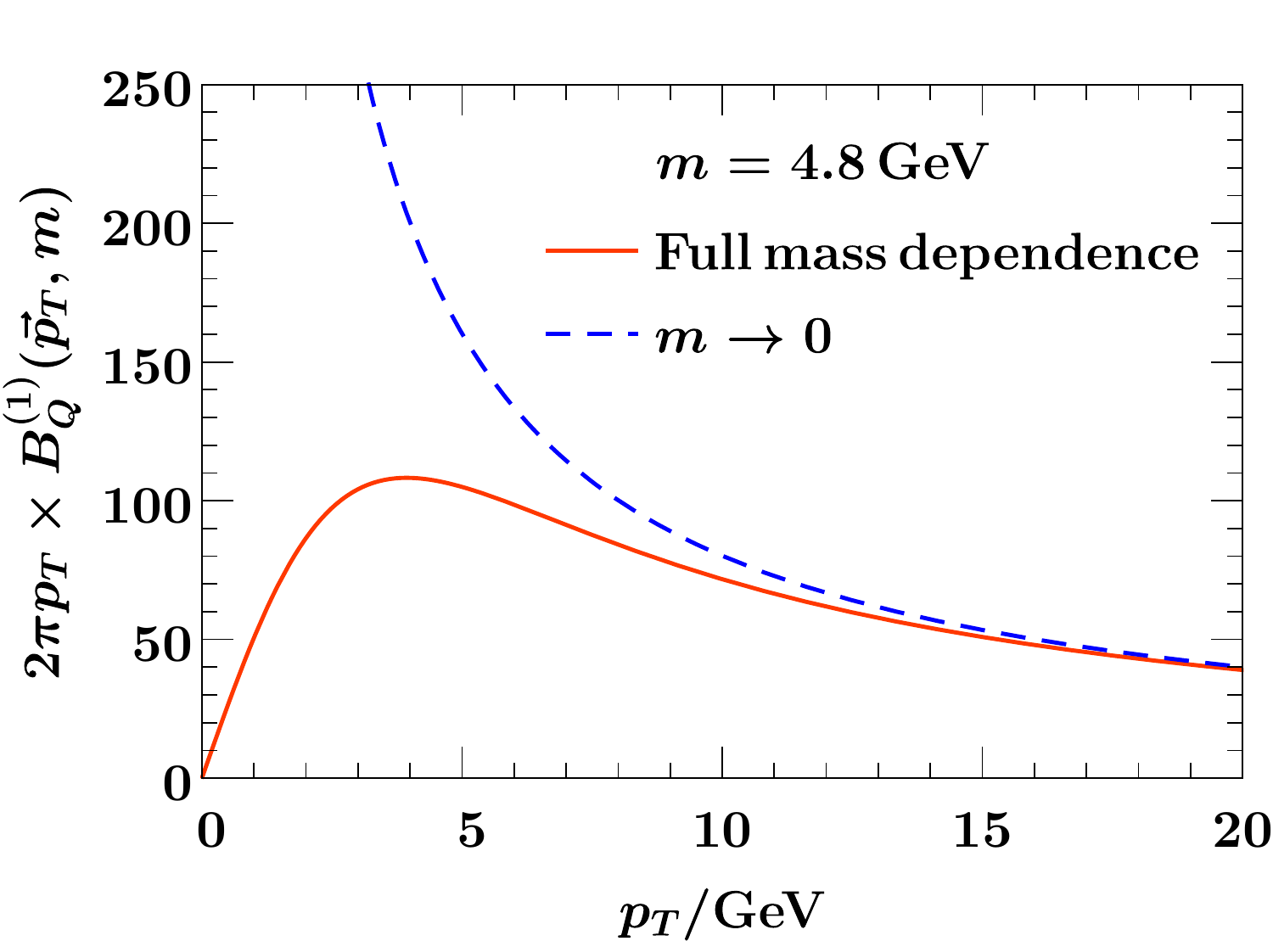}%
\hfill%
\includegraphics[height=5.3cm]{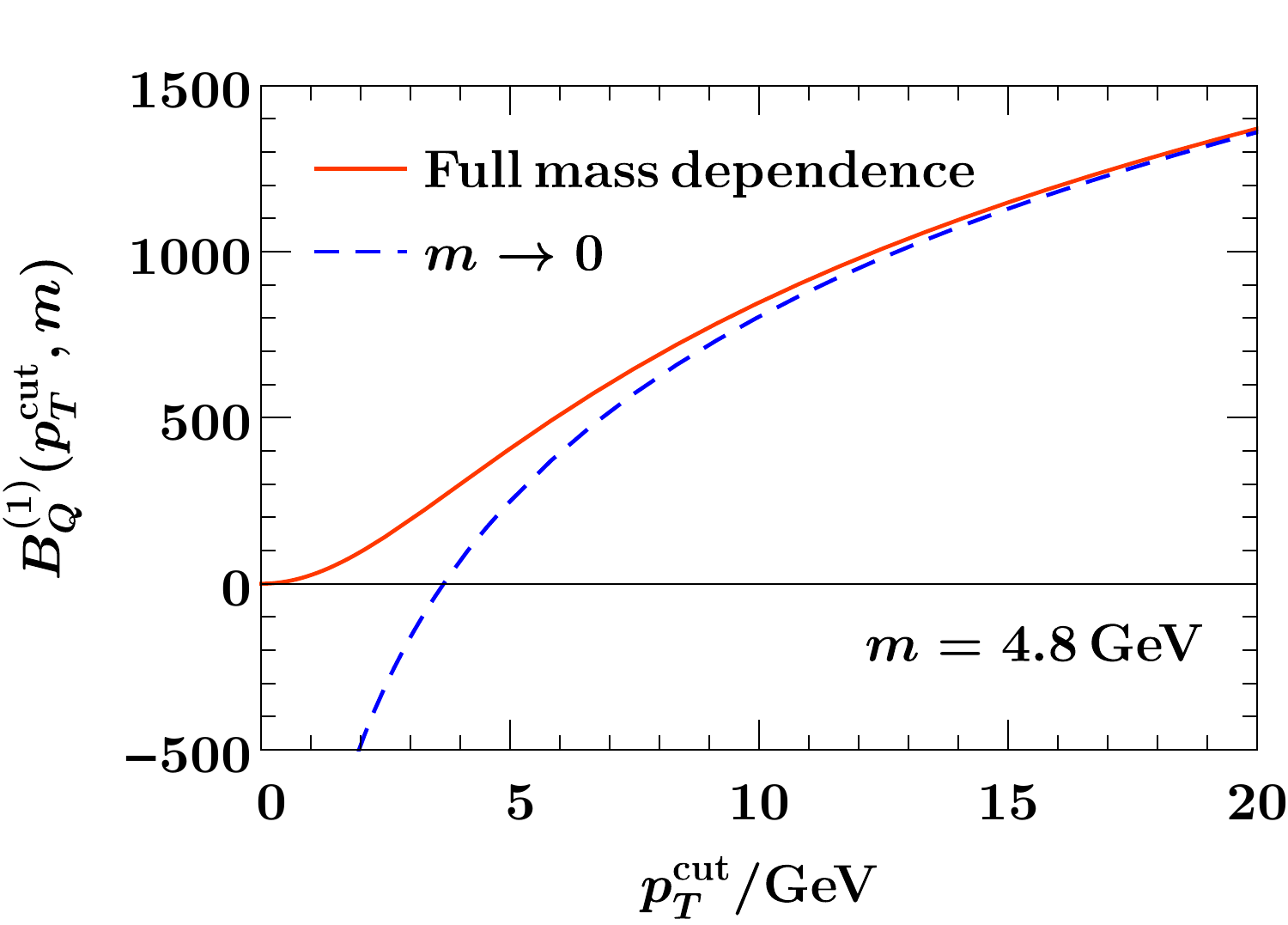}%
\caption{Massive $b$-quark beam function (left panel) and its cumulant (right panel) together with its $m\to 0$ limit. The input parameters are described in the text.}
\label{fig:massiveBeamfct}
\end{figure}

In \fig{massiveBeamfct} we show the result for the massive quark beam function $B^{(1)}_Q =\mathcal{I}^{(1)}_{Qg} \otimes_x f_g$ at $\mathcal{O}(\alpha_s)$ as function of $p_T$ using the full massive matching coefficient $\mathcal{I}^{(1)}_{Qg}$ (solid orange) and its small mass limit in \eq{IQg_ml}. Note that the results differential in $p_T$ are not explicitly $\mu$-dependent at $\mathcal{O}(\alpha_s)$. In the right panel we show the corresponding results for the cumulant
\begin{align}\label{eq:BQ_cum}
B_Q(p_T^{\rm cut},m) \equiv \int_{|\vec{p}_T| <p_T^{\rm cut}} \df^2 p_T
\, B_Q(\vec{p}_T,m)
\,,\end{align}
which also includes the $\delta^{(2)}(\vec p_T)$ constant contribution.
We can see that in both cases the small mass limit is correctly approached for $p_T^{(\rm cut)} \gg m_b$, while for $p_T^{(\rm cut)} \ll m_b$ the primary mass effects decouple with the result going to zero. The corrections to the small mass limit become sizeable for $p_T \sim m_b$ and vanish quite fast for larger $p_T$.

\begin{figure}
\includegraphics[height=5.3 cm]{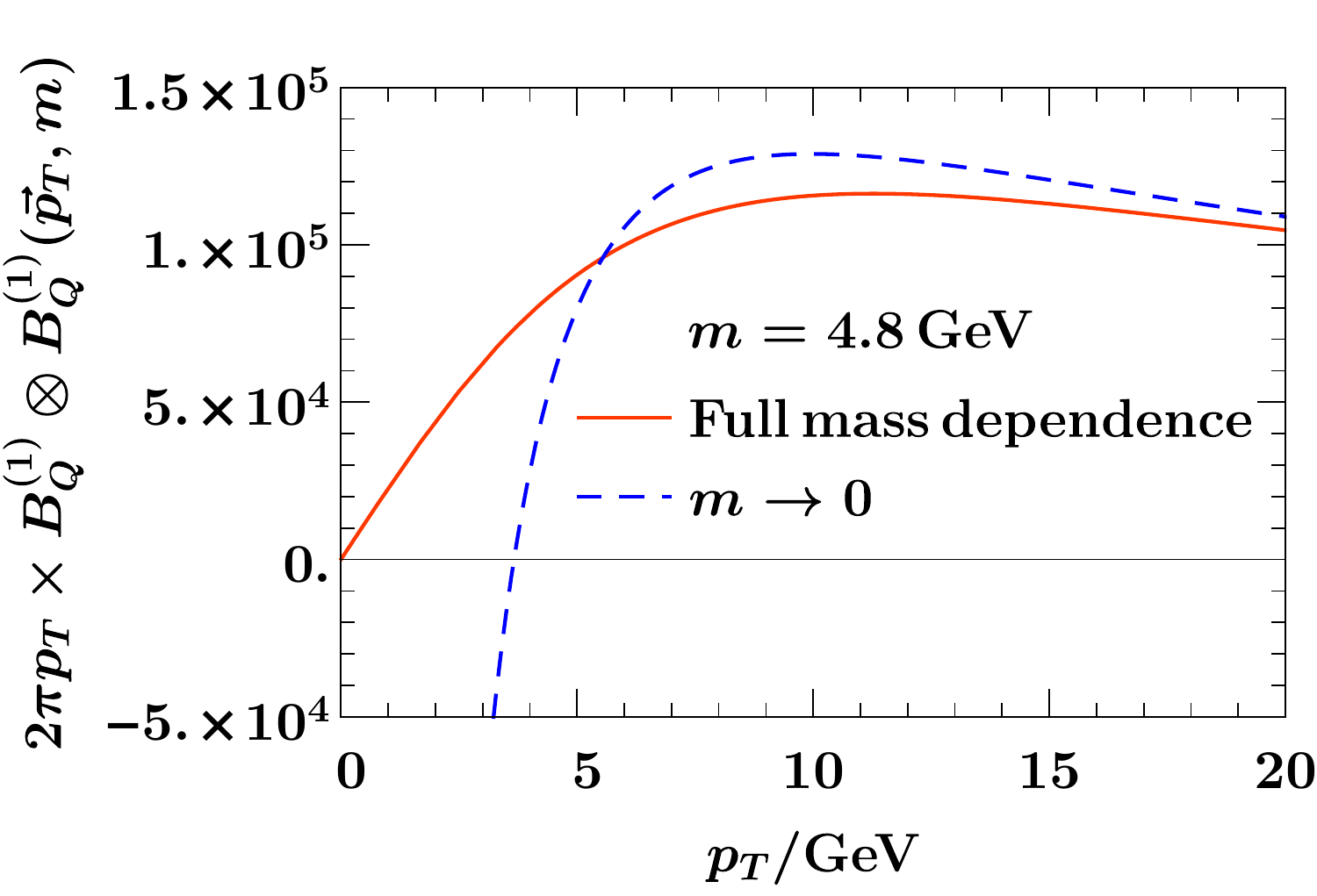}%
\hfill%
\includegraphics[height=5.3 cm]{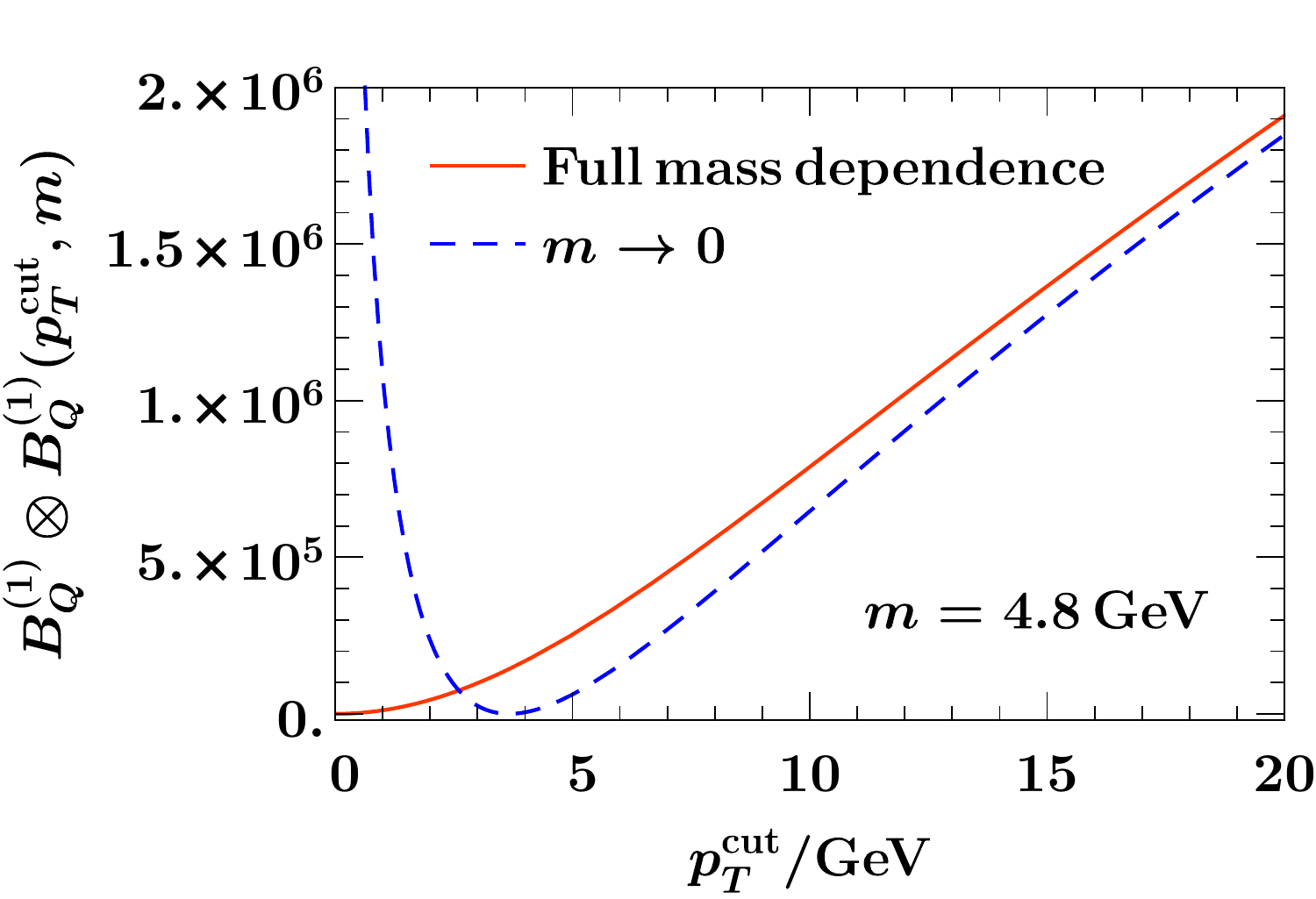}%
\caption{Convolution of two massive $b$-quark beam functions together with the result in the $m\to 0$ limit differential in the total $p_T\equiv|\vec p_T|$ (left panel) and the corresponding cumulant (right panel). This is proportional to the primary massive quark correction to the $Z$-boson spectrum at $\mathcal{O}(\alpha_s^2)$.}
\label{fig:doublebottomBeamfct}
\end{figure}
 
In \fig{doublebottomBeamfct} we show the result for the convolution between two massive quark beam functions,
\begin{align}
B^{(1)}_Q \otimes B^{(1)}_Q (\vec{p}_T,m) \equiv \int\! \df^2 p_T' \, B^{(1)}_Q(\vec{p}_T-\vec{p}^{\,\prime}_T,m) \, B^{(1)}_Q(\vec{p}^{\,\prime}_T,m) \,,
\end{align}
which enters the result for $Z$-boson production at $\mathcal{O}(\alpha_s^2 T_F^2)$ and NNLL$'$. The analytic expression for the convolution between the two one-loop mass-dependent coefficients is given in \eq{qTfo_primary}. We see that now the corrections to small-mass limit remain nonnegligible even for larger values of $p_T$. This is due to the fact that the $\vec{p}_T$-convolution generates a logarithmic dependence in the spectrum, such that the power corrections of $\mathcal{O}(m_b^2/p_T^2)$ become enhanced by logarithms $ \ln(p_T^2/m_b^2)$.

Next, we consider the secondary massive quark corrections at $\mathcal{O}(\alpha_s^2 C_F T_F)$. The result for the mass-dependent TMD beam function coefficient $\mathcal{I}^{(2,h)}_{qq}(\vec p_T, m, z)$ is given in \eq{TMD_Iqq2_massive}. In the decoupling limit $p_T\ll m$ all its terms without distributions in $\vec{p}_T$ give $\mathcal{O}(p_T^2/m^2)$ power-suppressed contributions. Combining its remaining distributional terms with the contributions arising from changing the $\alpha_s$ scheme from $n_l+1$ to $n_l$ flavors yields
\begin{align}
\mathcal{I}_{qq}^{(2,h)}\Bigl(\vec{p}_T,m,z,\mu,\frac{\nu}{\omega}\Bigr)-\frac{4}{3}L_m\mathcal{I}_{qq}^{(1)}\Bigl(\vec{p}_T,z,\mu,\frac{\nu}{\omega}\Bigr) =\delta^{(2)}(\vec{p}_T)\,\delta(1-z)\,H_c^{(2)}\Bigl(m,\mu,\frac{\nu}{\omega}\Bigr)+\mathcal{O}\Bigl(\frac{p_T^2}{m^2}\Bigr)
\,,\end{align}
confirming the relation in \eq{consistencyqT_beam1} at this order.
The massless one-loop coefficient $\mathcal{I}_{qq}^{(1)}$ and the collinear mass-mode function $H_c^{(2)}$ can be found in \eqs{TMD_Iqq1}{Hn2}, respectively. On the other hand, in the limit $m\ll p_T$ we get
\begin{align}\label{eq:Iqq_smallm}
 &\mathcal{I}_{qq}^{(2,h)}\Bigl(\vec{p}_T,m,z,\mu,\frac{\nu}{\omega}\Bigr)=\mathcal{I}^{(2,l)}_{qq}\Bigl(\vec{p}_T,z,\mu,\frac{\nu}{\omega}\Bigr)+\delta^{2}(\vec{p}_T) \,\mathcal{M}_{qq}^{(2)}(m,z,\mu)+\mathcal{O}\Bigl(\frac{m^2}{p_T^2}\Bigr)\;,
\end{align}
such that all infrared mass dependence is given by the PDF matching, as required by the relation in \eq{consistencyqT_beam2}. The results for the massless coefficient and the PDF matching coefficient are given in eqs.~\eqref{eq:TMD_Iqq2} and \eqref{eq:Mqq2}, respectively.

For the coefficient $\mathcal{I}_{qg}^{(2,h)}$ at $\mathcal{O}(\alpha_s^2 T_F^2)$ the limiting behavior is trivial, since it vanishes identically in the $(n_l)$-flavor scheme for $\alpha_s$, and in the $(n_l+1)$-flavor scheme for $\alpha_s$ it is exactly
\begin{align}
T_F\,\mathcal{I}_{qg}^{(2,h)}(\vec{p}_T,m,z,\mu) = \mathcal{I}_{qg}^{(1)}(\vec{p}_T,z,\mu)  \otimes_z \mathcal{M}^{(1)}_{gg}(m,z,\mu) \, .
\end{align}

The mass-dependent TMD soft function is given in \eq{pT_soft_massive_ren}. In the limit $p_T \ll m$ all its terms without distributions in $\vec{p}_T$ become $\mathcal{O}(p_T^2/m^2)$ power suppressed, just as for the beam function. Combining its remaining distributional terms with the contributions arising from changing the scheme of the strong coupling from $n_l+1$ to $n_l$ flavors yields
\begin{align}
 S^{(2,h)}(\vec{p}_T,m,\mu,\nu)-\frac{4}{3}L_m S^{(1)}(\vec{p}_T,\mu,\nu)=\delta^{(2)}(\vec{p}_T)\,H_s^{(2)}(m,\mu,\nu)+\mathcal{O}\Bigl(\frac{p_T^2}{m^2}\Bigr)
\,,\end{align}
confirming the relation in \eq{consistencyqT_soft1}.
The massless one-loop TMD soft function $S^{(1)}$ and the softmass-mode function $H_s^{(2)}$ are given in \eqs{TMD_soft_massless_1}{Hs2}, respectively. Since the soft function is free of IR singularities, the limit $m \ll p_T$ just yields the massless soft function in \eq{TMD_soft_massless_2},
\begin{align}\label{eq:S_smallm}
S^{(2,h)}(\vec{p}_T,m,\mu,\nu)= S^{(2,l)}(\vec{p}_T,\mu,\nu)+\mathcal{O}\Bigl(\frac{m^2}{p_T^2}\Bigr)
\,.\end{align}

\begin{figure}
\includegraphics[height=5.3cm]{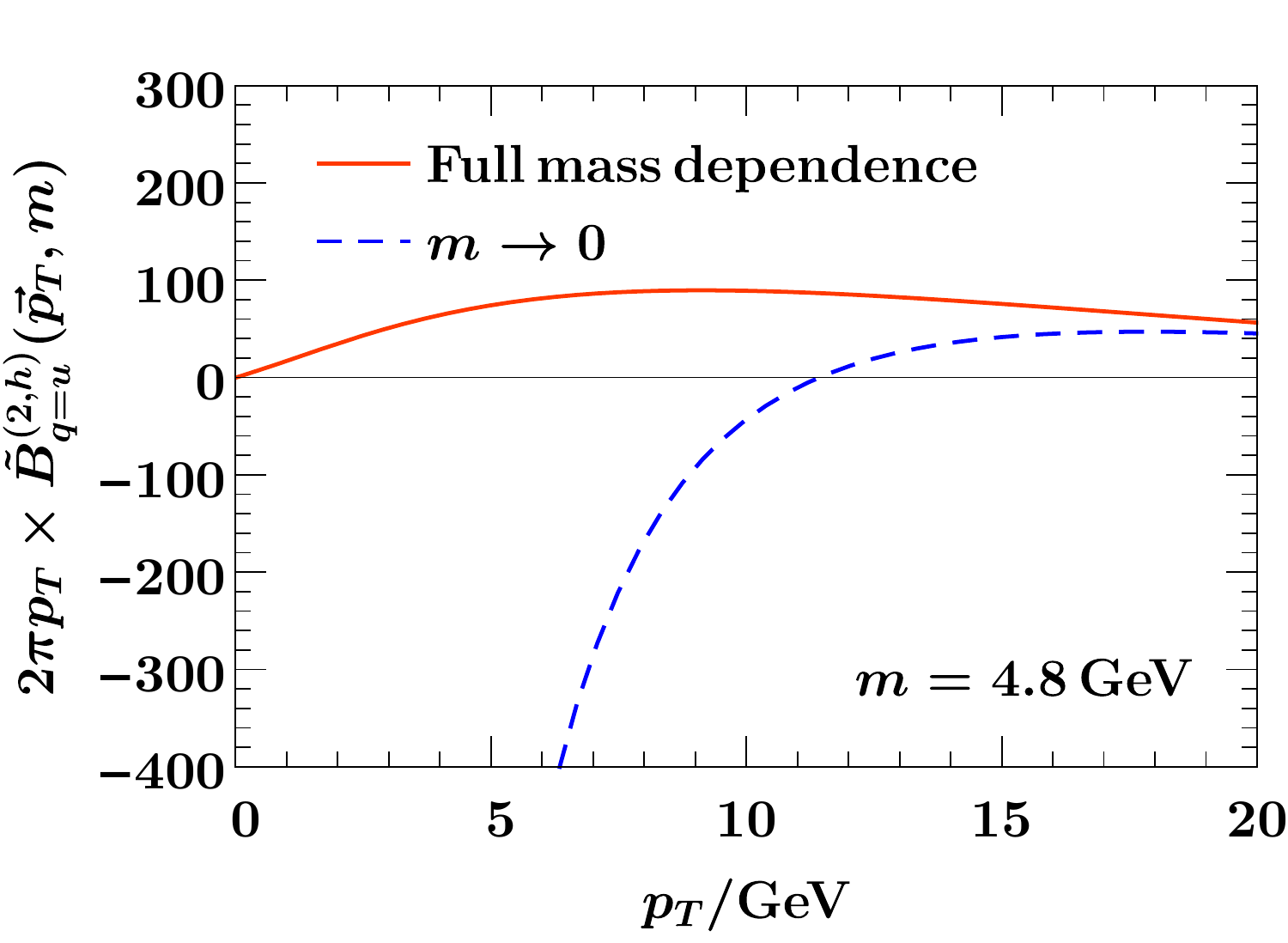}%
\hfill%
\includegraphics[height=5.2cm]{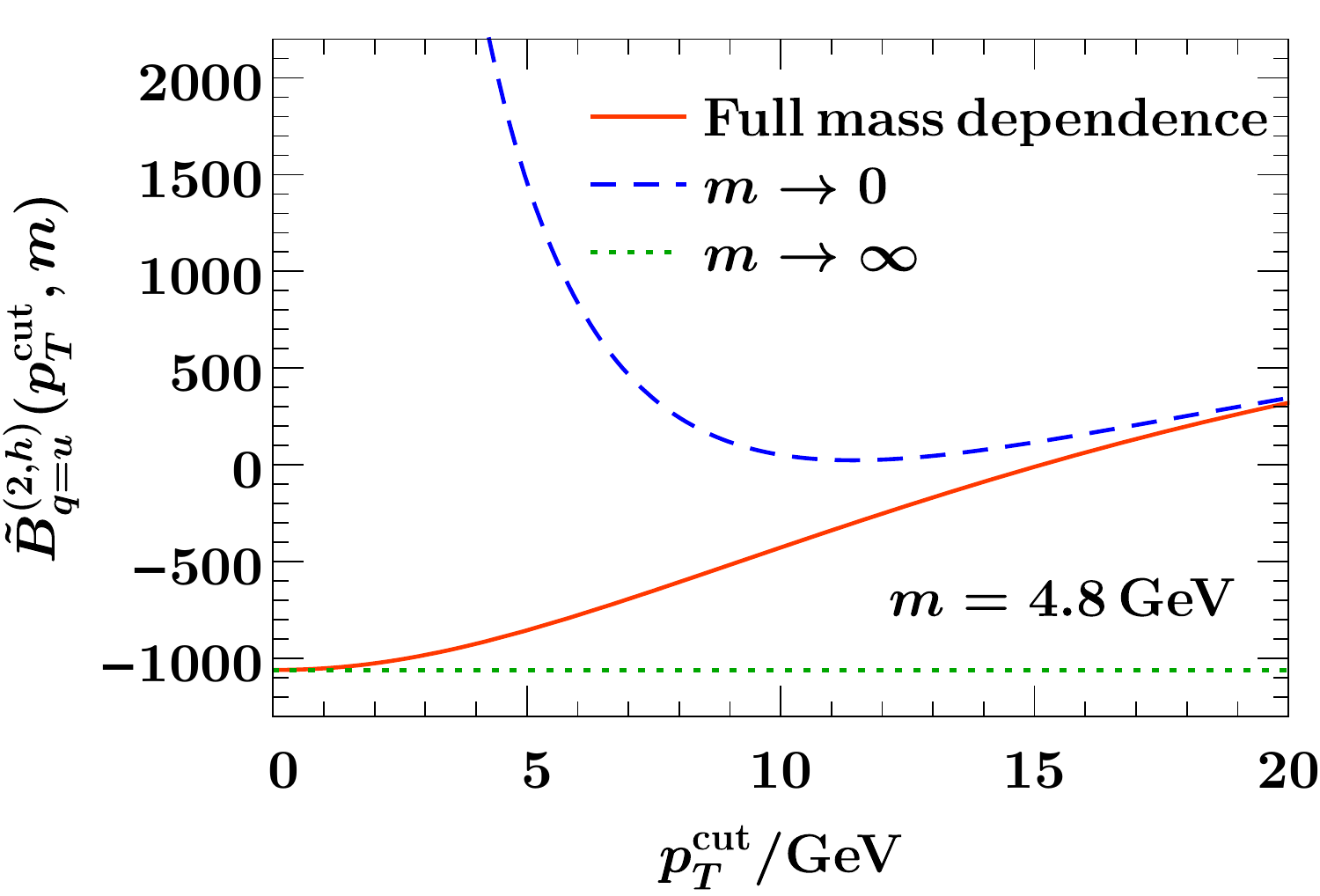}%
\caption{Secondary massive bottom quark corrections to the $u$-quark beam function (left panel) and its cumulant (right panel) at $\mathcal{O}(\alpha_s^2 C_F T_F)$ (including the square-root of the soft function here) for $\mu=m_b$ as a function of $q_T$.}
\label{fig:secondary_beamsoft}
\end{figure}

We now discuss the numerical impact of the $\mathcal{O}(m^2/p_T^2)$ terms from secondary mass effects.
Since the individual results for the beam and soft functions depend on the specific regularization scheme, we consider their symmetrized combination
\begin{align}
\tilde{B}_q(\vec{p}_{T},m,\omega,x,\mu) =  \int \df^2 p_T' \,  B_q\Bigl(\vec{p}_{T}-\vec{p}^{\,\prime}_{T},m,x,\mu,\frac{\nu}{\omega}\Bigr) \, \sqrt{S(\vec{p}^{\,\prime}_{T},m,\mu,\nu)}
\,,\end{align}
which is independent of $\nu$.%
\footnote{This combination is sometimes used as definition of a TMD PDF. $\tilde{B}_q$ contains large rapidity logarithms, which are resummed once the soft and beam functions are evaluated at their natural rapidity scales and evolved to a common scale $\nu$. For demonstrating the size of the power corrections here, we evaluate it at fixed order.}
The $\mathcal{O}(\alpha_s^2 C_F T_F)$ corrections explicitly depend on $\mu$ and the flavor-number scheme, but the difference between the full result and the small mass limits given in \eqs{Iqq_smallm}{S_smallm} do not.
In \fig{secondary_beamsoft} we show the result for the $\mathcal{O}(\alpha_s^2 C_F T_F)$ corrections (with $\alpha_s = \alpha_s^{(n_l+1)}$) to the $u$-quark beam function, both differential in $p_T$ and the corresponding cumulant.
We see that the full mass dependent results correctly reproduce the small and large mass limits.
The corrections to the massless are much larger than for the primary mass effects. In particular, they are still of
$\mathcal{O}(100\%)$ for $p_T^{(\rm cut)} \sim 10$ GeV. This clearly indicates that for secondary radiation involving two massive quarks in the final state the corrections are rather of $\mathcal{O}(4m^2/p_T^2)$, as one might expect.

\subsubsection{Limiting behavior for $\Tau$}

We carry out the discussion for beam thrust in close analogy. The virtuality-dependent massive quark beam function coefficient at one loop is given in \eq{virtuality_beam_coefficient}. In the limit $t \ll m^2$ the primary massive quarks correctly decouple,
\begin{align}
\mathcal{I}_{Qg}^{(1)}(t,m,z)=\mathcal{O}\Bigl(\frac{t}{m^2}\Bigr)
\,.\end{align}
In the opposite limit $m^2\ll t$ we get
\begin{align}
\mathcal{I}_{Qg}^{(1)}(t,m,z)&=T_F\,\theta(1-z)\theta(z)\biggl\{2P_{qg}(z) \,\frac{1}{\mu^2}\mathcal{L}_0\Bigl(\frac{t}{\mu^2}\Bigr) +\delta(t)\biggl[2P_{qg}(z)\biggl(-L_m+\ln\frac{1-z}{z}\biggr)
\nn \\ & \qquad
+4z(1-z)\biggr]\biggr\}+\mathcal{O}\left(\frac{m^2}{t}\right)
\nn \\
&=\mathcal{I}_{qg}^{(1)}(t,z,\mu)+\delta(t)\,\mathcal{M}_{Qg}^{(1)}(m,z,\mu)+\mathcal{O}\Bigl(\frac{m^2}{t}\Bigr)\;,
\end{align}
as required by the relation~\eqref{eq:consistencyTau_beam2}.
The massless one-loop matching coefficient $\mathcal{I}_{qg}^{(1)}$ and the PDF matching coefficient $\mathcal{M}_{Qg}^{(1)}$ are given in \eqs{Virt_Iqq1}{quark_gluon_PDF_matching}, respectively.

The secondary massive quark corrections to the virtuality-dependent beam function are given in \eq{Virt_Iqq2_massive}. In the decoupling limit $t\ll m^2$ all its nondistributional terms become $\mathcal{O}(t/m^2)$ power suppressed. Combining the remaining distributional terms in $t$ with the contributions arising from changing the scheme of the strong coupling from $n_l+1$ to $n_l$ flavors yields
\begin{align}
\mathcal{I}_{qq}^{(2,h)}\Bigl(t,m,z,\mu,\frac{\nu}{\omega}\Bigr) -\frac{4}{3}L_m\mathcal{I}_{qq}^{(1)}(t,z,\mu)=\delta(t)\delta(1-z)\, H_c^{(2)}\Bigl(m,\mu,\frac{\nu}{\omega}\Bigr)+\mathcal{O}\Bigl(\frac{t}{m^2}\Bigr)
\,,\end{align}
in agreement with \eq{consistencyTau_beam1}.
The massless result for $\mathcal{I}_{qq}^{(1)}$ and the collinear mass-mode function $H_c^{(2)}$ are given in \eqs{Virt_Iqq1}{Hn2}, respectively. In the limit $m^2\ll t$ we get
\begin{align}
&\mathcal{I}_{qq}^{(2,h)}\Bigl(t,m,z,\mu,\frac{\nu}{\omega}\Bigr)
\nn \\ & \quad
=\mathcal{I}^{(2,l)}_{qq}(t,z,\mu)+\delta(t)\,\mathcal{M}_{qq}^{(2)}(m,z,\mu)+\delta(1-z)\,\frac{1}{\omega}\,\mathcal{S}_c^{(2)}\Bigl(\frac{t}{\omega},m,\mu,\nu\Bigr)+\mathcal{O}\Bigl(\frac{m^2}{t}\Bigr)
\,.\end{align}
All infrared mass dependence is contained in the PDF matching coefficient and the csoft function, as required by \eq{consistencyTau_beam2}. The functions on the right-hand side are given in eqs.~\eqref{eq:Virt_Iqq2}, \eqref{eq:Mqq2}, and \eqref{eq:csoft}, respectively.

\begin{figure}
\includegraphics[height=5.3cm]{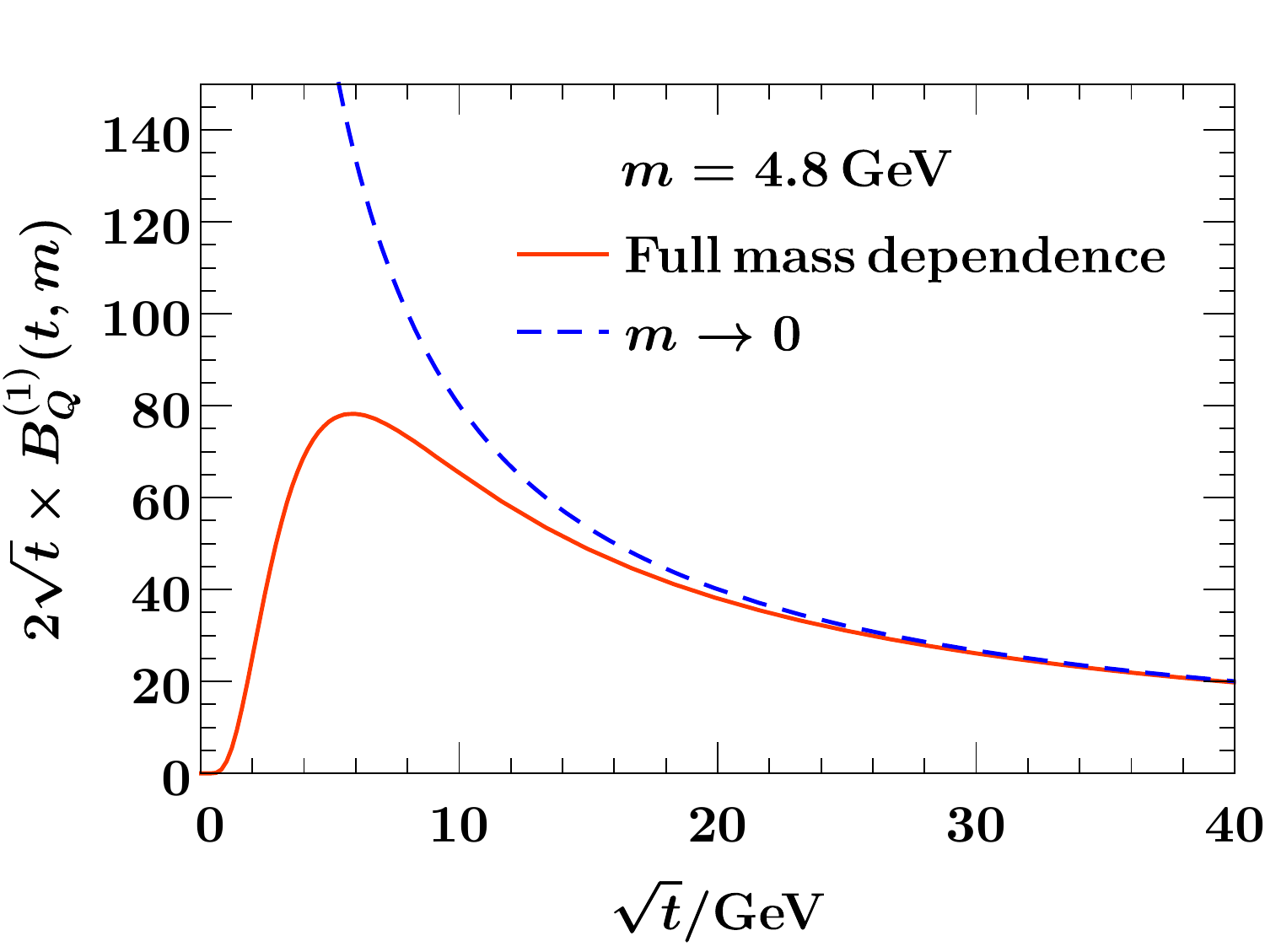}%
\hfill%
\includegraphics[height=5.3cm]{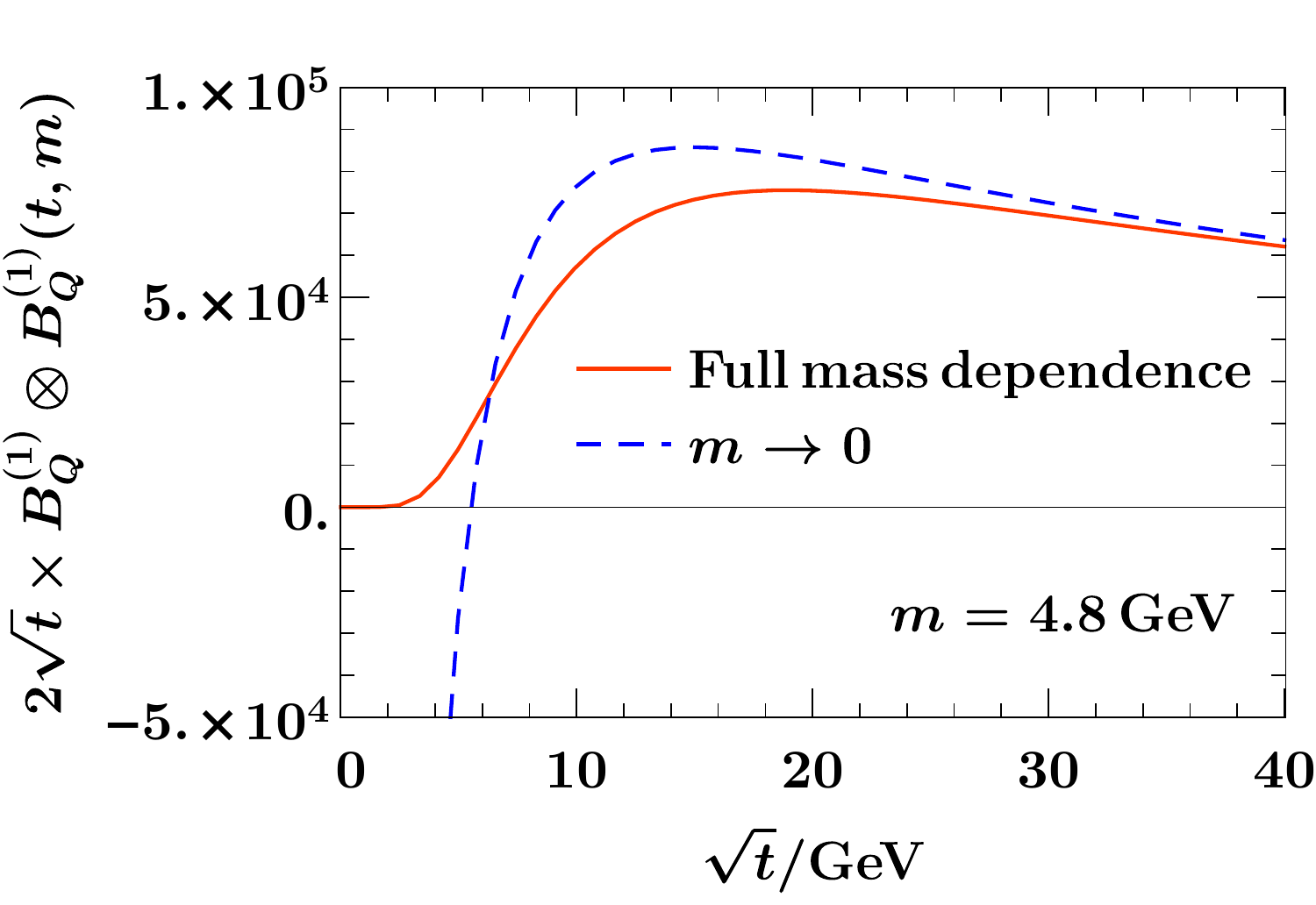}%
\caption{Massive $b$-quark beam function (left panel) and the convolution between two of these (right panel) together with the $m\to 0$ limit as a function of $\sqrt{t} \sim \sqrt{Q \Tau}$.}
\label{fig:massiveBeamfct_thrust}
\end{figure}

\begin{figure}
\includegraphics[height=5.3cm]{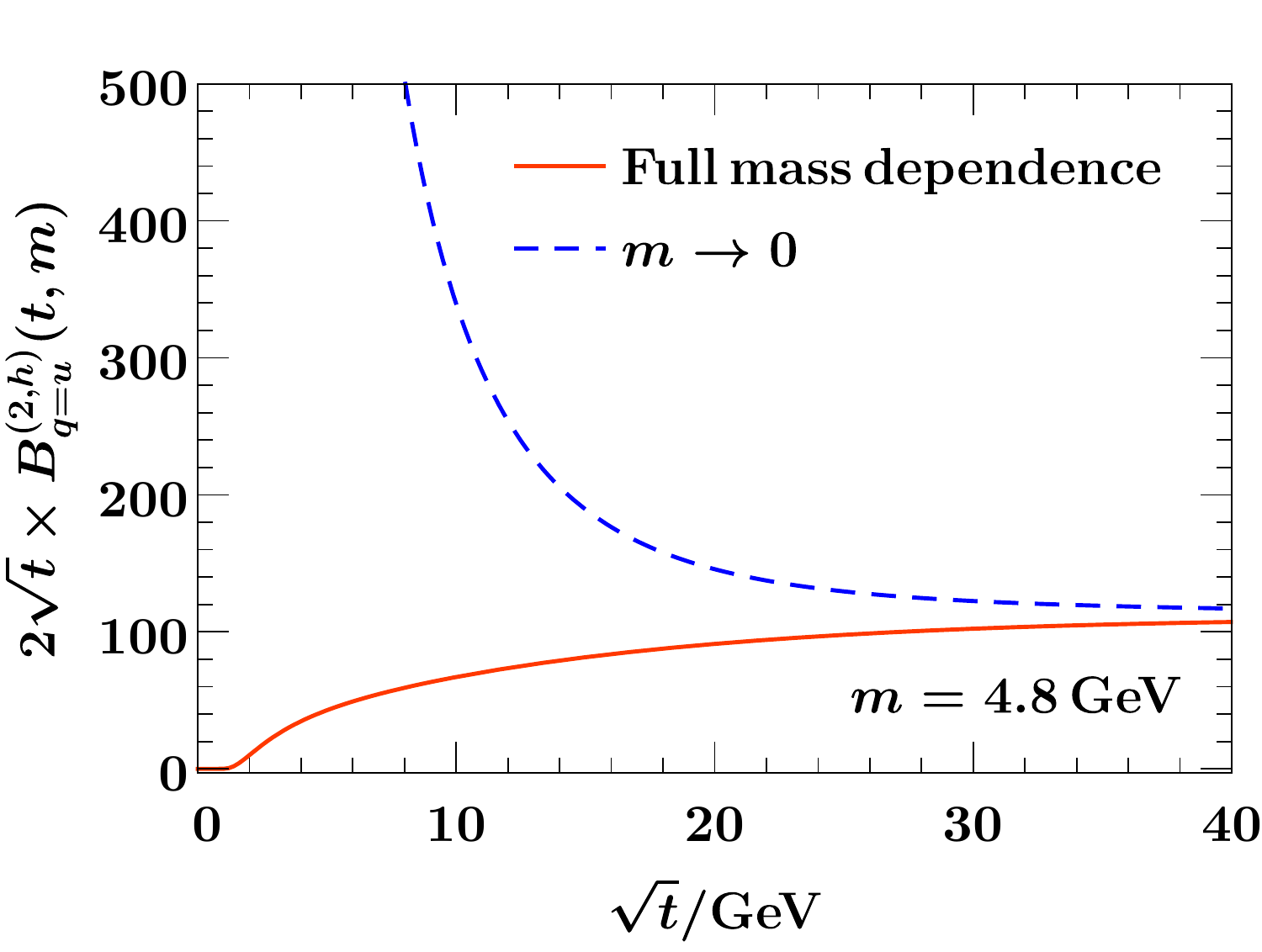}%
\hfill%
\includegraphics[height=5.3cm]{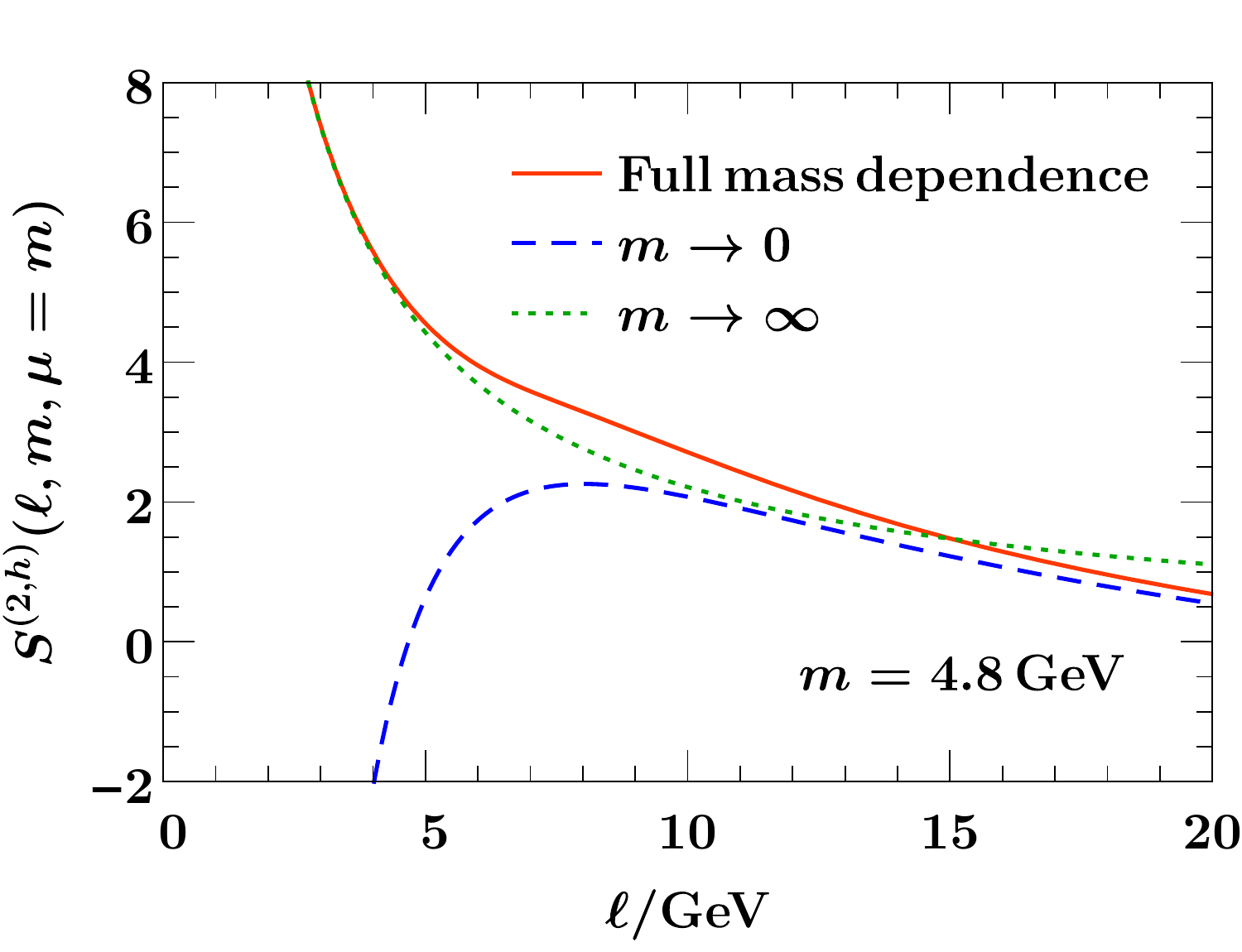}
\caption{Secondary massive $b$-quark corrections to the $u$-quark beam function for $Y=0$ (left panel) and the soft function (right panel) at $\mathcal{O}(\alpha_s^2 C_F T_F)$ for $\mu=m_b$ as functions of $\sqrt{t} \sim \sqrt{Q \Tau}$ and $\ell \sim \Tau$, respectively.}
\label{fig:secondary_beamsoft_thrust}
\end{figure}

The mass-dependent corrections to the (beam) thrust soft function are given in \eq{soft_tau}. In the limit $\ell \ll m$ all its nondistributional terms become  $\mathcal{O}(\ell^2/m^2)$ power suppressed. Combining the remaining distributional terms with the contributions arising from changing the scheme of the strong coupling from $n_l+1$ to $n_l$ flavors yields
\begin{align}
S^{(2,h)}(\ell,m,\mu) -\frac{4}{3}L_m  S^{(1)}(\ell,\mu)
=\delta(\ell)\,H_s^{(2)}(m,\mu,\nu)+\mathcal{S}_c^{(2)}(\ell,m,\mu,\nu)+\mathcal{O}\Bigl(\frac{\ell^2}{m^2}\Bigr)
\,,\end{align}
in agreement with \eq{consistencyTau_soft1}.
The massless one-loop thrust soft function $S^{(1)}$, the soft mass-mode function $H_s^{(2)}$, and the csoft function $\mathcal{S}_c^{(2)}$ can be found in eqs.~\eqref{eq:soft_massless_1}, \eqref{eq:Hs2}, and \eqref{eq:csoft}, respectively.
For $m \ll \ell$ the correct massless result is recovered,
\begin{align}
 S^{(2,h)}(\ell,m,\mu) = S^{(2,l)}(\ell,\mu)  +\mathcal{O}\Bigl(\frac{m^2}{\ell^2}\Bigr) \, ,
\end{align}
which was already checked in \refcite{Gritschacher:2013tza}.

In \fig{massiveBeamfct_thrust}, we show the numerical results for the one-loop massive beam function and the convolution between two of these (which is the leading order correction from primary massive quarks for the $Z$-boson production) as a function of $\sqrt{t} \sim \sqrt{ Q \Tau}$. The mass effects become relevant for $\sqrt{t} \sim m_b \sim 5 $ GeV (corresponding to $\Tau \lesssim 1 $ GeV for $Q=m_Z$). The corrections to the massless limit for the convolution of two beam functions is nonnegligible also for larger values.
In \fig{secondary_beamsoft_thrust}, we show the result for the secondary $\mathcal{O}(\alpha_s^2 C_F T_F)$ corrections to the beam and soft function. The corrections to the massless limit for the beam function remain sizeable even for $\sqrt{t} \gtrsim 2 m_b$. For the soft function, the mass effects are important for $\Tau \sim \ell \sim m_b$ and become small for $\ell >  10 \, {\rm GeV} \sim 2 m_b$.
Note that the small bump in the soft function in \fig{secondary_beamsoft_thrust} originates from the correction term $\Delta S_{\tau}$ in \eq{soft_tau}. The associated correction in the massless limit is fully contained in the $\delta(\ell)$ term.

\section{Rapidity evolution}
\label{sec:nu_evolution}

Here, we discuss the solutions of the rapidity RGEs in \eq{Hs_nuevolution}, or equivalently \eqs{BTau_nuevolution}{Csoft_nuevolution}, and in particular the rapidity evolution for the mass-dependent soft function in \eq{BSm_nuevolution} for $q_T \sim m$, where the massive quark corrections give rise to a different running than for massless flavors. Our primary aim here is to highlight the different features with respect to the massless case, while leaving the practical implementation for future work.

The rapidity evolution for the mass-mode matching functions $H_s$ and $H_c$ according to \eq{Hs_nuevolution} has been discussed in \refcite{Hoang:2015vua}. The evolution for the beam thrust beam function and csoft function according to \eqs {BTau_nuevolution}{Csoft_nuevolution} is completely analogous. For example, the $\nu$-evolved soft matching function $H_s$ is given by
\begin{align}
H_s(m,\mu,\nu) &= V(m,\mu,\nu,\nu_0) \, H_s(m,\mu,\nu_0)
\,,\\\nn
V(m, \mu, \nu , \nu_0) &= \exp \biggl\{\Big[ 4\eta_\Gamma^{(n_l)}(\mu_0(m),\mu) -4\eta_\Gamma^{(n_l+1)}(\mu_0(m),\mu)
 + \gamma_{\nu, H_s}(m,\mu_0(m))  \Big]  \ln \frac{\nu}{\nu_0} \biggr \}
\, .\end{align}
The evolution function $\eta_\Gamma$ is defined by
\begin{align}
\eta_\Gamma^{(n_f)}(\mu_0,\mu)
&= \int_{\mu_0}^\mu  \frac{\df\mu'}{\mu'}\, \Gamma^{(n_f)}_{\rm{cusp}}[\alpha_s^{(n_f)}(\mu^\prime)]
\,,
\end{align}
and resums the $\mu$-dependent logarithms inside the $\nu$ anomalous dimension as required by consistency with the
$\mu$ evolution to maintain the path independence in $\mu$-$\nu$-space~\cite{Chiu:2012ir}.
With the canonical scale choice
\begin{align}\label{eq:mu0_m}
\mu_0(m) =m \, ,
\end{align}
all logarithmic terms in the boundary condition $\gamma_{\nu,H_s}(m, \mu_0(m))$ are minimized.

The solution of the rapidity RGE for the soft function is substantially more involved due to its
two-dimensional convolution structure on $\vec{p}_T$.
The formal solution of the rapidity RGE for massless quarks in \eq{RGE_ml} is most conveniently found by Fourier transforming to impact parameter space with $b=|\vec{b}|$, where the rapidity RGE becomes multiplicative
\begin{align}\label{eq:nusoft_b}
\nu \frac{\df}{\df \nu} \,\tilde{S}^{(n_f)}(b,\mu,\nu) = \tilde{\gamma}^{(n_f)}_{\nu,S}(b,\mu) \, \tilde{S}^{(n_f)}(b,\mu,\nu)
\,.\end{align}
The consistency (path independence) between $\mu$ and $\nu$ evolution requires the rapidity anomalous dimension in Fourier space to satisfy
\begin{align}\label{eq:rapidity_mu_dependence}
\mu\frac{\df}{\df\mu}\,\tilde{\gamma}^{(n_f)}_{\nu,S}(b,\mu) = -4\,\Gamma_{\rm{cusp}}[\alpha_s^{(n_f)}(\mu)]
\,.\end{align}
Its solution is given by
\begin{align}\label{eq:resummed_gamma}
 \tilde{\gamma}^{(n_f)}_{\nu,S}(b,\mu) = -4\eta_\Gamma^{(n_f)}(\mu_0(b), \mu) + \tilde{\gamma}^{(n_f)}_{\nu,S}(b,\mu_0(b))
\,.
\end{align}
The logarithms of $\ln(\mu \, b \, e^{\gamma_E}/2)$ in the second boundary term are eliminated by the canonical scale choice
\begin{align} \label{eq:mu0_b}
 \mu^{(l)}_0(b) = \frac{2\,e^{-\gamma_E}}{b}
\,.\end{align}
With this choice, the $\nu$ evolution of the soft function in Fourier space at any given scale $\mu$ is given by
\begin{align}
\tilde S(b,\mu,\nu) = \tilde S(b,\mu,\nu_0) \exp\biggl[ \tilde{\gamma}^{(n_f)}_{\nu,S}(b,\mu)\,\ln \frac{\nu}{\nu_0} \biggr]
\,.\end{align}
As is well known, the rapidity evolution kernel becomes intrinsically nonperturbative at $1/b \ll \Lambda_{\rm QCD}$~\cite{Collins:1981uk, Collins:1981va, Collins:1984kg}. This nonperturbative sensitivity appears through the resummed rapidity anomalous dimension, which with the canonical scale choice in \eq{mu0_b} gets evaluated at $\alpha_s(1/b)$.
It is important to note that this is not an artefact of performing the evolution in  Fourier space.
Rather this is a physical effect, which also happens when the $\nu$ evolution is consistently performed in momentum space. As shown in \refcite{Ebert:2016gcn}, in this case the appropriate resummed result for $\gamma_{\nu,S}(\vec{p}_T,\mu)$ explicitly depends on $\alpha_s(p_T)$, which means it becomes nonperturbative for $p_T \lesssim \Lambda_{\rm QCD}$.

For the massive quark corrections in the regime $q_T \sim m$ the $\mu$ dependence of the rapidity anomalous dimension is the same as for the massless quarks, i.e.~eq.~\eqref{eq:rapidity_mu_dependence}, such that
\begin{align}
 \tilde{\gamma}^{(h)}_{\nu,S}(b,m,\mu)=4\eta_\Gamma^{(n_l)}(\mu_0(b,m),\mu) -
 4\eta_\Gamma^{(n_l+1)}(\mu_0(b,m),\mu) + \tilde{\gamma}^{(h)}_{\nu,S}(b,m,\mu_0(b,m)) \, .
\end{align}
Here $\tilde{\gamma}^{(h)}_{\nu,S}$ denotes only the contributions of the massive flavor to the full anomalous dimension.
The explicit mass dependence arises in the $\mu$-independent boundary contribution, which depends on both $b$ and $m$. From the relations in eqs.~\eqref{eq:consistencyqT_soft2} and \eqref{eq:consistencyqT_soft1} we can directly infer the limiting behavior to the anomalous dimension,
\begin{align}\label{eq:gammaS_limits}
\tilde{\gamma}_{\nu,S}(b,m,\mu) &=\tilde{\gamma}_{\nu,S}^{(n_l+1)}(b,\mu)+\mathcal{O}(m^2b^2)
\,,\nn \\
\tilde{\gamma}_{\nu,S}(b,m,\mu)&=\tilde{\gamma}_{\nu,S}^{(n_l)}(b,\mu)+\gamma_{\nu, H_s}(m,\mu)+\mathcal{O}\Bigl(\frac{1}{m^2b^2}\Bigr)
\,.\end{align}
This means that the massive quark corrections $\tilde{\gamma}^{(h)}_{\nu,S}$  are the same as for a massless flavor in the limit $m\ll1/b$ and are the same as the rapidity anomalous dimension of the soft mass mode function $H_s$ in the limit $1/b \ll m$, provided one uses the $(n_l+1)$ and $(n_l)$-flavor scheme for $\alpha_s$, respectively. To eliminate the logarithms inside $\tilde{\gamma}^{(h)}_{\nu,S}$, the canonical scale choice $\mu_0(b,m)$ should behave like the massless case for $m\ll  1/b$ and like the choice for the mass-mode matching functions for $m \gg 1/b$,
\begin{align}\label{eq:mu0_massive}
\mu^{(h)}_0(b,m)&\sim \mu^{(l)}_0(b) = \frac{2\,\mathrm{e}^{-\gamma_E}}{b}  && \text{for}\quad 1/b \to \infty
\, , \nn \\
\mu^{(h)}_0(b,m)&\sim m  && \text{for}\quad  1/b \to 0
\,.\end{align}
Since $\mu^{(h)}_0(b,m)$ freezes out naturally at the perturbative mass scale for $1/b \to 0$, the nonperturbative sensitivity in the $\nu$ evolution gets regulated by the quark mass for the massive quark contributions.

\begin{figure}
\centering
\includegraphics[width=0.5\textwidth]{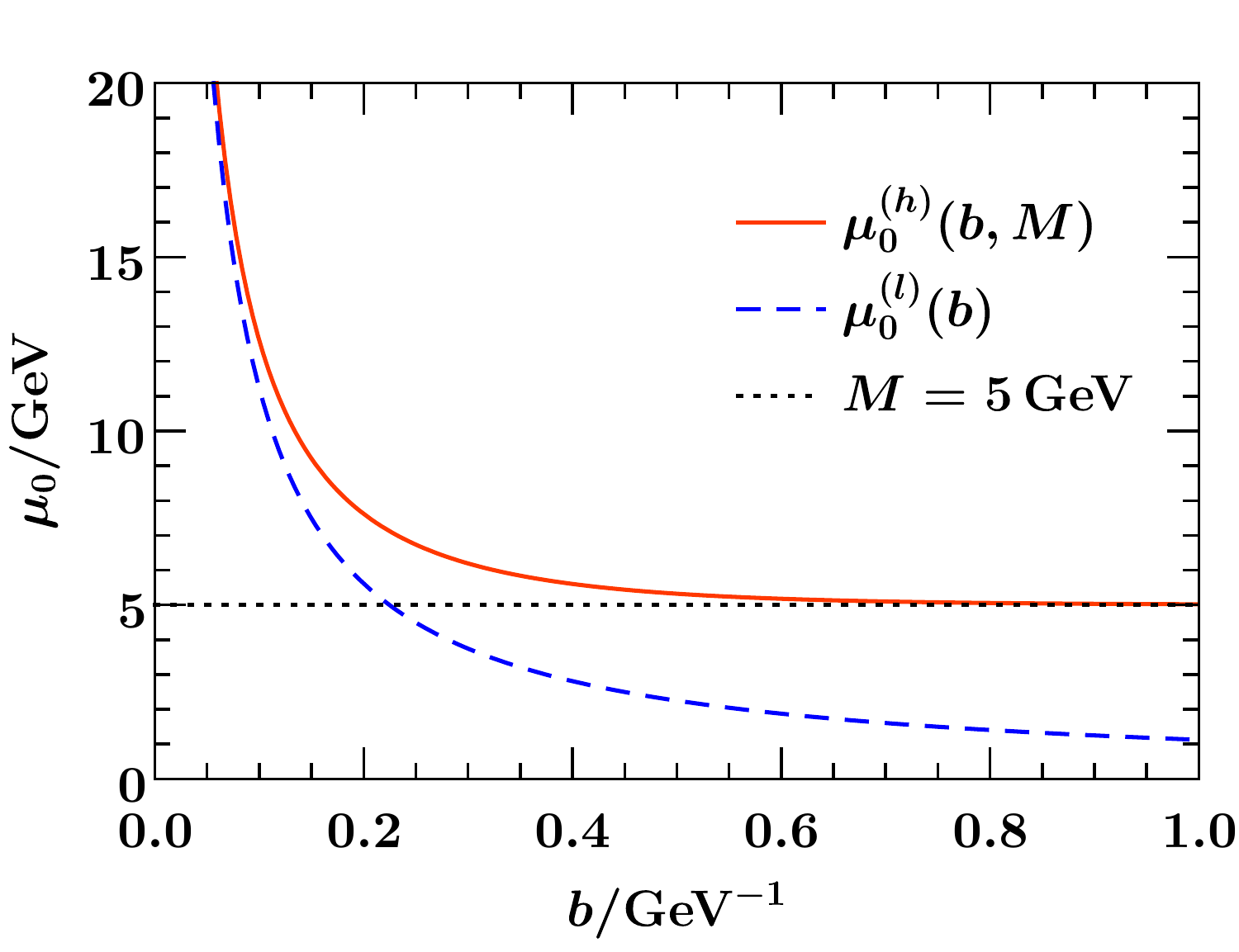}
\caption{The canonical scales $\mu^{(h)}_0(b,M)$ for the massive case (red, solid) and $\mu_0^{(l)}(b) = \mu_0(b,M=0)$ for the massless case (blue, dashed) with $M= 5$ GeV.}
\label{fig:M_scale}
\end{figure}

We first illustrate this behavior in a simple one-loop toy example: We consider the radiation of a massive gluon (with mass $M$) having the same couplings as a (massless) gluon in QCD, which exhibits the main features of the full results for secondary massive quarks. The associated corrections are obtained in the calculations of app.~\ref{sec:massive_gluon_TMD_soft} as intermediate results for the two-loop case. In $b$-space the one-loop rapidity anomalous dimensions for massless and massive gluons are given by
\begin{align} \label{eq:gammaS_M}
\tilde{\gamma}_{\nu,S}^{(1)}(b,\mu) &= -C_F \,8L_b
\,, \nn \\
\tilde{\gamma}_{\nu,S}^{(1)}(b,M,\mu) &=C_F \,\Bigl[ 8L_M + 16K_0(bM) \Bigr]
\,,\end{align}
where $K_0$ denotes the modified Bessel function of the second kind and
\begin{align}\label{eq:Lb}
L_b \equiv\ln\frac{b^2\mu^2\mathrm{e}^{2\gamma_E}}{4}
\, , \qquad
L_M \equiv \ln\frac{M^2}{\mu^2}
\,.\end{align}
The mass-dependent result has the limiting behavior
\begin{align}
\tilde{\gamma}_{\nu,S}^{(1)}(b,M,\mu) & =-C_F\,8L_b+\mathcal{O}(M^2b^2) \, ,\nn \\
\tilde{\gamma}_{\nu,S}^{(1)}(b,M,\mu) & =C_F\,8L_M+\mathcal{O}\Bigl(\frac{1}{M^2b^2}\Bigr)\;,
\end{align}
in close analogy to \eq{gammaS_limits}. A natural choice to eliminate any large terms in \eq{gammaS_M} in both limits is
\begin{align}\label{eq:mu0_M}
 \mu^{(h)}_0(b,M) = M\,e^{K_0(bM)}
\,.\end{align}
for which $\tilde{\gamma}^{(1)}_{\nu,S}(b,M,\mu^{(h)}_0(b,M))$ just vanishes. The behavior of this choice as a function of $b$ compared to the massless result is shown in fig.~\ref{fig:M_scale}.

For the full secondary massive quark corrections at $\mathcal{O}(\alpha_s^2)$ the Fourier transform of \eq{gammaS_m} reads (expanded in terms of $\alpha_s^{(n_l+1)}$ as in \eq{alphas_expansion})
\begin{align}
\tilde{\gamma}_{\nu,S}^{(2,h)}(b,m,\mu)
&=C_F\,\biggl\{-\frac{32}{3}L_bL_m-\frac{16}{3}L_m^2-\frac{160}{9}L_m-\frac{448}{27}
\nn \\ &\quad
+\frac{8\sqrt{\pi}}{3}\Bigl[ 2\,\MeijerG[\Big]{3}{0}{1}{3}{\frac{3}{2}}{0,0,0}{m^2b^2}+\MeijerG[\Big]{3}{0}{1}{3}{\frac{5}{2}}{0,0,1}{m^2b^2}\Bigr]\biggr\}
\,,\end{align}
where $G$ denotes a Meijer G function. This result has the limiting behavior
\begin{align}
\tilde{\gamma}_{\nu,S}^{(2,h)}(b,m,\mu)
&= C_F\,\biggl(\frac{16}{3}L_b^2+\frac{160}{9}L_b+\frac{448}{27}\biggr) + \mathcal{O}(m^2b^2)
\, , \nn \\
\tilde{\gamma}_{\nu,S}^{(2,h)}(b,m,\mu)-\frac{4}{3}L_m\tilde{\gamma}_{\nu,S}^{(1)}(b,\mu)
&=C_F\,\biggl(-\frac{16}{3}L_m^2-\frac{160}{9}L_m-\frac{448}{27}\biggr) + \mathcal{O}\Bigl(\frac{1}{m^2b^2}\Bigr)
\,.\end{align}
Hence, the correct massless limit is recovered, while in the large-mass limit one obtains the anomalous dimension in \eq{gammaHs}. Note that one needs to perform a change for the strong coupling between the $n_l+1$ and $n_l$ flavor schemes to obtain both limits correctly. To minimize the logarithms for any regime one should thus adopt a canonical scale choice that satisfies \eq{mu0_massive}, as for example in \eq{mu0_M}.

\section{Outlook: Phenomenological impact for Drell-Yan}\label{sec:outlook}

Our results can be applied to properly take into account bottom quark mass effects for the Drell-Yan $q_T$ spectrum at NNLL$'$.
While a full resummation analysis is beyond the scope of this paper, we can estimate the potential size of the
quark-mass effects by looking at the fixed-order $q_T$ spectrum.

\begin{figure}
\includegraphics[height=5.3cm]{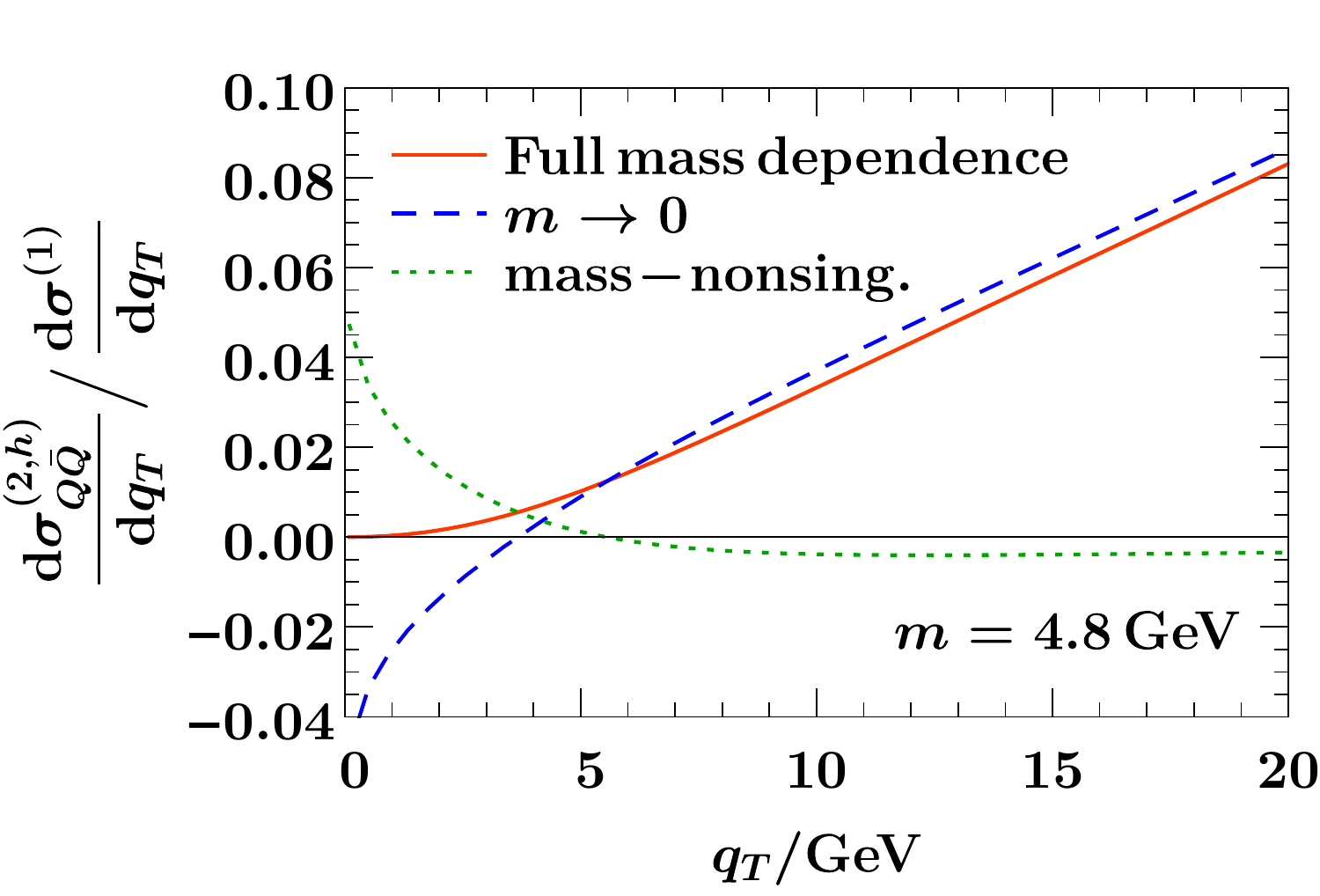}%
\hfill%
\includegraphics[height=5.3cm]{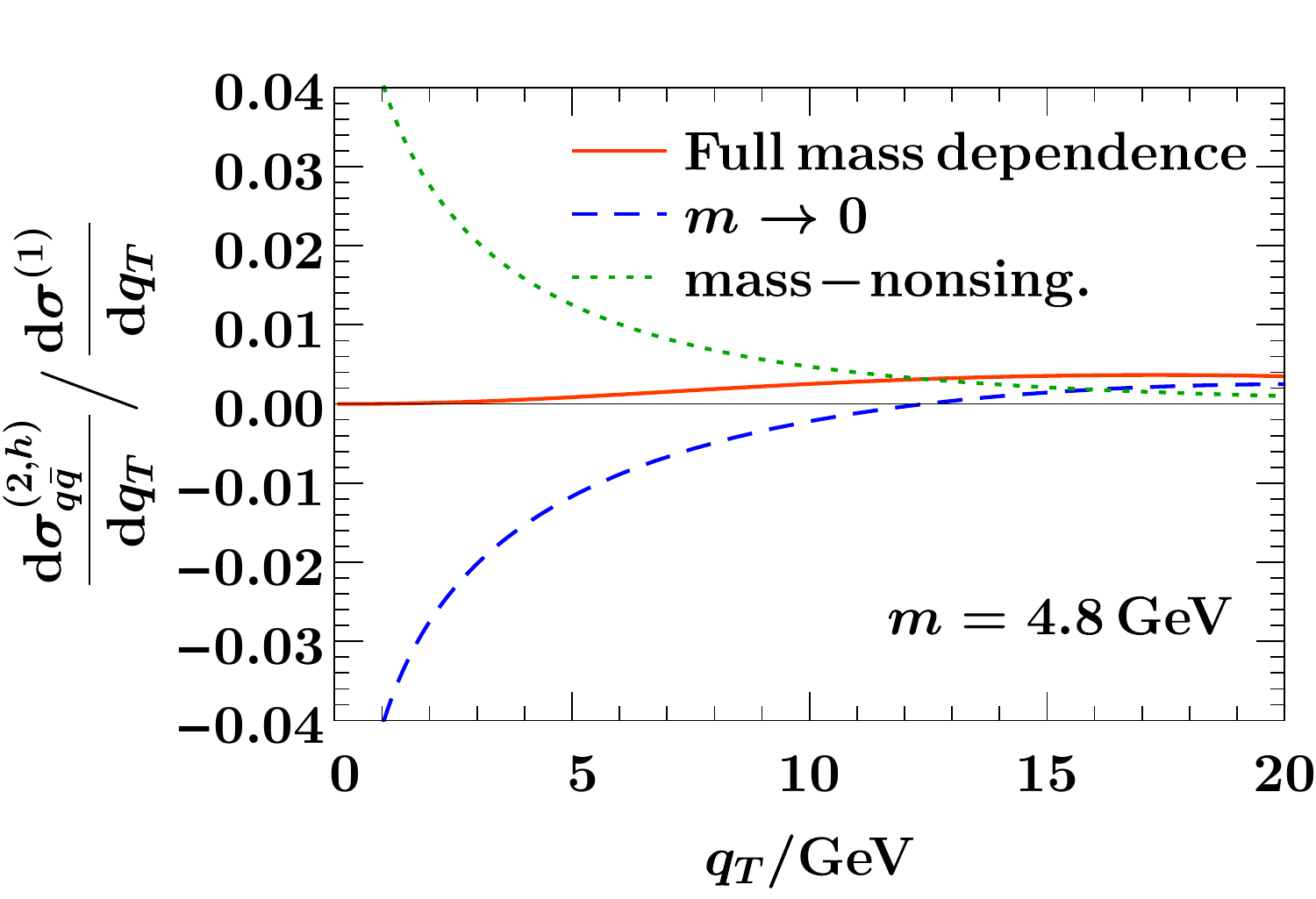}%
\caption{Primary (left panel) and secondary (right panel) massive bottom quark contributions
for the $Z$-boson $q_T$ spectrum at fixed $\mathcal{O}(\alpha_s^2 T_F^2)$ and $\mathcal{O}(\alpha_s^2 C_F T_F)$, respectively.
The results are given relative to the full $\ord{\alpha_s}$ result including all flavors.}
\label{fig:dsigmaZ}
\end{figure}

In \fig{dsigmaZ}, we show separately the contributions from primary and secondary massive quarks
to the cross section at $\mathcal{O}(\alpha_s^2)$, normalized to the $\ord{\alpha_s}$ spectrum $\df \sigma^{(1)}$ including all flavors (treating the charm as a massless flavor).
We utilize the MMHT2014 NNLO PDFs~\cite{Harland-Lang:2014zoa} and evaluate the contributions
for $\mu=m_b=4.8 $ GeV, $Q=m_Z$, $Y=0$, and $E_{\rm cm}=13$ TeV.
Note that the secondary mass contributions at $\ord{\alpha_s^2}$ are explicitly $\mu$-dependent and scheme-dependent,
the nonsingular mass correction, i.e.~the difference between the full massive result for $\mu \sim m_b$
and the massless limit (encoded partially in a massive PDF), is $\mu$ independent at this order.
As can be seen, the relative contribution of the $b\bar{b}$-initiated channel grows with larger $q_T$,
while the impact of the secondary contributions including the full mass dependence
is at the sub-percent level throughout the spectrum. As expected, the nonsingular mass corrections are
very small for $m_b \ll q_T$, but can reach the order of percent for $q_T \sim m_b$,
which roughly corresponds to the peak region of the distribution where the cross section is largest.

\begin{figure}
\includegraphics[height=5.5cm]{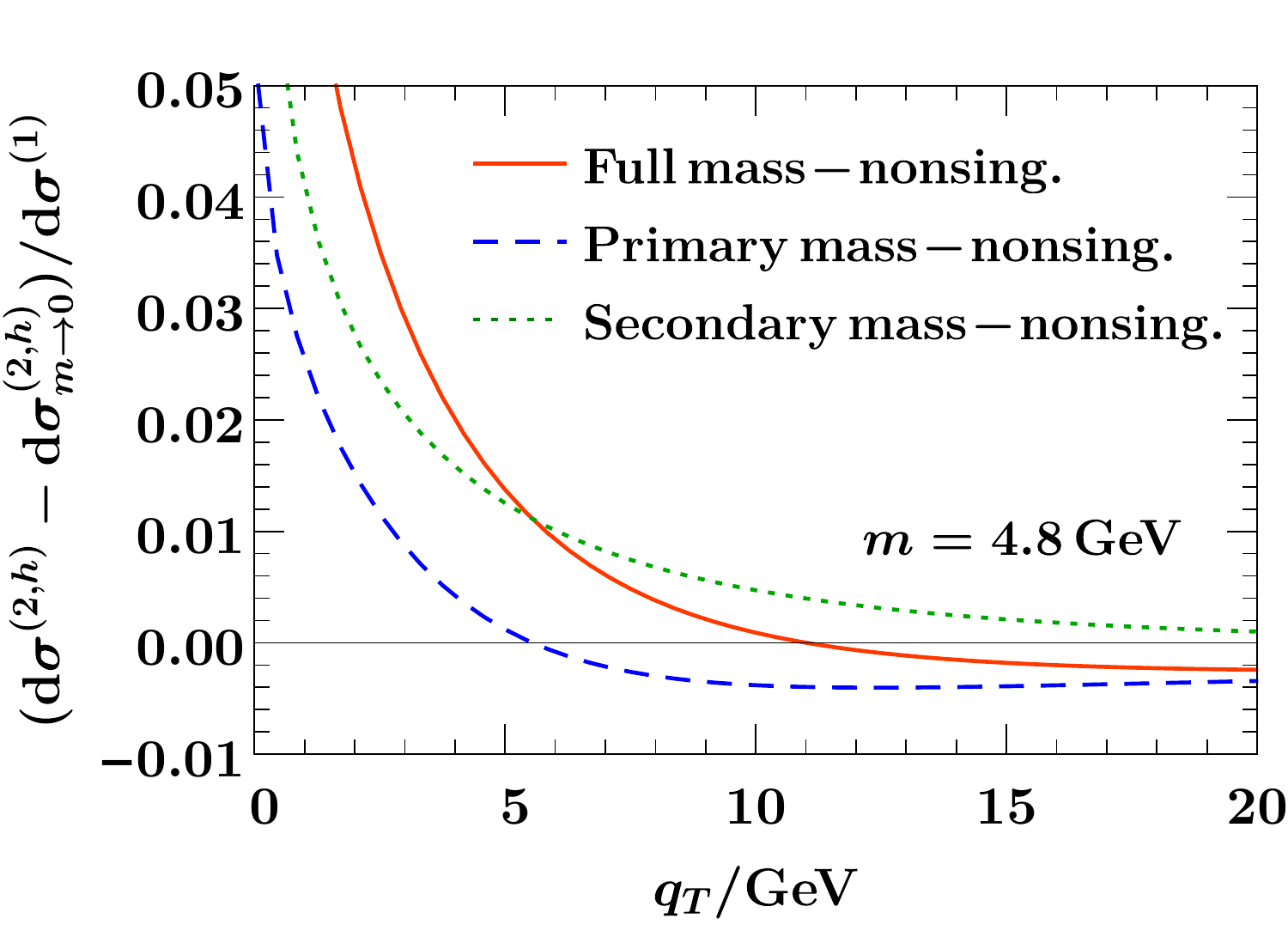}%
\hfill%
\includegraphics[height=5.5cm]{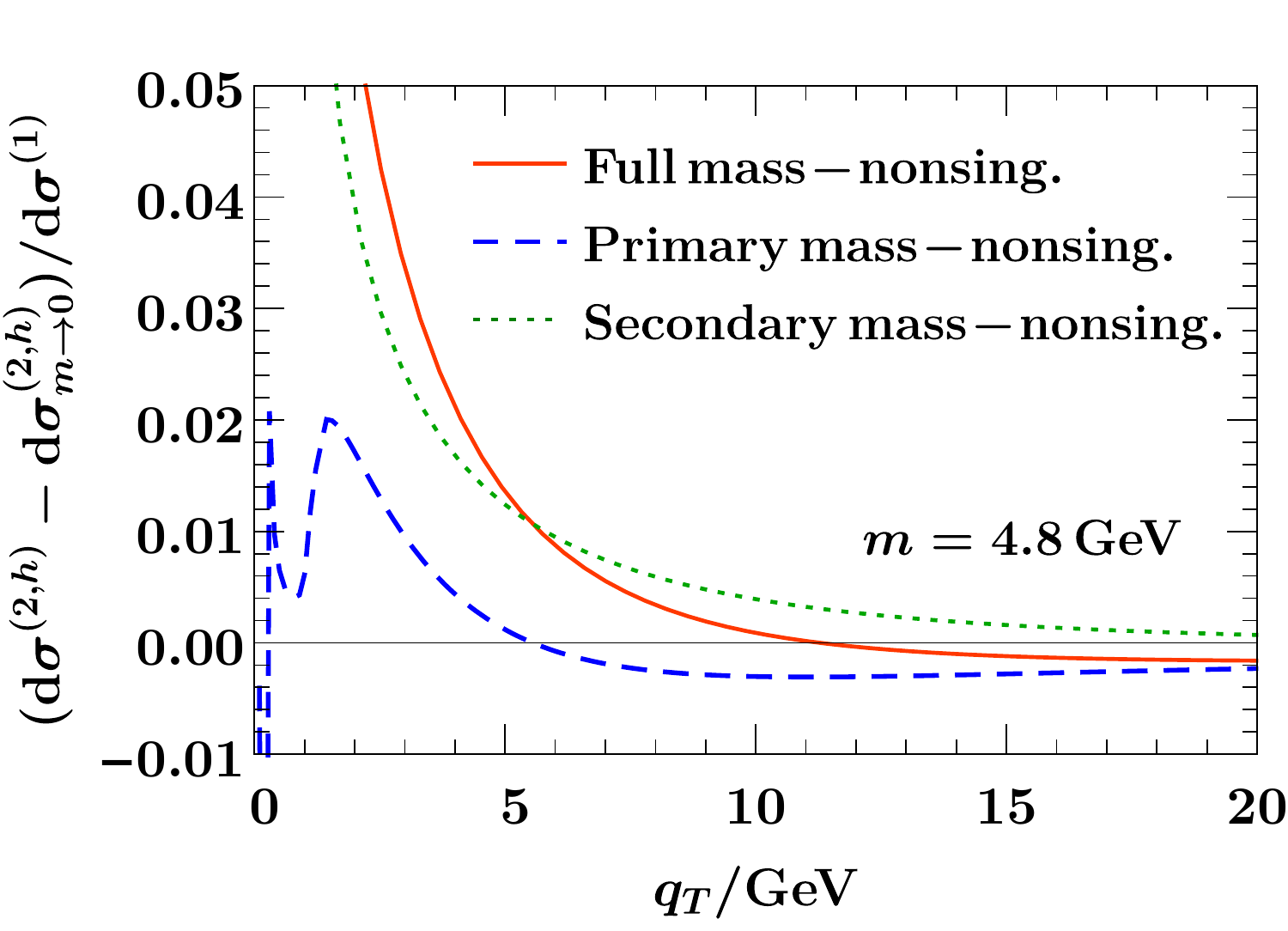}%
 \caption{Different types of mass nonsingular corrections for $Z$-boson production at $\mu=m_b$ (left panel)
 and $\mu=q_T$ (right panel).}
 \label{fig:dsigmaZ_total}
\end{figure}

The same can also be seen in \fig{dsigmaZ_total}, where we show the mass nonsingular corrections to the massless limit
for primary and secondary contributions as well as their sum. They are shown for $\mu =m_b$ on the left and for $\mu=q_T$
on the right. We see that these corrections are (at fixed order) indeed only weakly dependent on the value of $\mu$ (for $q_T \gtrsim 2$ GeV). All in all, the bottom quark mass can have a relevant effect for high precision predictions of the $q_T$-spectrum at the order of percent around the peak of the distribution ($\sim 5$ GeV).
Below the peak of the distribution the fixed-order result is of course not expected to give a reliable quantitative result,
and furthermore nonperturbative corrections become important in this regime.
Nevertheless, we expect the qualitative features like the sign and order of magnitude of the mass
effects to provide an indication for the behaviour of the full resummed result.

For $W$ production sizable corrections from bottom quark effects arise only through secondary contributions (due to the strong CKM suppression of the primary contributions), which have a similar impact as for $Z$-production. On the other hand, charm-initiated production plays an important role and enters already at $\mathcal{O}(\alpha_s)$. Estimating the nonsingular mass corrections for $q_T \sim m_c$ is more subtle, since higher-order corrections in the strong coupling and nonperturbative effects are likely to dominate the effect from the known beam function at $\mathcal{O}(\alpha_s)$ at these low scales. Thus, we do not attempt to determine their characteristic size here and leave this to future work. An analysis based on the leading-order matrix element and its potential impact on the determination of $m_W$ can be found in \refcite{Berge:2005rv}.

\section{Conclusions}
\label{sec:conclusions}

Massive quark effects provide a challenge for high-precision predictions at colliders.
Using a SCET-based factorization framework, we have discussed how to systematically incorporate
massive quark corrections into exclusive differential cross sections at the LHC,
using the measurement of the transverse momentum $q_T$ and beam thrust for Drell-Yan production
as prototypical examples. We have discussed the relevant factorization setup for the different
hierarchies between the mass scale and the other relevant kinematic scales.
We find that the presence of (secondary) massive quarks can lead to the emergence or alteration of rapidity logarithms
thus changing the resummation structure in a nontrivial way.

The generic framework for the description of mass effects generalizes to other exclusive cross sections
with different jet-resolution measurements and final-state kinematic cuts, which will require additional calculations
of the relevant factorization ingredients. Our results for the beam thrust spectrum allow for a systematic inclusion of massive quark effects into the Geneva Monte-Carlo program~\cite{Alioli:2012fc, Alioli:2015toa}
at NNLL$'+$NNLO in its underlying jet resolution variable.
Several of our results are also immediately relevant for other processes besides Drell-Yan.
The massive quark beam functions are relevant for any heavy-quark initiated
process, for example exclusive $b \bar{b} H$-production. The mass-dependent soft function and rapidity anomalous dimension at $\mathcal{O}(\alpha_s^2)$ satisfy Casimir scaling and can be therefore also utilized for the description of gluon-fusion processes, e.g.~the Higgs $q_T$-spectrum. 

An important application of our framework is to the precise theoretical description of the
Drell-Yan $q_T$ spectrum. To this end, we have computed all required mass-dependent beam
and soft functions up to $\mathcal{O}(\alpha_s^2)$ allowing for the description of massive quark
effects in the Drell-Yan $q_T$ spectrum at NNLL$'$.
In particular, our results provide an important ingredient for a detailed investigation of quark-mass effects in the ratio of $W$ and $Z$ boson spectra at small $q_T$, which is important for the precision measurement of the $W$-boson mass at the LHC.

\begin{acknowledgments}
We thank Markus Ebert and Maximilian Stahlhofen for useful discussions.
This work was supported by the German Science Foundation (DFG) through the Emmy-Noether Grant No.~TA 867/1-1 and the Collaborative Research Center (SFB) 676 Particles, Strings and the Early Universe. D.S. is supported by the Austrian Science Fund (FWF) under the Doctoral Program No. W1252-N27 Particles and Interactions. We also thank the Erwin-Schr{\"o}dinger Institute (ESI) for partial support in the framework of the ESI program Challenges and Concepts for Field Theory and Applications in the Era of LHC Run-2.
\end{acknowledgments}

\appendix
\addtocontents{toc}{\protect\setcounter{tocdepth}{1}}

\section{Results for massless quarks}

Here we summarize the relevant results with massless quarks for the hard, beam, and soft functions.

\subsection{Hard function}

The massless quark hard function is directly related to the QCD form factor and has been computed at $\mathcal{O}(\alpha_s^2)$ in ref.~\cite{Matsuura}. The $\mathcal{O}(\alpha_s)$ and $\mathcal{O}(\alpha_s^2C_FT_F)$ corrections read in an expansion in terms of $\alpha_s =\alpha_s^{(n_f)}(\mu)$ in analogy to \eq{alphas_expansion} (with $L_Q=\ln(Q^2/\mu^2)$)
\begin{align}\label{eq:hard_massless}
H^{(1)}(Q,\mu) & = H^{(0)}(Q) \,C_F\,\Bigl(-2L_Q^2+6L_Q-16+\frac{7\pi^2}{3} \Bigr)
\,, \\
H^{(2,l)}(Q,\mu) &= H^{(0)}(Q) \,C_F\,\biggl[-\frac{8}{9}L_Q^3+\frac{76}{9}L_Q^2-\Bigl(\frac{836}{27}-\frac{16\pi^2}{9}\Bigr)L_Q+\frac{4085}{81}-\frac{182\pi^2}{27}+\frac{8\zeta_3}{9}\biggr]
\nn\,,\end{align}
where $H^{(0)}$ is the tree-level contribution. Note that for a single quark flavor there is in addition a nonvanishing correction to the axial current contribution relevant for $Z$-boson production, but cancels within an isospin doublet for massless quarks.

The anomalous dimensions are
\begin{align}\label{eq:gammaH}
\gamma_{H}^{(1)}(Q,\mu) &=C_F\,(8L_Q-12)
\,, \nn \\
\gamma_{H}^{(2,l)}(Q,\mu) &=C_F\,\biggl(-\frac{160}{9}L_Q+\frac{520}{27}+\frac{8\pi^2}{3}\biggr)
\,.\end{align}

\subsection{Beam functions}

\subsubsection{TMD beam function}

The matching coefficients entering the TMD beam function have been computed at $\mathcal{O}(\alpha_s^2)$ in various schemes~\cite{Catani:2012qa,Gehrmann:2012ze,Gehrmann:2014yya,Echevarria:2016scs} and are obtained for the symmetric $\eta$-regulator in \refcite{Luebbert:2016itl}. The results at $\mathcal{O}(\alpha_s)$ are
\begin{align}\label{eq:TMD_Iqg1}
\mathcal{I}_{qg}^{(1)}(\vec{p}_T,z,\mu) &=\theta(z)\theta(1-z)\,T_F\,\Bigl[ 2P_{qg}(z)\mathcal{L}_0(\vec{p}_T,\mu)+4z(1-z)\delta^{(2)}(\vec{p}_T) \Bigr]
\,, \\
\mathcal{I}_{qq}^{(1)}\Bigl(\vec{p}_T,z,\mu,\frac{\nu}{\omega}\Bigr)
&= \theta(z)\,C_F\,\Bigl\{\mathcal{L}_0(\vec{p}_T,\mu)\Bigl[-\Bigl(4\ln\frac{\nu}{\omega}+3\Bigr)\delta(1-z)+2P_{qq}(z)\Bigr] \nn \\
& \quad+2\delta^{(2)}(\vec{p}_T) \,\theta(1-z)(1-z) \Bigr\}
\,. \label{eq:TMD_Iqq1}
\end{align}
The splitting functions are
\begin{align}
P_{qg}(z) = z^2+(1-z)^2 \, , \qquad P_{qq}(z) = 2\mathcal{L}_0(1-z)+\frac{3}{2}\delta(1-z)-\theta(1-z)(1+z) \, .
\end{align}
At $\mathcal{O}(\alpha_s^2 C_F T_F)$ the massless matching coefficient is given by
\begin{align}\label{eq:TMD_Iqq2}
\mathcal{I}_{qq}^{(2,l)}\Bigl(\vec{p}_T,z,\mu,\frac{\nu}{\omega}\Bigr)&=\theta(z)\,C_F\biggl\{\mathcal{L}_1(\vec{p}_T,\mu)\biggl[\frac{16}{3}\mathcal{L}_0(1-z)-\frac{16}{3}\ln\frac{\nu}{\omega}\,\delta(1-z)-\frac{8}{3}\theta(1-z)(1+z)\biggr] 
\nn \\& \quad 
+\mathcal{L}_0(\vec{p}_T,\mu)\biggl[-\frac{80}{9}\mathcal{L}_0(1-z)+\frac{80}{9}\ln\frac{\nu}{\omega}\,\delta(1-z)
\nn \\ & \qquad
+\theta(1-z)\biggl(-\frac{8}{3}\frac{1+z^2}{1-z}\ln z+\frac{16}{9}+\frac{64z}{9}\biggr)\biggr]
\nn \\  & \quad
+ \delta^{(2)}(\vec{p}_T)\biggl[\frac{224}{27}\mathcal{L}_0(1-z)-\frac{224}{27}\ln\frac{\nu}{\omega}\,\delta(1-z)
\nn \\ &\qquad
+\theta(1-z)\biggl(\frac{2}{3}\frac{1+z^2}{1-z}\ln^2 z + \frac{20}{9}\frac{1+z^2}{1-z}\ln z-\frac{148}{27}-\frac{76z}{27}\biggr)\biggr]
\biggr\}
\,.\end{align}
The anomalous dimensions of the massless quark TMD beam function, as defined in \eq{RGE_ml}, are given at  $\mathcal{O}(\alpha_s)$ and $\mathcal{O}(\alpha_s^2C_FT_F)$ by
\begin{align}\label{eq:gammaB} 
\gamma_{B}^{(1)}\Bigl(\frac{\nu}{\omega}\Bigr)
&=C_F\,\Bigl( 8\ln\frac{\nu}{\omega}+6 \Bigr)
\,, \nn\\
\gamma_{B}^{(2,l)}\Bigl(\frac{\nu}{\omega}\Bigr)
&=C_F\,\Bigl( -\frac{160}{9}\ln\frac{\nu}{\omega}-\frac{4}{3}-\frac{16\pi^2}{9} \Bigr)
\,, \nn\\
\gamma_{\nu,B}^{(1)}(\vec{p}_T,\mu) &=-C_F \,4\mathcal{L}_0(\vec{p}_T,\mu)
\,, \nn\\
\gamma_{\nu,B}^{(2,l)}(\vec{p}_T,\mu)&=C_F\,\biggl[-\frac{16}{3}\mathcal{L}_1(\vec{p}_T,\mu)+\frac{80}{9}\mathcal{L}_0(\vec{p}_T,\mu)-\frac{224}{27}\delta^{(2)}(\vec{p}_T)\biggr]
\,.\end{align}

\subsubsection{Virtuality-dependent beam function}

The virtuality-dependent beam functions for massless quarks are known to two loop order \cite{Gaunt:2014xga,Gaunt:2014cfa}. The matching coefficients at $\mathcal{O}(\alpha_s)$ read
\begin{align}
\mathcal{I}_{qg}^{(1)}(t,z,\mu)
&= \theta(z)\theta(1-z)\,T_F\, \biggl\{2P_{qg}(z) \frac{1}{\mu^2}\mathcal{L}_0\Bigl(\frac{t}{\mu^2}\Bigr) +\delta(t)\Bigl[2P_{qg}(z)\ln\frac{1-z}{z} +4z(1-z)\Bigr]\biggr\}
\,, \nn \\
\mathcal{I}_{qq}^{(1)}(t,z,\mu)
&= \theta(z)\,C_F\, \biggl\{
 \frac{4}{\mu^2}\mathcal{L}_1\Bigl(\frac{t}{\mu^2}\Bigr) \delta(1-z)
+ \frac{1}{\mu^2}\mathcal{L}_0\Bigl(\frac{t}{\mu^2}\Bigr) \Bigl[2P_{qq}(z)-3 \delta(1-z)\Bigr]
\nn \\ & \quad
+ \delta(t)\biggl[4\mathcal{L}_1(1-z)-\frac{\pi^2}{3}\delta(1-z)
\nn \\ & \qquad
+\theta(1-z) \Bigl[ 2(1-z -2(1+z)\ln (1-z)- 2\frac{1+z^2}{1-z}\ln z \Bigr] \biggr]
\biggr\}
\label{eq:Virt_Iqq1}
\,.\end{align}
The massless matching coefficient at order $\mathcal{O}(\alpha_s^2C_FT_F)$ for one quark flavor reads
\begin{align}\label{eq:Virt_Iqq2}
&\mathcal{I}_{qq}^{(2,l)}(t,z,\mu)
\\ \quad
&= \theta(z)\,C_F \biggl\{
\frac{8}{3} \frac{1}{\mu^2} \mathcal{L}_2\Bigl(\frac{t}{\mu^2}\Bigr)\,\delta(1-z)
\nn \\ & \quad
+ \frac{1}{\mu^2} \mathcal{L}_1\Bigl(\frac{t}{\mu^2}\Bigr) \biggl[\frac{16}{3}\mathcal{L}_0(1-z)-\frac{80}{9}\delta(1-z)-\frac{8}{3}\theta(1-z)(1+z)\biggr]
\nn \\ & \quad
+ \frac{1}{\mu^2} \mathcal{L}_0\Bigl(\frac{t}{\mu^2}\Bigr)
\biggl[\frac{16}{3}\mathcal{L}_1(1-z)-\frac{80}{9}\mathcal{L}_0(1-z) +\delta(1-z)\Bigl(\frac{224}{27}-\frac{8 \pi ^2}{9}\Bigr)
\nn \\ & \qquad
+\theta(1-z)\biggl( -\frac{8}{3}(1+z)\ln(1-z)-\frac{16(1+z^2)}{3(1-z)}\ln z+\frac{16}{9}+\frac{64 z}{9}\biggr)\biggr]
\nn \\ & \quad
+ \delta(t)\biggl[\frac{8}{3}\mathcal{L}_2(1-z)-\frac{80}{9}\mathcal{L}_1(1-z)+\mathcal{L}_0(1-z)\Bigl(\frac{224}{27}-\frac{8\pi^2}{9}\Bigr)
\nn \\ & \qquad
 +\delta(1-z)\Bigl(-\frac{656}{81}+\frac{10 \pi ^2}{9}+\frac{40 \zeta_3}{9}\Bigr) +\theta(1-z) \biggl( -\frac{8(1+z^2)}{3(1-z)}\,\mathrm{Li}_2(1-z) 
\nn \\ & \qquad
-\frac{4}{3}(1+z)\ln^2(1-z)-\frac{16(1+z^2)}{3(1-z)}\ln(1-z)\ln z+\frac{10 \left(1+z^2\right)}{3 (1-z)}\ln^2 z 
\nn \\ & \qquad
+\Bigl(\frac{16}{9}+\frac{64z}{9}\Bigr)\ln(1-z)+\frac{4 (5-2z+7z^2)}{3 (1-z)}\ln z -\frac{148}{27}-\frac{76z}{27}+\frac{4\pi^2}{9}(1+z) \biggr) \biggr]
\biggr\}
\nn \,.\end{align}
The anomalous dimension of the massless quark beam function at order $\mathcal{O}(\alpha_s)$ and $\mathcal{O}(\alpha_s^2C_FT_F)$ are given by
\begin{align}\label{eq:gammaB_thrust}
\gamma_{B}^{(1)}(t,\mu)
&= C_F \biggl[-\frac{8}{\mu^2}\mathcal{L}_0\Bigl(\frac{t}{\mu^2}\Bigr) + 6\delta(t)\biggr]
\,, \nn \\
\gamma_{B}^{(2,l)}(t,\mu)
&=C_F\,\biggl[\frac{160}{9} \frac{1}{\mu^2}\mathcal{L}_0\Bigl(\frac{t}{\mu^2}\Bigr)
 + \delta(t)\Bigl(-\frac{484}{27}-\frac{8\pi^2}{9}\Bigr)\biggr]
\,.\end{align}

\subsection{Soft functions}

\subsubsection{TMD soft function}

The TMD soft function for massless quarks with the symmetric $\eta$-regulator has been computed at two loops in ref.~\cite{Luebbert:2016itl}. At $\mathcal{O}(\alpha_s)$ and $\mathcal{O}(\alpha_s^2C_FT_F)$ it is given by
\begin{align}\label{eq:TMD_soft_massless_1}
S^{(1)}(\vec{p}_T,\mu,\nu)
&=C_F\,\biggl[
-4\mathcal{L}_1(\vec{p}_T,\mu)
+8\ln\frac{\nu}{\mu}\, \mathcal{L}_0(\vec{p}_T,\mu)
-\frac{\pi^2}{3}\delta^{(2)}(\vec{p}_T)
\biggr]
\,, \\
\label{eq:TMD_soft_massless_2}
S^{(2,l)}(\vec{p}_T,\mu,\nu)
&=C_F\,\biggl[
-\frac{16}{3}\mathcal{L}_2(\vec{p}_T,\mu)
+\mathcal{L}_1(\vec{p}_T,\mu)\Bigl(\frac{32}{3}\ln\frac{\nu}{\mu}+\frac{80}{9}\Bigr)
-\mathcal{L}_0(\vec{p}_T,\mu)\Bigl(\frac{160}{9}\ln\frac{\nu}{\mu}+\frac{8\pi^2}{9}\Bigr)
\nn \\ & \quad
+ \delta^{(2)}(\vec{p}_T)\Bigl(\frac{448}{27}\ln\frac{\nu}{\mu}-\frac{656}{81}+\frac{10\pi^2}{9}-\frac{8\zeta_3}{9}\Bigr)
\biggr]
\,.\end{align}
The corresponding anomalous dimensions are
\begin{align}\label{eq:SpT}
\gamma^{(1)}_{S}(\mu,\nu) &=-C_F\, 16\ln\frac{\nu}{\mu}
\,, \nn \\
\gamma_{S}^{(2,l)}(\mu,\nu)&=C_F\,\biggl(\frac{320}{9}\ln\frac{\nu}{\mu}-\frac{448}{27}+\frac{8\pi^2}{9}\biggr)
\,, \nn \\
\gamma^{(1)}_{\nu,S}(\vec{p}_T,\mu) &=C_F\, 8\mathcal{L}_0(\vec{p}_T,\mu)
\,, \nn \\
\gamma_{\nu, S}^{(2,l)}(\vec{p}_T,\mu)&=C_F\,\biggl[
\frac{32}{3}\mathcal{L}_1(\vec{p}_T,\mu)
-\frac{160}{9}\mathcal{L}_0(\vec{p}_T,\mu)
+ \frac{448}{27}\delta^{(2)}(\vec{p}_T)
\biggr]
\,.\end{align}

\subsubsection{Thrust soft function}

The thrust soft function is known to two loops~\cite{Kelley:2011ng, Monni:2011gb}.
At $\mathcal{O}(\alpha_s)$ and $\mathcal{O}(\alpha_s^2C_FT_F)$ it is given by
\begin{align}\label{eq:soft_massless_1}
S^{(1)}(\ell,\mu)
&= C_F\, \biggl[- 16\frac{1}{\mu}\mathcal{L}_1\Bigl(\frac{\ell}{\mu}\Bigr)  +\frac{\pi^2}{3}\delta(\ell) \biggr]
\,, \\
S^{(2,l)}(\ell,\mu)
&=C_F\,\biggl[
- \frac{64}{3}\frac{1}{\mu}\mathcal{L}_2\Bigl(\frac{\ell}{\mu}\Bigr)
+\frac{320}{9} \frac{1}{\mu}\mathcal{L}_1\Bigl(\frac{\ell}{\mu}\Bigr)
+ \frac{1}{\mu}\mathcal{L}_0\Bigl(\frac{\ell}{\mu}\Bigr) \Bigl( -\frac{448}{27}+\frac{16\pi^2}{9}\Bigr)
\nn \\ & \quad
+ \delta(\ell)\Bigr(\frac{80}{81}+\frac{74\pi^2}{27}-\frac{232}{9}\zeta_3\Bigr)
\biggr]
\label{eq:soft_massless_2}
\,.\end{align}
The corresponding $\mu$ anomalous dimension is given by
\begin{align}\label{eq:gamma_soft}
\gamma_{S}^{(1)}(\ell,\mu) &= 16C_F\,\frac{1}{\mu}\mathcal{L}_0\Bigl(\frac{\ell}{\mu}\Bigr)
\,, \nn \\
\gamma_{S}^{(2,l)}(\ell,\mu) &= C_F \biggl[-\frac{320}{9} \frac{1}{\mu}\mathcal{L}_0\Bigl(\frac{\ell}{\mu}\Bigr)
+ \delta(\ell)\Bigl(\frac{448}{27}-\frac{8\pi^2}{9}\Bigr)\biggr]
\,.\end{align}

\section{Calculations of massive quark corrections}
\label{app:calculations}

We calculate the quark mass dependent beam and soft functions for primary and secondary contributions at one and two loops, respectively. The final renormalized results are given and discussed in \sec{results}. For the computation of the collinear massive quark corrections we use the Feynman rules determined from the collinear massive quark Lagrangian~\cite{Rothstein:2003wh,Leibovich:2003jd,Fleming:2007qr}. For the secondary corrections we use in practice regular QCD Feynman rules, since the collinear sector is essentially just a boosted version of QCD. (The interactions of the massive quarks in the soft sector are anyway given by the usual QCD Feynman rules.)
First, we calculate the massive quark beam function in \sec{Calc_primarymassive}, before discussing the computation of the secondary corrections for the massless quark beam and soft functions in secs.~\ref{sec:dispersion} -- \ref{sec:csoft}. All computations are carried out in Feynman gauge.

\subsection{Massive quark beam function at $\mathcal{O}(\alpha_s)$}
\label{sec:Calc_primarymassive}

The massive quark beam function operator for a measurement function $\mathcal{M}$ is defined as (see e.g.~\refcites{Stewart:2009yx,Stewart:2010qs,Jain:2011iu,Chiu:2012ir})
\begin{align}
 \mathcal{O}_{Q}(\{\mathcal{M}\},\omega,m)  & = \overline{\chi}_{n,m}(0) \mathcal{M}(\mathcal{P}^\mu,\hat{p}^+)\frac{\slashed{\bar{n}}}{2}\bigl[\delta(\omega-\overline{\mathcal{P}}_n)\chi_{n,m}(0)\bigr] \, , 
\end{align}
where $\chi_{n,m}$ indicates a massive collinear quark field, $\mathcal{P}^\mu$ is the label momentum operator, and $\hat{p}^+$ extracts the residual momentum component $n \cdot k$. For the transverse momentum dependent (TMD), virtuality dependent, and fully differential case the measurement functions are
\begin{align}
\mathcal{M}_{\perp}= \delta^{(2)}(\vec{p}_T-\vec{\mathcal{P}}_{\perp}) \, , \quad  \mathcal{M}_{p^+}= \delta(t-\omega  \hat{p}^+) \, , \quad \mathcal{M}_{\perp,p^+} = \delta^{(2)}(\vec{p}_T-\vec{\mathcal{P}}_{\perp})\,\delta(t-\omega  \hat{p}^+) \, .
\end{align}
For convenience we discuss also the fully differential case here, from which the other two cases can be obtained by an integration over the respective other variable. The beam functions are proton matrix elements of the operators $\mathcal{O}_Q$. To compute the (perturbative) matching coefficients onto the PDFs, we take matrix elements with partonic states, denoting e.g.
\begin{align}\label{eq:beamfct_matrixelement}
 B_{Qg}\Bigl(\{\mathcal{M}\},m,z=\frac{\omega}{p^-}\Bigr) \equiv  \langle g_n(p)\vert \mathcal{O}_Q(\{\mathcal{M}\},\omega,m) \vert g_n(p)\rangle\, ,
\end{align}
for an initial collinear gluon state with momentum $p^\mu =p^- n^\mu/2$. 

 \begin{figure}
\centering
\includegraphics[scale=0.5]{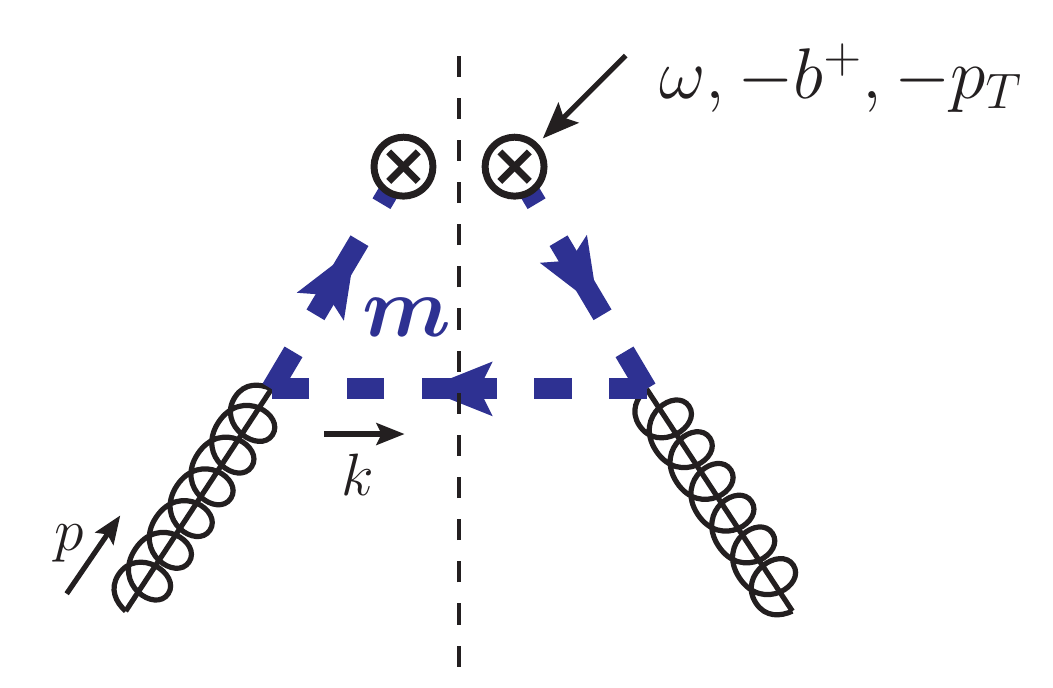}
\caption{One-loop diagram contributing to the massive quark beam function.}
\label{fig:BeamPrimary}
\end{figure}

At $\mathcal{O}(\alpha_s)$ the only contribution to the massive quark beam function originates from an initial collinear gluon splitting into a heavy quark-antiquark pair. The corresponding diagram is given in fig.~\ref{fig:BeamPrimary}. The kinematics of the on-shell final state is fully constrained at one loop, so that the diagram can be evaluated without performing any integration. For the fully differential case we obtain
\begin{align}
 B_{Qg}(t,\vec{p}_T,m,z)\Big|_{\mathcal{O}(\alpha_s)}&=8\pi\alpha_sT_F \,\theta(\omega)\,\theta(t) \int\frac{\mathrm{d}^4k}{(2\pi)^4}\,\frac{|\vec{k}_{\perp}|^2\big[(p^-)^2-2(p^--k^-)k^-\big]+m^2(p^-)^2}{(p^-)^2(k^+)^2k^-} \nn \\
 & \quad \times \delta(\omega-p^-+k^-)\,\delta(b^+-k^+)\,\delta^{(2)}(\vec{p}_T-\vec{k}_{\perp}) \,2\pi\delta(k^2-m^2)\nn \\
 &=\frac{\alpha_sT_F}{4\pi^2}\,\theta(z)\,\theta(t)\,\delta\Bigl(p_T^2-\frac{t(1-z)}{z}+m^2\Bigr)\frac{2}{t}\Bigl(P_{qg}(z)+\frac{2m^2z^2}{t}\Bigr) \nn \\
 & = \frac{\alpha_s}{4\pi}\, \mathcal{I}^{(1)}_{Qg}(t,\vec{p}_T,m,z) \, ,  \label{eq:fullydif_primary}
\end{align}
where $P_{qg}(z)=z^2+(1-z)^2$ is the leading-order gluon-quark splitting function. The correction $B_{Qg}$ at $\mathcal{O}(\alpha_s)$ is UV and IR finite. It corresponds directly to the matching coefficient $\mathcal{I}^{(1)}_{Qg}$, given as the one-loop coefficient in an expansion in terms of $\alpha_s$ as in \eq{alphas_expansion}. The matching coefficients for the TMD and virtuality-dependent beam functions can be obtained here by a trivial integration of this result,
\begin{align}
\mathcal{I}_{Qg}^{(1)}(\vec{p}_T,m,z)= \int \mathrm{d}t\,\mathcal{I}_{Qg}^{(1)}(t,\vec{p}_T,m,z) \, , \qquad \mathcal{I}_{Qg}^{(1)}(t,m,z)=\int \mathrm{d}^2p_T\,\mathcal{I}_{Qg}^{(1)}(t,\vec{p}_T,m,z)
\,,\end{align}
which yields the results in \eqs{TMD_beam_coefficient}{virtuality_beam_coefficient}.
Note that in general, this integration has to be performed for the {\it bare} result with the full dependence on the UV and rapidity regulator. However, in this case all matrix elements are finite and do not require any renormalization at this order.

\subsection{Dispersive technique for secondary massive quark corrections}
\label{sec:dispersion}

For observables where only the sum over the final-state hadronic momenta enters the measurement, one can use dispersion relations to obtain the results for secondary massive quark radiation at $\mathcal{O}(\alpha_s^2)$ from
the corresponding results for “massive gluon” radiation at $\mathcal{O}(\alpha_s)$. This has been discussed in detail in ref.~\cite{Pietrulewicz:2014qza}. The key relation is that the insertion of a vacuum polarization function for massive quarks $\Pi_{\mu \nu}(m^2,p^2)$ between two gluon propagators can be written as
\begin{align}\label{eq:propagatorunsubtracted}
  \frac{-\,i\,g^{\mu\rho}}{p^2 +i \epsilon}\, \Pi_{\rho\sigma}(m^2,p^2)
 \, \frac{-\,i\,g^{\sigma\nu}}{p^2 +i \epsilon} &=\frac{1}{\pi} \int\frac{\df M^2}{M^2}\,\frac{-\,i\,\Big(g^{\mu\nu}-\frac{p^\mu p^\nu}{p^2}\Big)}
 {p^2-M^2+i \epsilon} \,\mathrm{Im}\! \left[\Pi(m^2,M^2)\right] \nn \\
 & \quad-\frac{-\,i\,\Big(g^{\mu\nu}-\frac{p^\mu p^\nu}{p^2}\Big)}{p^2+i \epsilon}\,\Pi(m^2,0) \, .
\end{align}
The first term contains a gluon propagator with effective mass $M$ and the absorptive part of the vacuum polarization function, which reads in $d=4-2\eps$ dimensions
\begin{align}\label{eq:Im_Pi}
&\mathrm{Im}\!\left[\Pi(m^2,p^2)\right] = \theta(p^2-4m^2) \, \frac{\alpha_s T_F}{4 \pi} \,\frac{ (4\mu^{2}e^{\gamma_E})^{\e}\pi^{3/2}}{\Gamma(\frac{5}{2}-\eps)} 
\Big(1-\eps+\frac{2m^2}{p^2}\Big) \,(p^2)^{-\eps}\, \Big(1-\frac{4m^2}{p^2}\Big)^{1/2-\eps}  \, .
\end{align}
To obtain the first term on the right-hand side in \eq{propagatorunsubtracted} the vacuum polarization function (and thus the strong coupling) was renormalized in the on-shell scheme, i.e., with $n_l$ active quark flavors. The second term in \eq{propagatorunsubtracted} translates back to an unrenormalized strong coupling and consists of a massless gluon propagator and the $\mathcal{O}(\alpha_s)$ vacuum polarization function at zero momentum transfer, which is given by
\begin{align}
\Pi(m^2,0)  = \frac{\alpha_s T_F}{4\pi} \,\frac{4}{3} \,\Gamma(\eps)\bigg(\frac{\mu^2 e^{\gamma_E}}{m^2}\bigg)^{\eps} \equiv \frac{\alpha_s T_F}{4\pi} \,\Pi^{(1)}(m^2,0) 
 \, .
\label{eq:vacpolzero}
\end{align}

In the following we will first carry out the computation of the beam and  soft functions at $\mathcal{O}(\alpha_s)$ for the radiation of a ``massive gluon" and in a second step use the relation in \eq{propagatorunsubtracted} to obtain the associated results for massive quarks at $\mathcal{O}(\alpha_s^2 C_F T_F)$. In our calculations we drop the contributions from the terms proportional to $p^\mu p^\nu$, which vanish in total due to gauge invariance. 

\subsection{Secondary mass effects in light-quark beam functions}\label{sec:Calc_secondarymassive}

We compute the massive quark corrections to the TMD and virtuality-dependent light-quark beam function at $\mathcal{O}(\alpha_s^2 C_F T_F)$ starting with the massive gluon case at $\mathcal{O}(\alpha_s)$. Only the contributions to the matching coefficient $\mathcal{I}_{qq}$ are nontrivial, so we consider only diagrams with a quark in the initial state.

\subsubsection{Quark beam function with a massive gluon at $\mathcal{O}(\alpha_s)$}

\begin{figure}
\hfill\subfigure[]{\includegraphics[scale=0.5]{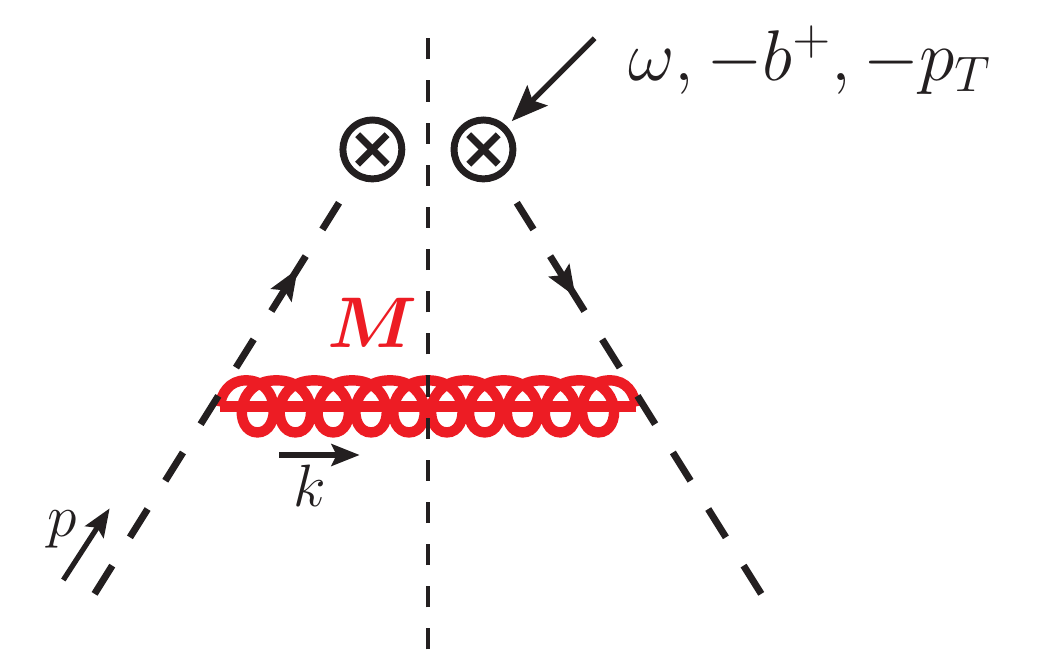}\label{fig:beam_a}}%
\hfill\subfigure[]{\includegraphics[scale=0.5]{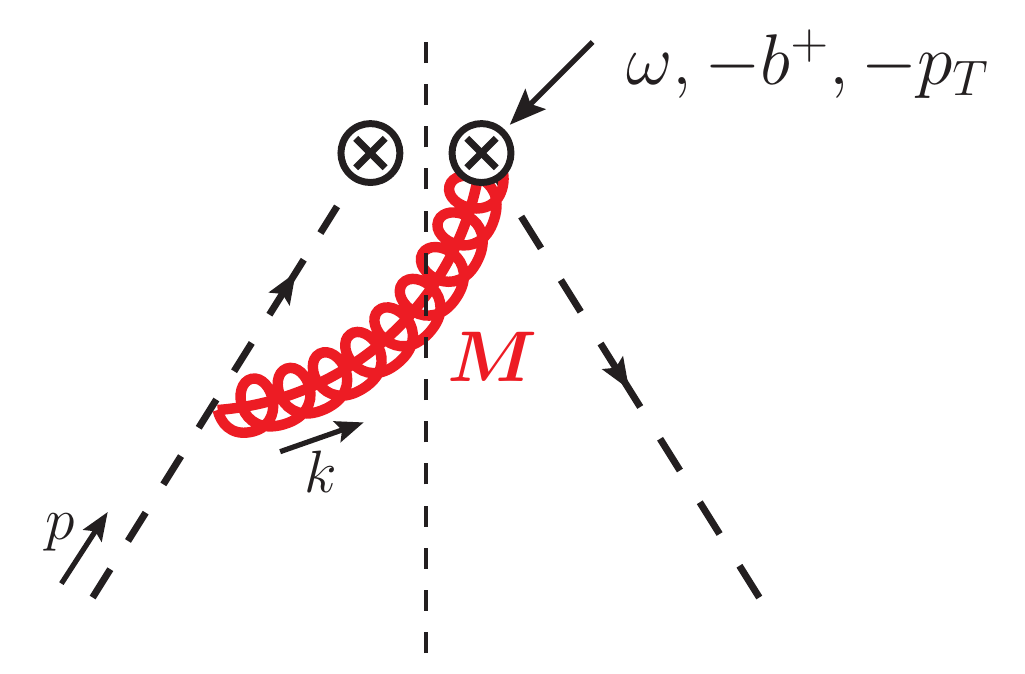}\label{fig:beam_b}}%
\hfill\subfigure[]{\includegraphics[scale=0.5]{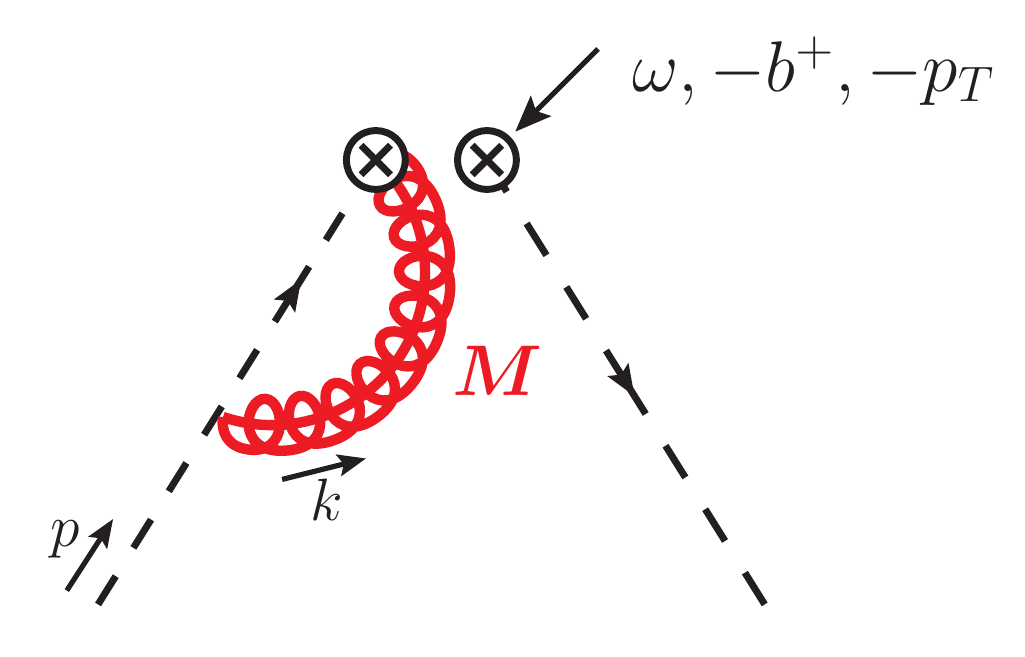}\label{fig:beam_c}}%
\hspace*{\fill}
\caption{Light-quark beam function diagrams for massive gluon radiation at one loop. In addition, also the wave function renormalization correction and the mirror diagrams for (b) and (c) have to be included in the calculation.}
\label{fig:beam}
\end{figure}

\paragraph{Contributions to the fully-differential beam function}

As in \sec{Calc_primarymassive} we start also here with the computation of the corrections for the fully-differential beam function. The contributing one-loop diagrams to the matrix element $B_{qq}$ with massless quarks in the initial state, defined in analogy to \eq{beamfct_matrixelement}, are displayed in fig.~\ref{fig:beam}. They consist of a purely virtual and a real-radiation part,
\begin{align}
B^{(1,\rm bare)}_{qq}(t,\vec{p}_T,M,\omega,z)=\delta(1-z)\,\delta(t)\,\delta^{(2)}(\vec{p}_T)\, B^{(1,\rm bare)}_{qq,\rm{virt}}(M,\omega)+B^{(1,\rm bare)}_{qq,\rm{real}}(t,\vec{p}_T,M,\omega,z)\;. \label{eq:fully_differential_general}
\end{align}

The virtual massive gluon contributions in fig.~\ref{fig:beam_c} are the same as for other collinear quark operators like the current or the PDF and have been computed e.g.~in ref.~\cite{Chiu:2012ir}. Including the wave function renormalization diagrams the $d$-dimensional result reads~\cite{Hoang:2015iva}
\begin{align}
 B^{(1,\rm{bare})}_{qq,\rm{virt}}(M,\omega)=C_F\,\Bigl(\frac{\mu^2\mathrm{e}^{\gamma_E}}{M^2}\Bigr)^{\e}\Gamma(\e)\,\biggl\{\frac{4}{\eta}+4\ln\frac{ \nu}{\omega}+4H_{1-\e}-\frac{2(1-\e)}{2-\e}+\mathcal{O}(\eta)\biggr\}\;, \label{eq:beam_virt}
\end{align}
where $H_{\alpha}=\psi(1+\alpha)+\gamma_E$ is the Harmonic number. Here the rapidity divergences have been regulated using the symmetric $\eta$ regulator acting on the Wilson lines \cite{Chiu:2012ir, Chiu:2011qc}, while UV divergences are regulated with dimensional regularization as usual. Furthermore, the gluon mass provides an IR cutoff.

The real radiation contributions in figs.~\ref{fig:beam_a} and \ref{fig:beam_b} can be easily evaluated, since all momentum components are fully determined by the measurement. For the first diagram we get
\begin{align}
 B^{(a)}&=8\pi\alpha_sC_F p^-\theta(\omega)\,\theta(t)\int\frac{\mathrm{d}^4k}{(2\pi)^4}\,\frac{|\vec{k}_{\perp}|^2}{\big[(p-k)^2+i \eps\big]^2} \,  \delta(\omega-p^-+k^-)\, \delta(t-\omega k^+) \nn \\
 &\hspace{4cm}\times \delta^{(2)}(\vec{p}_T-\vec{k}_{\perp}) \, 2\pi\delta(k^2-M^2)\nn \\
 &=\frac{\alpha_sC_F}{4\pi^2}\,\theta(z)\,\theta(t)\,\delta\Bigl(p_T^2-\frac{t(1-z)}{z}+M^2\Bigr)\,\frac{2\bigl(t(1-z)-zM^2\bigr)}{(t-zM^2)^2}\;. 
\end{align}
Since UV divergences do not appear for the real radiation corrections and the gluon mass regulates all IR divergences we do not need to employ dimensional regularization here. The second diagram in fig.~\ref{fig:beam_b} yields
\begin{align}
 B^{(b)}&=-8\pi\alpha_sC_F p^-\theta(\omega)\,\theta(t)\int\frac{\mathrm{d}^4k}{(2\pi)^4}\,\frac{(p^--k^-)}{(p-k)^2+i \epsilon}\, \frac{\nu^\eta}{(k^-)^\eta}\, \delta(\omega-p^-+k^-)\, \delta(t-\omega k^+)  \nn\\
 &\hspace{4cm}\times \delta^{(2)}(\vec{p}_T-\vec{k}_{\perp}) \, 2\pi\delta(k^2-M^2)\nn \\
 &=\frac{\alpha_sC_F}{4\pi^2} \,\theta(z)\,\theta(t)\,\delta\Bigl(p_T^2-\frac{t(1-z)}{z}+M^2\Bigr)\Bigl(\frac{\nu}{\omega}\Bigr)^{\eta}\frac{2z^{1-\eta}}{(t-zM^2)(1-z)^{\eta}} \, .
 \end{align}
 While the fully-differential quark beam function itself does not contain any rapidity divergences, we have included here the $\eta$ regulator, since we will use this result to obtain the TMD beam function by integrating over the virtuality, which results in  rapidity divergences for this real radiation correction. The full real radiation contributions at one loop yield
\begin{align}
 \frac{\alpha_s}{4\pi}B_{qq,\rm{real}}^{(1,\rm{bare})}(t,\vec{p}_T,M,\omega,z)&=B^{(a)}+2B^{(b)} \nn \\
 &=\frac{\alpha_sC_F}{4\pi}\,\frac{1}{\pi}\,\theta(z)\,\theta(t)\,\delta\Bigl(p_T^2-\frac{t(1-z)}{z}+M^2\Bigr)\nn \\
 &\quad \times\frac{2}{t-zM^2}\biggl[\Bigl(\frac{\nu}{\omega}\Bigr)^{\eta}\frac{2z^{1-\eta}}{(1-z)^{\eta}}+\frac{t(1-z)-zM^2}{t-zM^2}\biggr]
 \,.\label{eq:beam_real}
\end{align}
For both virtual and real radiation corrections all soft-bin subtractions are parametrically power suppressed or scaleless and therefore do not contribute.

\paragraph{Contributions to the TMD beam function}

The corrections for the TMD beam function with a massive gluon can be obtained by integrating the fully-differential beam function in \eq{fully_differential_general} over the virtuality $t$. We write them again as
\begin{align}
 B_{qq}^{(1,\rm{bare})}(\vec{p}_T,M,z)=\delta(1-z)\,\delta^{(2)}(\vec{p}_T)\,B^{(1,\rm{bare})}_{qq,\rm{virt}}(M,\omega)+B^{(1,\rm{bare})}_{qq,\rm{real}}(\vec{p}_T,M,\omega,z)\;,
\end{align}
where $B^{(1,\rm{bare})}_{qq,\rm{virt}}$ is given in \eq{beam_virt} and
\begin{align}
 B_{qq,\rm{real}}^{(1,\rm{bare})}(\vec{p}_T,M,\omega,z)&=\int\mathrm{d}t\,B_{qq,\rm{real}}^{(1,\rm{bare})}(t,\vec{p}_T,M,z) \\
 &=C_F\,\theta(z)\,\theta(1-z)\,\frac{1}{\pi}\,\frac{2}{p_T^2+zM^2}\biggl[\frac{p_T^2(1-z)}{p_T^2+zM^2}+\frac{2z^{1-\eta}}{(1-z)^{1+\eta}}\Bigl(\frac{\nu}{\omega}\Bigr)^{\eta}\biggr]\;. \nn
\end{align}
Here it is necessary to keep a nonvanishing value for $\eta$ in the second term to regularize the rapidity divergence for $z \to 1$. Expanding for $\eta \to 0$ we get
\begin{align}
 B_{qq,\rm{real}}^{(1,\rm{bare})}(\vec{p}_T,M,\omega,z)&=C_F\,\theta(z)\,\frac{1}{\pi}\,\biggl\{\frac{4}{p_T^2+M^2}\Bigl[-\delta(1-z)\Bigl(\frac{1}{\eta}+\ln\frac{\nu}{\omega}\Bigr)+\mathcal{L}_0(1-z)\Bigr]\nn \\
 &+\theta(1-z)\frac{2p_T^2}{p_T^2+zM^2}\Bigl[\frac{1-z}{p_T^2+zM^2}-\frac{2}{p_T^2+M^2}\Bigr]\biggr\}+\mathcal{O}(\eta)\;. \label{eq:beam_pTreal}
\end{align}

\paragraph{Contributions to the virtuality-dependent beam function}

The virtuality-dependent beam function with a massive gluon can be obtained by integrating the results for the fully-differential beam function over $\vec{p}_T$. We decompose the corrections again into a virtual and real radiation part,
\begin{align}\label{eq:BM_Tau}
 B_{qq}^{(1,\rm{bare})}(t,M,z)=\delta(1-z)\,\delta(t)\, B^{(1,\rm{bare})}_{qq,\rm{virt}}(M,\omega)+B^{(1,\rm{bare})}_{qq,\rm{real}}(t,M,z)\;,
\end{align}
where $B^{(1,\rm{bare})}_{qq,\rm{virt}}$ is given in \eq{beam_virt} and
\begin{align}
 B_{qq,\rm{real}}^{(1)}(t,M,z)&=\int\mathrm{d}^2\vec{p}_T\,B_{qq,\rm{real}}^{(1,\rm{bare})}(t,\vec{p}_T,M,\omega,z) \\
 &=C_F\,\theta(z)\,\theta(t)\, \theta\Bigl(\frac{t(1-z)}{z}-M^2\Bigr)\,\frac{2}{t-zM^2}\Bigl(\frac{2z}{1-z}+\frac{t(1-z)-zM^2}{t-zM^2}\Bigr)\,, \nn
\end{align}
with the fully-differential real radiation contributions in \eq{beam_real}.
Here the $\eta$ regulator has already been dropped, since for the virtuality-dependent beam function no rapidity divergences arise from the real radiation contributions.

\subsubsection{Secondary massive quark effects in the TMD beam function}

To obtain the secondary massive quark corrections from the one-loop results with a massive gluon, we first convolve the one-loop results with the imaginary part of the vacuum polarization function according to \eq{propagatorunsubtracted} and define
\begin{align}\label{eq:dispersion_beam}
\frac{\alpha_sT_F}{4\pi}B^{(2,h,\rm{bare})}_{qq,\rm{virt}}(m,\omega) &=\frac{1}{\pi}\int\frac{\mathrm{d}M^2}{M^2}\,{\rm{Im}}\bigl[\Pi(m^2,M^2)\bigr]\times B^{(1,\rm{bare})}_{qq,\rm{virt}}(M,\omega)\;, \nn \\
\frac{\alpha_sT_F}{4\pi}B^{(2,h,\rm{bare})}_{qq,\rm{real}}(\vec{p}_T,m,\omega,z) &=\frac{1}{\pi}\int\frac{\mathrm{d}M^2}{M^2}\,{\rm{Im}}\bigl[\Pi(m^2,M^2)\bigr]\times B^{(1,\rm{bare})}_{qq,\rm{real}}(\vec{p}_T,M,\omega,z)\;.
\end{align}
The results from these dispersion integrations are
\begin{align}
B^{(2,h,\rm{bare})}_{qq,\rm{virt}}(m,\omega) &=C_F\,\biggl\{\Bigl(\frac{1}{\eta}+\ln\frac{\nu}{\omega}\Bigr)\biggl[\frac{8}{3\e^2}-\frac{1}{\e}\Bigl(\frac{16}{3}L_m+\frac{40}{9}\Bigr)+\frac{16}{3}L_m^2+\frac{80}{9}L_m+\frac{224}{27} \nn \\
& \quad +\frac{4\pi^2}{9}+\mathcal{O}(\e) \biggr]+\frac{2}{\e^2}-\frac{1}{\e}\Bigl(4L_m+\frac{1}{3}+\frac{4\pi^2}{9}\Bigr)+4L_m^2+\Bigl(\frac{2}{3}+\frac{8\pi^2}{9}\Bigr)L_m \nn \\
& \quad +\frac{73}{18} +\frac{29\pi^2}{27}-\frac{8\zeta_3}{3}\biggr\}\;, \label{eq:bare_beam_virt} \\
B^{(2,h,\rm{bare})}_{qq,\rm{real}}(\vec{p}_T,m,\omega,z) &=C_F \,\frac{1}{\pi p_T^2}\biggl\{\frac{16}{9\eta}\,\delta(1-z)\biggl[5-12\hat{m}^2-3c(1-2\hat{m}^2)\ln\Bigl(\frac{c+1}{c-1}\Bigr)\biggr] \nn \\
& \quad +b_{\rm{real}}^{q_T}\Bigl(\frac{m^2}{p_T^2},z,\frac{\nu}{\omega}\Bigr)\biggr\}\;, \label{eq:bare_beam_real}
\end{align}
with
\begin{align}
&b_{\rm{real}}^{q_T}\Bigl(\hat{m}^2,z,\frac{\nu}{\omega}\Bigr)=\theta(z)\frac{16}{9}\biggl\{ \biggl[5-12\hat{m}^2-3c(1-2\hat{m}^2)\ln\Bigl(\frac{c+1}{c-1}\Bigr)\biggr]\nn \\
&\qquad \times\Bigl(\delta(1-z)\ln\frac{\nu}{\omega}-\mathcal{L}_0(1-z)\Bigr) \nn \\
& \qquad+\theta(1-z) \biggl[\frac{3}{2d(1-z)} \bigl[1+z^2+2\hat{m}^2z(1+z^2)+4\hat{m}^4z^2(-5+6z-5z^2)\bigr]\,\ln\Bigl(\frac{d+1}{d-1}\Bigr)\nn \\
& \qquad-\frac{3c(1-2\hat{m}^2)}{1-z}\ln\Bigl(\frac{c+1}{c-1}\Bigr)+1+4z+3\hat{m}^2(-4+z-5z^2)\biggr]\biggr\}\;,\label{eq:bm_real}
\end{align}
and $\hat{m}$, $c$, $d$ defined in eq.~\eqref{eq:mhat}. Using \eq{dispersion_beam} entails that the massive quark corrections to the strong coupling are renormalized in the on-shell scheme, i.e., the expansion is in terms of $\alpha_s=\alpha_s^{(n_l)}$. Since the beam function matrix element has to be renormalized entirely in the $n_l+1$ flavor theory, we need to account for the second term in \eq{propagatorunsubtracted} (which switches back to an unrenormalized $\alpha_s$) and renormalize the massive quark corrections to the strong coupling in the $\MS$ scheme, such that the expansion is in terms of $\alpha_s=\alpha_s^{(n_l+1)}$. The beam function operator is renormalized according to
\begin{align}
\mathcal{O}^{(\rm bare)}_q(\vec{p}_T,m,\omega) = \int \df ^2 p_T' \, Z_B\Bigl(\vec{p}_T-\vec{p}^{\,\prime}_T ,m,\mu,\frac{\nu}{\omega}\Bigr) \,\mathcal{O}_q(\vec{p}^{\,\prime}_T ,m,\omega,\mu,\nu) \, ,
\end{align}
where the counterterm encodes also the rapidity divergences. This yields for the renormalized matrix element with initial state quarks at $\mathcal{O}(\alpha_s^2 C_F T_F)$ in terms of $\alpha_s = \alpha_s^{(n_l+1)}$
\begin{align}\label{eq:Bqq_renormalized}
B_{qq}^{(2,h)}\Bigl(\vec{p}_T,m,z,\mu,\frac{\nu}{\omega}\Bigr) &=\delta^{(2)}(\vec{p}_T)\,\delta(1-z)\,B^{(2,h,\rm{bare})}_{qq,\rm{virt}}(m,\omega)+B^{(2,h,\rm{bare})}_{qq,\rm{real}}(\vec{p}_T,m,\omega,z) \\
& \quad  -\Bigl(\Pi^{(1)}(m^2,0) - \frac{4}{3\eps}\Bigr) \underbrace{B_{qq}^{(1,\rm bare)}(\vec{p}_T,\omega,z)}_{ \makebox[0pt]{\scriptsize $=B^{(1)}_{qq}(\vec{p}_T,z,\mu,\frac{\nu}{\omega}) + Z_B^{(1)}(\vec{p}_T,\mu,\frac{\nu}{\omega})\, \delta(1-z)$}} -\delta(1-z)Z_{B}^{(2,h)}\Bigl(\vec{p}_T,m,\mu,\frac{\nu}{\omega}\Bigl) \, . \nn
\end{align}
where the (bare) vacuum polarization function $\Pi^{(1)}(m^2,0) $ is given in \eq{vacpolzero}. The one-loop counterterm reads
\begin{align}\label{eq:ZB1}
&Z_{B}^{(1)}\Bigl(\vec{p}_T,\mu,\frac{\nu}{\omega}\Bigr)=C_F\,\biggl\{\delta^{(2)}(\vec{p}_T)\biggl[\frac{1}{\eta}\Bigl(\frac{4}{\e}+\mathcal{O}(\e)\Bigr)+\frac{1}{\e}\Bigl(3+4\ln\frac{\nu}{\omega}\Bigr)\biggr]-\frac{1}{\eta}\Bigl(4+\mathcal{O}(\e)\Bigr)\mathcal{L}_0(\vec{p}_T,\mu)\biggr\}\;.
\end{align}
The two-loop counterterm $Z_B^{(2)}$ absorbs all remaining UV and rapidity divergences in \eq{Bqq_renormalized} and is given by
\begin{align}\label{eq:ZB2}
&Z_{B}^{(2,h)}\Bigl(\vec{p}_T,m,\mu,\frac{\nu}{\omega}\Bigr)=C_F\,\biggl\{\delta^{(2)}(\vec{p}_T)\biggl[\frac{1}{\eta}\Bigl(\frac{8}{3\e^2}-\frac{40}{9\e}+\frac{8}{3}L_m^2+\frac{80}{9}L_m+\frac{224}{27}+\mathcal{O}(\e)\Bigr)\nn \\
&\qquad+\frac{1}{\e^2}\Bigl(2+\frac{8}{3}\ln\frac{\nu}{\omega}\Bigr)-\frac{1}{\e}\Bigl(\frac{1}{3}+\frac{4\pi^2}{9}+\frac{40}{9}\ln\frac{\nu}{\omega}\Bigr)\biggr]-\frac{1}{\eta}\Bigl(\frac{16}{3}L_m+\mathcal{O}(\e)\Bigr)\mathcal{L}_0(\vec{p}_T,\mu) \nn \\
&\qquad+\frac{1}{\eta} \,\frac{16}{9\pi p_T^2}\biggl[5-12\hat{m}^2-3c(1-2\hat{m}^2)\ln\frac{c+1}{c-1}\biggr]\biggr\}
\,.\end{align}
This yields the anomalous dimensions in \eq{anomdim_TMD}. The renormalized one-loop partonic beam function $B_{qq}^{(1)}$ still contains IR divergences, so its exact form depends on the choice of the IR regulator.

The beam function matching coefficient $\mathcal{I}_{qq}$ as defined in~\eqref{eq:Bm1_qT} can be now easily obtained. Note that the PDFs are renormalized in an $n_l$-flavor theory with $\alpha_s=\alpha_s^{(n_l)}$ in contrast to the beam function. Thus, there is a contribution coming from the scheme change of $\alpha_s$ to $n_l+1$ flavors for the (renormalized) one-loop PDF correction, i.e.
\begin{align}
\mathcal{I}_{qq}^{(2,h)}\Bigl(\vec{p}_T,m,z,\mu,\frac{\nu}{\omega}\Bigr) &=B_{qq}^{(2,h)}\Bigl(\vec{p}_T,m,z,\mu,\frac{\nu}{\omega}\Bigr)-\delta^{(2)}(\vec{p}_T)\,\frac{4}{3}L_mf_{qq}^{(1)}(z,\mu) \nn \\
& =\delta^{(2)}(\vec{p}_T)\,\delta(1-z)B^{(2,h,\rm{bare})}_{qq,\rm{virt}}(m,\omega)+B^{(2,h,\rm{bare})}_{qq,\rm{real}}(\vec{p}_T,m,\omega,z) \nn \\
&\quad -\delta(1-z)\biggl[\Bigl(\Pi^{(1)}(m^2,0) - \frac{4}{3\eps}\Bigr)\, Z_{B}^{(1)}\Bigl(\vec{p}_T,\mu,\frac{\nu}{\omega}\Bigr)+Z_{B}^{(2,h)}\Bigl(\vec{p}_T,m,\mu,\frac{\nu}{\omega}\Bigl)\biggr] \nn \\
&\quad+\frac{4}{3}L_m\underbrace{\biggl(B^{(1)}_{qq}\Bigl(\vec{p}_T,z,\mu,\frac{\nu}{\omega}\Bigr)-\delta^{(2)}(\vec{p}_T)f_{qq}^{(1)}(z,\mu)\biggr)}_{=\,\mathcal{I}_{qq}^{(1)}(\vec{p}_T,z,\mu,\frac{\nu}{\omega})} \, .
\end{align}
Here the IR divergences cancel between the one-loop beam function and the PDF to give the finite one-loop matching coefficient $\mathcal{I}_{qq}^{(1)}$, which is given in eq.~\eqref{eq:TMD_Iqq1}. Using eqs.~(\ref{eq:bare_beam_virt}),~(\ref{eq:bare_beam_real}),~(\ref{eq:ZB1}) and~(\ref{eq:ZB2}) we obtain the full result for the secondary massive quark corrections to the beam function matching coefficient given in eq.~\eqref{eq:TMD_Iqq2_massive}.

\subsubsection{Secondary massive quark effects in the virtuality-dependent beam function}\label{app:beamfunc_thrust_secondary}

We proceed with the virtuality-dependent beam function. While the virtual contributions are the same as for the TMD beam function given in \eq{bare_beam_virt}, the dispersion integration for the real radiation terms yields
\begin{align} \label{eq:Breal_Tau}
\frac{\alpha_sT_F}{4\pi}B^{(2,h)}_{qq,\rm{real}}(t,m,z) &=\frac{1}{\pi}\int\frac{\mathrm{d}M^2}{M^2}\,{\rm{Im}}\bigl[\Pi(m^2,M^2)\bigr]\times B^{(1)}_{qq,\rm{real}}(t,M,z)
\nn \\
&= \frac{\alpha_sT_F}{4\pi}\, \frac{C_F}{t}\,b_{\rm{real}}^{\Tau}\Bigl(\frac{m^2}{t},z\Bigr)
\,,\end{align}
with
\begin{align}
b_{\rm{real}}^{\Tau}(\hat{m}^2,z)
&=\theta(z)\,\theta(v)\,\frac{8}{9(1-z)}\biggl\{-\frac{3}{u} \ln\frac{u-v}{u+v} \biggl[1+z^2-2\hat{m}_t^2 z(1+z^2)
-4 \hat{m}_t^4 z^2(2-3z+5z^2)\biggr]
\nn\\ &\quad
-2v\biggl[4-3z+4z^2 +\frac{z(11-21z+29z^2-15z^3)}{1-z}\,\hat{m}_t^2\biggr]\biggr\}\,, \label{eq:bm_realTau}
\end{align}
and $\hat{m}_t$, $u$, $v$ as in \eq{u_and_v}.

To obtain the quark mass dependent matching coefficient $\mathcal{I}^{(2,h)}_{qq}$ we carry out our calculation using a gluon mass $\Lambda \ll \sqrt{Q \Tau} \sim m$ as IR regulator. Although the result is independent of the regulator, this is technically most convenient, since this allows us to match two \SCETb theories with each other in a straightforward way.%
\footnote{Alternatively, one can also perform the matching between theories where the fluctuations related to the $n_l$ massless flavors are described within a \SCETa theory. In this setup, there is no csoft function on the right-hand side of the matching relation in contrast to \eq{beam_matching2}. However, in this case the zero-bin subtractions for the collinear fields with respect to the ultrasoft modes in the \SCETa $n_l$ flavor theory yield a nontrivial contribution to the beam-function matrix element on the left-hand side of the matching relation. Their contribution is equivalent to the inverse of the csoft function in \eq{beam_matching2}, such that the resulting matching coefficient $\mathcal{I}_{qq}$ is the same.}
While the \SCETb theory with $n_l+1$ flavors (i.e.~above the mass scale) contains collinear modes, the \SCETb theory with $n_l$ flavors (i.e.~below the mass scale) contains collinear and csoft modes like in the mode setup of \subsec{fact_Tau2}. The matching relation reads
\begin{align}\label{eq:beam_matching2}
\mathscr{B}_{qq}^{(n_l+1)}\Bigl(t,m,z,\mu,\frac{\nu}{\omega}\Bigr)=\int\mathrm{d} \ell\,\mathcal{I}_{qq}\Bigl(t-\omega \ell,m,z,\mu,\frac{\nu}{\omega}\Bigr) \otimes_z f^{(n_l)}_{qq}(z,\mu) \,\mathscr{S}^{(n_l)}(\ell,\mu,\nu)\;,
\end{align}
where $\mathscr{B}_{qq}^{(n_l+1)}$ corresponds to the pure \SCETb beam function matrix element and $\mathscr{S}^{(n_l)}$ represents the csoft matrix element.

In close analogy to \eq{Bqq_renormalized}  the renormalized \SCETb matrix element $\mathscr{B}_{qq}^{(n_l+1)}$ is given at $\mathcal{O}(\alpha_s^2 C_F T_F)$ by
\begin{align}
\mathscr{B}_{qq}^{(2,h)} \Bigl(t,m,z,\mu,\frac{\nu}{\omega}\Bigr) & =\delta(t)\,\delta(1-z)\,B^{(2,h,\rm{bare})}_{qq,\rm{virt}}(m,\omega)+B^{(2,h)}_{qq,\rm{real}}(t,m,z)  \\
& \quad -\Bigl(\Pi^{(1)}(m^2,0) - \frac{4}{3\eps}\Bigr) \underbrace{B_{qq}^{(1,\rm bare)}(t,z)}_{ \makebox[0pt]{\scriptsize $=\mathscr{B}^{(1)}_{qq}(t,z,\mu,\frac{\nu}{\omega}) + Z_\mathscr{B}^{(1)}(t,\mu,\frac{\nu}{\omega})\, \delta(1-z)$}} -\,\delta(1-z)Z_{\mathscr{B}}^{(2,h)}\Bigl(t,m,\mu,\frac{\nu}{\omega}\Bigl) \,.\nn 
\end{align}
To separate UV, rapidity, and IR divergences properly from each other, we also employ the \SCETb-type IR regulator (here a gluon mass $\Lambda$) for the one-loop expressions, and at this stage the renormalized matrix elements and the counterterms still depend on this IR regulator. The matching coefficient $\mathcal{I}_{qq}$ can now be calculated as (in an expansion in terms of $\alpha_s^{(n_l+1)}$)
\begin{align}\label{eq:Iqq_renormalized}
\mathcal{I}^{(2,h)}_{qq}\Bigl(t,m,z,\mu,\frac{\nu}{\omega}\Bigr)&= \mathscr{B}_{qq}^{(2,h)} \Bigl(t,m,z,\mu,\frac{\nu}{\omega}\Bigr) -\frac{4}{3}L_m \biggl[\delta(t) f_{qq}^{(1)}(z,\mu)+\delta(1-z) \frac{1}{\omega}\mathscr{S}^{(1)}\Bigl(\frac{t}{\omega},\mu,\nu\Bigr)\biggr] \nn \\
& = \delta(t)\delta(1-z)B^{(2,h,\rm{bare})}_{qq,\rm{virt}}(m,\omega)+B^{(2,h)}_{qq,\rm{real}}(t,m,z)- Z_\mathscr{B}^{(2,h)}\Bigl(t,m,\mu,\frac{\nu}{\omega}\Bigr) \delta(1-z) \nn \\
& \quad  -\Bigl(\Pi^{(1)}(m^2,0) - \frac{4}{3\eps}\Bigr) Z_\mathscr{B}^{(1)}\Bigl(t,\mu,\frac{\nu}{\omega}\Bigr) \delta(1-z) \nn \\
& \quad +\frac{4}{3}L_m\underbrace{\biggl[\mathscr{B}^{(1)}_{qq}(t,z,\mu)-\delta(t)f_{qq}^{(1)}(z,\mu)-\delta(1-z) \frac{1}{\omega}\mathscr{S}^{(1)}\Bigl(\frac{t}{\omega},\mu,\nu\Bigr)\biggr]}_{= \mathcal{I}_{qq}^{(1)}(t,z,\mu) }\,.
\end{align}
Here the IR divergences cancel between the one-loop beam function, the PDF, and the csoft matrix element and yield the finite one-loop matching coefficient $\mathcal{I}_{qq}^{(1)}$ given in eq.~\eqref{eq:Virt_Iqq1}. The counterterm $Z_{\mathscr{B}}$ in \SCETb is defined via
\begin{align}
\mathscr{B}_{qq}^{(\rm bare)}(t,m,z)= \int \df t'\, Z^{(n_l+1)}_{\mathscr{B}}\Bigl(t-t',m,\mu,\frac{\nu}{\omega}\Bigr) \, \mathscr{B}^{(n_l+1)}_{qq} \Bigl(t',m,z,\mu,\frac{\nu}{\omega}\Bigr)
\,.\end{align}
Using the results in \eqs{BM_Tau}{beam_virt} for a massive gluon gives the associated expression for $Z_{\mathscr{B}}^{(1)}$ (expanded in $\eta$ and $\eps$)
\begin{align}\label{eq:ZB1_Tau}
Z_{\mathscr{B}}^{(1)}\Bigl(t,\mu,\frac{\nu}{\omega}\Bigr)= C_F\,\delta(t) \biggl\{\frac{4}{\eta}\biggl[\frac{1}{\eps}-\ln \frac{\Lambda^2}{\mu^2} +\mathcal{O}(\eps)\biggr]+\frac{1}{\eps}\biggl[4\ln \frac{\nu}{\omega}+3 \biggr]\biggr\} \, .
\end{align}
The two-loop counterterm $Z^{(2,h)}_{\mathscr{B}}$ cancels all divergences in \eq{Iqq_renormalized} and reads\footnote{While the $1/\eta$-divergences in the counterterm of the beam function matrix element still contain IR sensitivity, this also happens for the counterterm of the csoft matrix element in \eq{beam_matching2}, such that the resulting rapidity anomalous dimension for the running at the boundary between the $n_l+1$ and $n_l$ theory is IR finite.}
\begin{align}\label{eq:ZB2_Tau}
Z^{(2,h)}_{\mathscr{B}}\Bigl(t,m,\mu,\frac{\nu}{\omega}\Bigr) &=C_F\,\delta(t)\,\biggl\{\frac{1}{\eta} \Bigl(\frac{8}{3\e^2}-\frac{40}{9 \eps}-\frac{16}{3}L_m \ln \frac{\Lambda^2}{\mu^2}+\frac{8}{3} L_m^2+\frac{80}{9}L_m+\frac{224}{27}+\mathcal{O}(\e)\Bigr)\nn \\
&\qquad  +\frac{1}{\e^2} \Bigl(\frac{8}{3}\ln \frac{\nu}{\omega}+2\Bigr)- \frac{1}{\eps}\Bigl(\frac{40}{9}\ln \frac{\nu}{\omega}+\frac{1}{3}+\frac{4\pi^2}{9}\Bigr)\biggr\}\;.
\end{align}
Using eqs.~(\ref{eq:beam_virt}),~(\ref{eq:Breal_Tau}),~(\ref{eq:ZB1_Tau}),~(\ref{eq:ZB2_Tau}) and~\eqref{eq:Virt_Iqq1} in \eq{Iqq_renormalized} we obtain the full two-loop result for the matching coefficient in eq.~\eqref{eq:Virt_Iqq2_massive}.

\subsection{Secondary mass effects in the TMD soft function}
\label{app:Calc_softfct}

The TMD soft function is defined as
\begin{align}
S(\vec{p}_T)=\frac{1}{N_c}\,\mathrm{tr}\,\bra{0}\overline{\mathrm{T}}\big[S_n^\dagger(0)S_{\bar{n}}(0)\big]\delta^{(2)}(\vec{p}_T-\vec{\mathcal{P}}_{\perp})\mathrm{T}\big[S_{\bar{n}}^{\dagger}(0)S_n(0)\big]\ket{0}\;,
\end{align}
with the soft Wilson line $S_n$ given by~\cite{Chiu:2012ir}
\begin{align}  \label{eq:regulator_S} 
S_{n}=&\sum_{\text{perms}}\mathrm{exp}\biggl[-\frac{g}{n\cdot \mathcal{P}}\frac{\nu^{\eta/2}}{|2\mathcal{P}_{3}|^{\eta/2}} \,n\cdot A_s\biggr] \,, 
\end{align}
and in analogy for the others. Again we will first calculate the one-loop corrections to the soft function with a massive gluon, which is used in a second step to obtain the corrections from secondary massive quarks at $\mathcal{O}(\alpha_s^2 C_F T_F)$.

\subsubsection{TMD soft function with a massive gluon at $\mathcal{O}(\alpha_s)$}
\label{sec:massive_gluon_TMD_soft}

We decompose the soft function with a massive gluon at one loop in terms of virtual and real radiation corrections,
\begin{align}
 S^{(1)}(\vec{p}_T,M,\mu,\nu)=\delta^{(2)}(\vec{p}_T)\,S^{(1)}_{\rm{virt}}(M,\mu,\nu)+S_{\rm{real}}^{(1)}(\vec{p}_T,M,\nu)\;.
\end{align}

\begin{figure}
\hfill%
\subfigure[]{\includegraphics[scale=0.5]{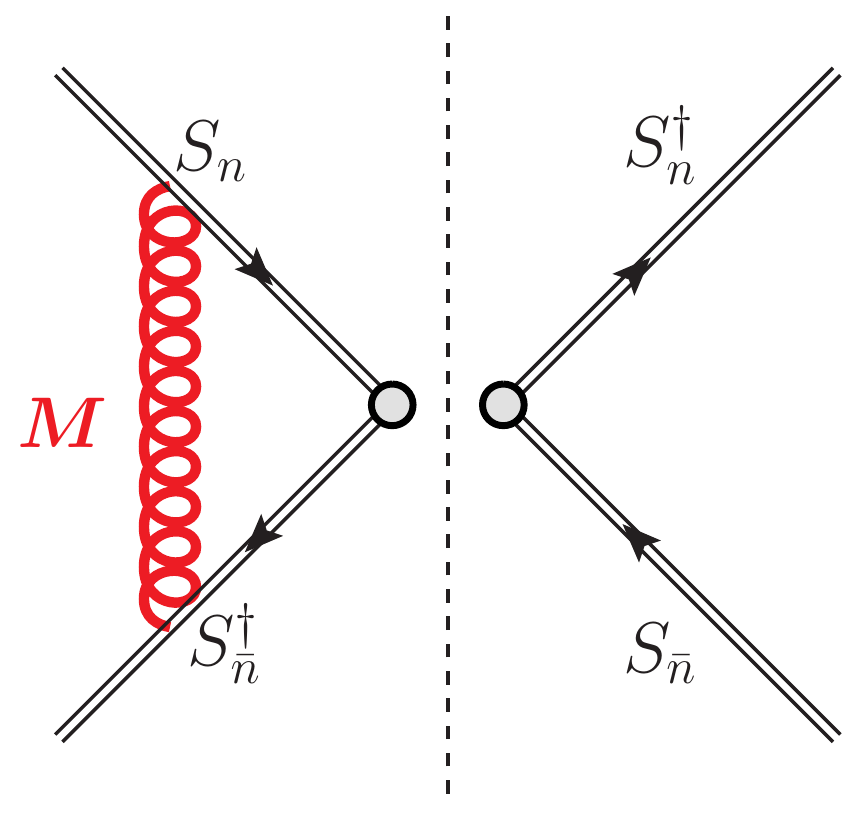}\label{fig:soft_function_virt}}%
\hfill%
\subfigure[]{\includegraphics[scale=0.5]{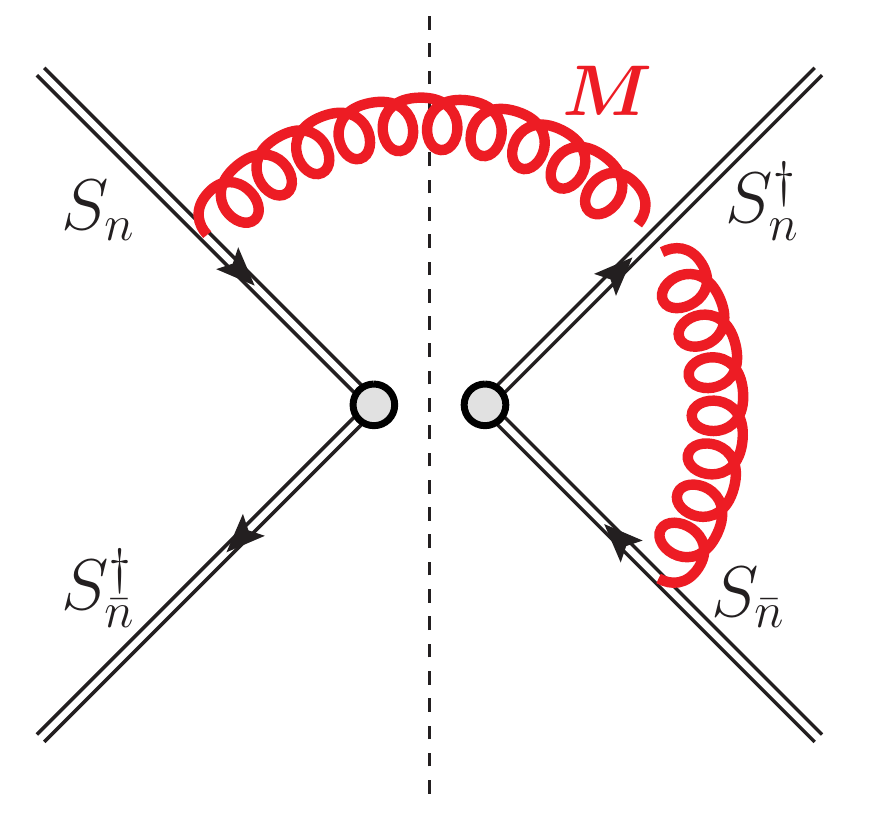}\label{fig:soft_function_real}}%
\hspace*{\fill}
\caption{Soft function corrections for a massive gluon at one-loop. The associated mirror diagrams need to be included in addition.}\label{fig:soft_function}
\end{figure}

The virtual contributions from the diagram in fig.~\ref{fig:soft_function_virt} (and its mirror diagram) are the same as for the Sudakov form factor computed in \refcite{Chiu:2012ir} and yield
\begin{align}
 S^{(1,\rm{bare})}_{\rm{virt}}(M)=C_F\,\Bigl(\frac{\mu^2\mathrm{e}^{\gamma_E}}{M^2}\Bigr)^{\e}\Gamma(\e)\,\Bigl[-\frac{8}{\eta}-8\ln\frac{\nu}{M}-4H_{\e-1}\Bigr]+\mathcal{O}(\eta)
\,.\end{align}
The UV-finite and IR-finite real radiation diagram in Fig.~\ref{fig:soft_function_real} gives
\begin{align}
 S^{(b)}&=8\pi\alpha_sC_F\int\frac{\mathrm{d}^4k}{(2\pi)^4}\,\frac{1}{k^+k^-} \, \frac{\nu^{\eta}}{|k^+-k^-|^\eta}\,\delta^{(2)}(\vec{p}_T-\vec{k}_{\perp})\,2\pi\delta(k^2-M^2)\nn \\
 &=\frac{\alpha_sC_F}{4\pi}\frac{2\,\Gamma(\frac{\eta}{2})\Gamma(\frac{1-\eta}{2})}{\pi^{\frac{3}{2}}(p_T^2+M^2)}\biggl(\frac{\nu}{2\sqrt{\smash[b]{p_T^2+M^2}}}\biggr)^{\eta}
\,.\end{align}
After expanding in $\eta$ and adding the mirror diagram, the real radiation contribution to the TMD soft function at one loop then reads
\begin{align}
 \frac{\alpha_s}{4\pi}S_{\rm{real}}^{(1,\rm{bare})}(\vec{p}_T,M)&=2S^{(b)} =\frac{\alpha_sC_F}{4\pi}\,\frac{4}{\pi(p_T^2+M^2)}\biggl[\frac{2}{\eta}+\ln\Bigl(\frac{\nu^2}{p_T^2+M^2}\Bigr)\biggr]+\mathcal{O}(\eta)
\,.\end{align}

\subsubsection{Secondary corrections at $\mathcal{O}(\alpha_s^2 C_F T_F)$}

To obtain the secondary massive quark corrections from the one-loop results with a massive gluon, we first convolve the one-loop results with the imaginary part of the vacuum polarization function,
\begin{align}\label{eq:dispersion_soft}
\frac{\alpha_sT_F}{4\pi}S^{(2,h,\rm{bare})}_{\rm{virt}}(m) &=\frac{1}{\pi}\int\frac{\mathrm{d}M^2}{M^2}\,{\rm{Im}}\bigl[\Pi(m^2,M^2)\bigr]\times S^{(1,\rm{bare})}_{\rm{virt}}(M)\;, \nn \\
\frac{\alpha_sT_F}{4\pi}S^{(2,h,\rm{bare})}_{\rm{real}}(\vec{p}_T,m) &=\frac{1}{\pi}\int\frac{\mathrm{d}M^2}{M^2}\,{\rm{Im}}\bigl[\Pi(m^2,M^2)\bigr]\times S^{(1,\rm{bare})}_{\rm{real}}(\vec{p}_T,M)\;.
\end{align}
The results from these dispersion integrations are
\begin{align}\label{eq:bare_soft_virt}
S^{(2,h,\rm{bare})}_{\rm{virt}}(m) & =C_F\,\biggl\{\biggl[-\frac{16}{3\e^2}+\frac{1}{\e}\Bigl(\frac{32}{3}L_m+\frac{80}{9}\Bigr)-\frac{32}{3}L_m^2-\frac{160}{9}L_m-\frac{448}{27}-\frac{8\pi^2}{9}+\mathcal{O}(\e)\biggr]
\nn \\ &\quad \times
\Bigl(\frac{1}{\eta}+\ln\frac{\nu}{\mu}\Bigr)+\frac{4}{\e^3}-\frac{1}{\e^2}\Bigl(\frac{16}{3}L_m+\frac{20}{9}\Bigr)+\frac{1}{\e}\Bigl(\frac{8}{3}L_m^2-\frac{112}{27}+\frac{2\pi^2}{3}\Bigr)+\frac{40}{9}L_m^2
\nn \\ &\quad
+\Bigl(\frac{448}{27}-\frac{8\pi^2}{9}\Bigr)L_m+\frac{656}{27}-\frac{10\pi^2}{27}-8\zeta_3\biggr\}
\,, \\
S^{(2,h,\rm{bare})}_{\rm{real}}(\vec{p}_T,m) & =\frac{C_F}{\pi p_T^2}\biggl\{\frac{32}{9\eta}\biggl[-5+12\hat{m}^2+3c(1-2\hat{m}^2)\ln\frac{c+1}{c-1}\biggr]+s_{\rm{real}}^{q_T}\Bigl(\frac{m^2}{p_T^2},\frac{\nu}{m}\Bigr)\biggr\}
\,, \label{eq:bare_soft_real}
\end{align}
with
\begin{align}
s_{\rm{real}}^{q_T}\Bigl(\hat{m}^2,\frac{\nu}{m}\Bigr) & =\frac{16}{9}\biggl\{2\biggl[-5+12\hat{m}^2+3c(1-2\hat{m}^2)\ln\frac{c+1}{c-1}\biggr]\ln\frac{\nu}{m}
\nn \\ & \quad
+3c(1-2\hat{m}^2)\biggl[\Li_2\left(\frac{(c-1)^2}{(c+1)^2}\right)+2\ln\frac{c+1}{c-1} \biggl(\ln\frac{c+1}{2c} + \ln\hat{m}\biggr)-\frac{\pi^2}{6}\biggr]
\nn \\ &\quad
+8\hat{m}^2+c(5-16\hat{m}^2)\ln\frac{c+1}{c-1}\biggr\}
\,, \label{eq:sm_real}
\end{align}
and $\hat{m}$ and $c$ as in eq.~\eqref{eq:mhat}. Using \eq{dispersion_soft} entails that the massive quark corrections to the strong coupling are renormalized in the on-shell scheme, i.e., the expansion is in terms of $\alpha_s=\alpha_s^{(n_l)}$. Since the soft function matrix element has to be renormalized entirely in the $n_l+1$ flavor theory, we need to account for the second term in \eq{propagatorunsubtracted} (which switches back to an unrenormalized $\alpha_s$) and renormalize the massive quark corrections to the strong coupling in the $\MS$ scheme, such that the expansion is in terms of $\alpha_s=\alpha_s^{(n_l+1)}$. The soft function is renormalized according to
\begin{align}
S^{(\rm bare)}(\vec{p}_T,m) = \int \df ^2 p_T' \, Z_S\Bigl(\vec{p}_T-\vec{p}^{\,\prime}_T ,m,\mu,\nu\Bigr) \,S(\vec{p}^{\,\prime}_T ,m,\mu,\nu) \, .
\end{align}
This yields for the renormalized matrix element with initial state quarks at $\mathcal{O}(\alpha_s^2 C_F T_F)$ in terms of $\alpha_s = \alpha_s^{(n_l+1)}$
\begin{align}\label{eq:S_renormalized}
S^{(2,h)}(\vec{p}_T,m,\mu,\nu) & =\delta^{(2)}(\vec{p}_T)\,S^{(2,h,\rm{bare})}_{\rm{virt}}(m)+S^{(2,h,\rm{bare})}_{\rm{real}}(\vec{p}_T,m)  \\
& \quad -\Bigl(\Pi^{(1)}(m^2,0) - \frac{4}{3\eps}\Bigr) \underbrace{S^{(1,\rm bare)}(\vec{p}_T,\mu,\nu)}_{ \makebox[0pt]{\scriptsize $=S^{(1)}(\vec{p}_T,\mu,\nu) + Z_S^{(1)}(\vec{p}_T,\mu,\nu)$}} -\delta(1-z)\,Z_{S}^{(2,h)}(\vec{p}_T,m,\mu,\nu)
\,, \nn
\end{align}
where the (bare) vacuum polarization function $\Pi^{(1)}(m^2,0) $ is given in \eq{vacpolzero} and the renormalized one-loop soft function $S^{(1)}$ is given in eq.~\eqref{eq:TMD_soft_massless_1}. The one-loop counterterm reads
\begin{align}\label{eq:ZS1}
  &Z_{S}^{(1)}(\vec{p}_T,\mu,\nu)=C_F\,\biggl\{\delta^{(2)}(\vec{p}_T)\biggl[\frac{1}{\eta}\Bigl(-\frac{8}{\e}+\mathcal{O}(\e)\Bigr)+\frac{4}{\e^2}-\frac{8}{\e}\ln\frac{\nu}{\mu}\biggr]+\frac{1}{\eta}\Bigl(8+\mathcal{O}(\e)\Bigr)\mathcal{L}_0(\vec{p}_T,\mu)\biggr\}\;,
\end{align}
The two-loop counterterm $Z_S^{(2)}$ absorbs all remaining UV and IR divergences in \eq{S_renormalized} and is given by
\begin{align}\label{eq:ZS2}
&Z_{S}^{(2,h)}(\vec{p}_T,m,\mu,\nu)=C_F\,\biggl\{\delta^{(2)}(\vec{p}_T)\biggl[\frac{1}{\eta}\Bigl(-\frac{16}{3\e^2}+\frac{80}{9\e}-\frac{16}{3}L_m^2-\frac{160}{9}L_m-\frac{448}{27}+\mathcal{O}(\e)\Bigr)\nn \\
&\qquad+\frac{4}{\e^3}-\frac{1}{\e^2}\Bigl(\frac{20}{9}+\frac{16}{3}\ln\frac{\nu}{\mu}\Bigr)+\frac{1}{\e}\Bigl(-\frac{112}{27}+\frac{2\pi^2}{9}+\frac{80}{9}\ln\frac{\nu}{\mu}\Bigr)\biggr] +\frac{1}{\eta}\Bigl(\frac{32}{3}L_m+\mathcal{O}(\e)\Bigr)\mathcal{L}_0(\vec{p}_T,\mu)\nn \\
&\qquad+\frac{1}{\eta} \,\frac{32}{9\pi p_T^2}\biggl(-5+12\hat{m}^2+3c(1-2\hat{m}^2)\ln\frac{c+1}{c-1} \biggr)\biggr\}
\,.\end{align}
This yields the anomalous dimensions in \eq{gammaS_m}. Using eqs.~(\ref{eq:bare_soft_virt}),~(\ref{eq:bare_soft_real}),~(\ref{eq:ZS1}), and~(\ref{eq:ZS2}) we obtain the full result for the secondary massive quark corrections to the TMD soft function in eq.~\eqref{eq:pT_soft_massive_ren}.

\subsection{Csoft function at two loops}
\label{sec:csoft}

We compute the csoft function $\mathcal{S}_c$ for beam thrust appearing in the hierarchy $\Tau \ll m \ll \sqrt{Q\Tau}$. As in the computation for the beam function matching coefficient in \app{beamfunc_thrust_secondary} we carry out the calculation using a \SCETb  IR regulator (a gluon mass $\Lambda \ll m$). In this context the csoft function is the matching coefficient between the csoft matrix elements in the $n_l+1$ and $n_l$ flavor \SCETb theories, 
\begin{align}\label{eq:csoft_matching2}
\mathscr{S}^{(n_l+1)}(\ell,m,\mu,\nu)=\int\mathrm{d}\ell^\prime\,\mathcal{S}_{c}(\ell-\ell^\prime,m,\mu,\nu) \,\mathscr{S}^{(n_l)}(\ell^\prime,\mu,\nu)\;.
\end{align}
The latter are defined for any direction $n$ as
\begin{align}\label{eq:csoft_def}
\mathscr{S}(\ell,m)=\frac{1}{N_c}\,\mathrm{tr}\,\bra{0}\overline{\mathrm{T}}\big[X_{n}^{(0)\dagger}(0)V^{(0)}_{n}(0)\big]\delta(\ell-n\cdot \hat{p})\mathrm{T}\big[V_{n}^{(0)\dagger}(0)X^{(0)}_{n}(0)\big]\ket{0}\;,
\end{align}
with the csoft Wilson lines given by (see e.g.~\refcites{Bauer:2011uc,Procura:2014cba}) 
\begin{align}  \label{eq:regulator_XV} 
X_{n}=&\sum_{\text{perms}}\mathrm{exp}\biggl[-\frac{g}{n\cdot \mathcal{P}}\frac{\nu^{\eta/2}}{(\bar{n} \cdot \mathcal{P})^{\eta/2}} \,n\cdot A_{cs}\biggr] \,, \quad V_{n}=\sum_{\text{perms}}\mathrm{exp}\biggl[-\frac{g}{\bar{n}\cdot \mathcal{P}}\frac{\nu^{\eta/2}}{(\bar{n} \cdot \mathcal{P})^{\eta/2}} \,\bar{n}\cdot A_{cs}\biggr] \,,
\end{align}
Besides replacing the soft fields by csoft fields we have also expanded the $\eta$ regulator according to the soft scaling as in ref.~\cite{Hoang:2015iva}.\footnote{If the regulator is not expanded, nonvanishing soft-bin subtraction appear which eliminate the overlap with soft mass mode momentum regions, see ref.~\cite{Lustermans:2016nvk}.}

\subsubsection{Csoft function with a massive gluon at $\mathcal{O}(\alpha_s)$}

We will first calculate the one-loop corrections to the csoft matrix elements $\mathscr{S}$ with a massive gluon, that can then be used to obtain the two-loop corrections with secondary massive quarks using the dispersion technique described in \sec{dispersion}. The one-loop results for the csoft matrix elements can be written as
\begin{align}
\mathscr{S}^{(1,\rm bare)}(\ell,M)=\delta(\ell) \, \mathscr{S}^{(1,\rm bare)}_{\rm{virt}}(M)+ \mathscr{S}_{\rm{real}}^{(1,\rm bare)}(\ell,M)\;.
\end{align}
The relevant contributions at one loop are displayed in the diagrams in fig.~\ref{fig:soft_function}, with the soft Wilson lines $S_n$ and $S_{\bar{n}}$ replaced by the csoft Wilson lines $X_n$ and $V_n$. With the choice of regularization in \eq{regulator_XV} the virtual diagram leads to a scaleless integral, such that $ \mathscr{S}^{(1,\rm{bare})}_{\rm{virt}}=0$. The real radiation diagram corresponding to fig.~\ref{fig:soft_function_real} yields
\begin{align}
\mathscr{S}^{(b)}&=8\pi\alpha_sC_F\tilde{\mu}^{2\e}\int\frac{\mathrm{d}^dk}{(2\pi)^d}\,\frac{1}{k^-k^+} \, \Bigl(\frac{\nu}{k^-}\Bigr)^{\eta}\,\delta(\ell-k^+)\,2\pi\delta(k^2-M^2) \nn\\
&=\frac{\alpha_sC_F}{4\pi}\frac{2\,\Gamma(\e+\eta)}{\Gamma(1+\eta)}\Bigl(\frac{\mu^2\mathrm{e}^{\gamma_E}}{M^2}\Bigr)^\e\Bigl(\frac{\nu}{M^2}\Bigr)^\eta\,\frac{\theta(\ell)}{\ell^{1-\eta}}\;.
\end{align}
Including also the mirror diagram and expanding in $\eta$ the total real radiation contribution to the csoft matrix element with a massive gluon is
\begin{align}\label{eq:Sc_real}
 \frac{\alpha_s}{4\pi} \mathscr{S}_{\rm real}^{(1,\rm bare)}(\ell,M) &=2\mathscr{S}^{(b)}  \\
 &=\frac{\alpha_sC_F}{4\pi}\Bigl(\frac{\mu^2\mathrm{e}^{\gamma_E}}{M^2}\Bigr)^\e\Gamma(\e)\,\biggl[\delta(\ell)\Bigl(\frac{1}{\eta}-\ln\frac{M^2}{\mu^2}+H_{\e-1}\Bigr)+\frac{\nu}{\mu^2}\mathcal{L}_0\Bigl(\frac{\ell\,\nu}{\mu^2}\Bigr) +\mathcal{O}(\eta)\biggr]\,. \nn
\end{align}

\subsubsection{Csoft function at $\mathcal{O}(\alpha_s^2)$}

We convolve the one-loop results with the imaginary part of the vacuum polarization function, which yields for the nonvanishing contributions
\begin{align}\label{eq:dispersion_csoft}
&\frac{\alpha_sT_F}{4\pi} \mathscr{S}^{(2,h,\rm{bare})}_{\rm{real}}(\ell,m)=\frac{1}{\pi}\int\frac{\mathrm{d}M^2}{M^2}\,{\rm{Im}}\bigl[\Pi(m^2,M^2)\bigr]\times \mathscr{S}^{(1,\rm{bare})}_{\rm{real}}(\ell,M)\;.
\end{align}
The result of this dispersion integral is
\begin{align}\label{eq:bare_csoft_real}
\mathscr{S}^{(2,h,\rm{bare})}_{\rm{real}}(\ell,m) &=C_F\,\biggl\{\biggl[\frac{8}{3\e^2}-\frac{1}{\e}\Bigl(\frac{16}{3}L_m+\frac{40}{9}\Bigr)+\frac{16}{3}L_m^2+\frac{80}{9}L_m+\frac{224}{27}+\frac{4\pi^2}{9}+\mathcal{O}(\e)\biggr]
\nn \\ &\qquad \times
\biggl[\frac{1}{\eta}\delta(\ell)+\frac{\nu}{\mu^2}\mathcal{L}_0\Bigl(\frac{\ell\,\nu}{\mu^2}\Bigr)\biggr] +\delta(\ell)\biggl[-\frac{4}{\e^3}+\frac{1}{\e^2}\Bigl(\frac{16}{3}L_m+\frac{20}{9}\Bigr) \nn \\
&\quad  +\frac{1}{\e}\Bigl(-\frac{8}{3}L_m^2+\frac{112}{27}-\frac{2\pi^2}{3}\Bigr)-\frac{40}{9}L_m^2+\Bigl(-\frac{448}{27}+\frac{8\pi^2}{9}\Bigr)L_m \nn \\
& \quad-\frac{656}{27}+\frac{10\pi^2}{27}+8\zeta_3\biggr]\biggr\}
\,.\end{align}
Using \eq{dispersion_csoft} entails that the massive quark corrections to the strong coupling are renormalized in the on-shell scheme, i.e., the expansion is in terms of $\alpha_s=\alpha_s^{(n_l)}$. To obtain the csoft function $\mathcal{S}_c$ we need to switch to $\alpha_s^{(n_l+1)}$ and furthermore subtract the correction $\mathscr{S}^{(2,n_l)}$ (with a strong coupling in the $n_l$ flavor scheme) according to \eq{csoft_matching2}. All purely massless contributions cancel each other and we obtain for the $\mathcal{O}(\alpha_s^2)$ corrections in an expansion in terms of $\alpha_s^{(n_l+1)}$
\begin{align}\label{eq:Sc_renormalized}
\mathcal{S}_c^{(2)}(\ell,m,\mu,\nu) &= \mathscr{S}^{(2,h)}(\ell,m,\mu,\nu)-\frac{4}{3}L_m \mathscr{S}^{(1)}(\ell,\mu,\nu)  \\
&=  \mathscr{S}^{(2,h,\rm{bare})}_{\rm{real}}(\ell,m)-\Bigl(\Pi^{(1)}(m^2,0) - \frac{4}{3\eps}\Bigr)\mathscr{S}^{(1,\rm bare)}(\ell) \nn \\
& \qquad - Z_{\mathscr{S}}^{(2,h)}(\ell,m,\mu,\nu)-\frac{4}{3}L_m \mathscr{S}^{(1)}(\ell,\mu,\nu)  \nn \\
& = \mathscr{S}^{(2,h,\rm{bare})}_{\rm{real}}(\ell,m)-\Bigl(\Pi^{(1)}(m^2,0) - \frac{4}{3\eps}\Bigr)Z_{\mathscr{S}}^{(1)}(\ell,\mu,\nu)- Z_{\mathscr{S}}^{(2,h)}(\ell,m,\mu,\nu) \;. \nn
\end{align}
Here the \SCETb counterterm is defined via
\begin{align}
\mathscr{S}^{(\rm bare)}(\ell,m) = \int \df \ell'\, Z^{(n_l+1)}_{\mathscr{S}}(\ell-\ell',m,\mu,\nu) \, \mathscr{S}^{(n_l+1)} (\ell',m,\mu,\nu)
\,.\end{align}
Employing a gluon mass the associated expression for $Z_{\mathscr{S}}^{(1)}$ can be read off from \eq{Sc_real} and is given by (expanded in $\eta$ and $\eps$)
\begin{align}\label{eq:ZSc1}
Z_{\mathscr{S}}^{(1)}(\ell,\mu,\nu)= 4 C_F \biggl\{\delta(\ell)\biggl[\frac{1}{\eta}\Bigl(\frac{1}{\eps}-\ln \frac{\Lambda^2}{\mu^2} +\mathcal{O}(\eps)\Bigr)-\frac{1}{\eps^2} \biggr]+ \frac{\nu}{\mu^2} \mathcal{L}_0\Bigl(\frac{\ell \nu}{\mu^2}\Bigr)\frac{1}{\eps}\biggr\} \, .
\end{align}
The counterterm $Z^{(2,h)}_{\mathscr{S}}$ absorbs all divergences and is given by\footnote{The anomalous dimension for the csoft function $\mathcal{S}_c $ can be obtained from the ratio of $Z^{(n_l+1)}_{\mathscr{S}}$ and $Z^{(n_l)}_{\mathscr{S}}$, upon which the IR sensitivity cancels.}
\begin{align}\label{eq:ZSc2}
Z^{(2,h)}_{\mathscr{S}}(\ell,m,\mu,\nu) &=C_F\,\biggl\{\delta(\ell)\biggl[\frac{1}{\eta} \Bigl(\frac{8}{3\e^2}-\frac{40}{9\e}-\frac{16}{3}L_m \ln\frac{\Lambda^2}{\mu^2}+\frac{8}{3}L_m^2+\frac{80}{9}L_m+\frac{224}{27}+\mathcal{O}(\e)\Bigr) \notag \\
& \quad  -\frac{4}{\e^3}+\frac{20}{9\e^2}+\frac{1}{\e}\Bigl(\frac{112}{27}-\frac{2\pi^2}{9}\Bigr)\biggr]+ \frac{\nu}{\mu^2}\mathcal{L}_0\Bigl(\frac{\ell\,\nu}{\mu^2}\Bigr)\biggl[\frac{8}{3\e^2}-\frac{40}{9\e}\biggr]\biggr\}\;.
\end{align}
Using eqs.~(\ref{eq:bare_csoft_real}),~(\ref{eq:ZSc1}) and~(\ref{eq:ZSc2}) in \eq{Sc_renormalized} we obtain the full result for the renormalized csoft function at two loops in eq.~\eqref{eq:csoft}.

\section{Massive quark effects at fixed order}

The factorization formulae in the \secs{qT}{Tau} contain together all information about the singular massive quark corrections to the differential cross sections in  QCD at fixed order (for any given hierarchy between the mass and $q_T$/$\Tau$). Here we provide the results at $\mathcal{O}(\alpha_s^2)$ for Drell-Yan for both primary and secondary corrections. We write for each of these contributions ($e=q_T^2, \Tau$)
\begin{align}
\frac{\df\sigma}{\df e \,\df Q^2 \, \df Y} (e,Q,m,x_a,x_b) = \sum_{i,j =q,\bar{q},g} \int \frac{\df z_a}{z_a}\,\frac{\df z_b}{z_b}\, \frac{\df\hat{\sigma}_{ij}}{\df e \,\df Q^2 \, \df Y} (e,Q,m,z_a,z_b,\mu)  \, f_i \Bigl(\frac{x_a}{z_a},\mu\Bigr) \,  f_j\Bigl(\frac{x_b}{z_b},\mu\Bigr) \, ,
\end{align}
and expand the partonic result in the $n_l$-flavor scheme for $\alpha_s$ as 
\begin{align}
\frac{\df\hat{\sigma}_{ij}}{\df e \,\df Q^2 \, \df Y} &= \frac{\df \hat{\sigma}^{(0)}_{ij}}{\df Q^2 \,\df Y}\, \delta(q_T^2)+\frac{\alpha^{(n_l)}_s(\mu)}{4\pi}\, \frac{\df\hat{\sigma}^{(1)}_{ij}}{\df e \,\df Q^2 \, \df Y}  \nn \\
& \quad+\biggl(\frac{\alpha^{(n_l)}_s(\mu)}{4\pi}\biggr)^2 \biggl[ T_F n_l\, \frac{\df\hat{\sigma}^{(2,l)}_{ij}}{\df e \,\df Q^2 \, \df Y}  + T_F \,\frac{\df\hat{\sigma}^{(2,h)}_{ij}}{\df e \,\df Q^2 \, \df Y} + \dots \biggr] +\mathcal{O}(\alpha_s^3) \, ,
\end{align}
where $\df \sigma^{(0)}_{q\bar{q}}/(\df Q^2 \df Y)$ denotes the Born cross section for the corresponding Drell-Yan process $q \bar{q} \to Z/\gamma^* \to \ell \bar{\ell}$. In this context $\df \sigma^{(0)}_{Q\bar{Q}}/(\df Q^2 \df Y)$ indicates the Born cross section for a massless quark $q$ with the same charge and isospin as the heavy quark $Q$.

\subsection{Fixed-order result for the $q_T$ spectrum}

The singular fixed-order corrections for the $q_T$-spectrum (i.e.~for $q_T \ll Q$) at $\mathcal{O}(\alpha_s^2 C_F T_F)$ consist of the virtual (full QCD) contributions encoded in \eqs{f_QCD}{h_axial} and the secondary collinear and soft real radiation corrections contained in \eqs{TMD_Iqq2_massive}{pT_soft_massive_ren}. Setting common scales $\mu=\mu_H=\mu_B=\mu_S$ and $\nu=\nu_B=\nu_S$ yields for the corrections to virtual photon production 
\begin{align}\label{eq:qTfo_secondary}
 \frac{\df \hat{\sigma}^{(2,h)}_{q\bar{q}}}{\df q_T^2 \,\df Q^2 \, \df Y}  & =  \frac{\df \hat{\sigma}^{(0)}_{q\bar{q}}}{\df Q^2 \,\df Y} \, C_F
\biggl\{ h_{\rm virt}\Bigl(\frac{m^2}{Q^2}\Bigr)  \,\delta(q_T^2) \,\delta(1-z_a) \, \delta(1-z_b) \nn \\
& \quad + \frac{1}{q_T^2}\, b^{q_T}_{\rm{real}}\Bigl(\frac{m^2}{q_T^2},z_a,\frac{\nu}{\omega_a}\Bigr)  \, \delta(1-z_b)  +\frac{1}{q_T^2} \, b^{q_T}_{\rm{real}}\Bigl(\frac{m^2}{q_T^2},z_b,\frac{\nu}{\omega_b}\Bigr) \,  \delta(1-z_a)  \nn \\
& \quad + \frac{1}{q_T^2}\,s^{q_T}_{{\rm real}}\Bigl(\frac{m^2}{q_T^2},\frac{\nu}{m}\Bigr) \, \delta(1-z_a) \, \delta(1-z_b)  +\mathcal{O}\Bigl(\frac{q_T}{Q}\Bigr)\biggr\}\, ,
\end{align}
where $h_{\rm virt}$, $b^{q_T}_{\rm{real}}$ and $s^{q_T}_{{\rm real}}$ are given in eqs.~(\ref{eq:f_QCD}),~(\ref{eq:bm_real}) and~(\ref{eq:sm_real}). For $Z$-boson production one has to include in addition the anomalous axial current correction in \eq{h_axial} as contribution to the $\delta(q_T^2)$-term (which gives in conjunction with the isospin partner a $\mu$-independent result).
 Writing out the nontrivial terms in the spectrum explicitly we get
\begin{align}\label{eq:qTfo_secondary2}
&\frac{\df\hat{\sigma}^{(2,h)}_{q\bar{q}}}{\df q_T^2 \,\df Q^2 \, \df Y}  (q_T^2,Q,m,z_a,z_b)  = \frac{\df \hat{\sigma}^{(0)}_{q\bar{q}}}{\df Q^2 \,\df Y}\,\theta(z_a) \,\theta(z_b) \, C_F
\biggl\{ h_{\rm virt}\Bigl(\frac{m^2}{Q^2}\Bigr) \,\delta(q_T^2) \,\delta(1-z_a) \,\delta(1-z_b)  \nn \\
& \quad  + \frac{\delta(1-z_b)}{q_T^2}\biggl[\biggl(-\frac{80}{9} + \frac{64}{3}\hat{m}^2 +\frac{16}{3}(1-2\hat{m}^2)\ln\frac{c+1}{c-1} \biggr)\Bigl(\mathcal{L}_0(1-z_a) + \delta(1-z_a)\ln\frac{Q}{m}\Bigr) \nn \\
& \qquad \quad+ \theta(1-z_a) \biggl(\frac{8}{3d_a(1-z_a)}  \bigl[1+z_a^2+2\hat{m}^2z_a(1+z_a^2)+4\hat{m}^4z_a^2(-5+6z_a-5z_a^2)\bigr]\,\ln\frac{d_a+1}{d_a-1}
\nn \\  & \qquad \quad
-\frac{16c(1-2\hat{m}^2)}{3(1-z_a)}\ln\frac{c+1}{c-1} + \frac{16}{9}+\frac{64}{9}z_a+ \frac{16}{3}\hat{m}^2(-4+z_a-5z_a^2) \biggr)
\nn \\ & \qquad \quad
+ \delta(1-z_a)  \biggl(\frac{8}{3} c(1-2\hat{m}^2)\biggl[\Li_2\Bigl(\frac{(c-1)^2}{(c+1)^2}\Bigr)+2\ln\frac{c+1}{c-1} \ln\frac{\hat{m}(c+1)}{2c} -\frac{\pi^2}{6}\biggr]
\nn \\ & \qquad \quad
+\frac{8}{9}c(5-16\hat{m}^2)\ln\frac{c+1}{c-1} + \frac{64}{9}\hat{m}^2\biggr)
 \biggr] \nn \\ & \quad
 + \frac{\delta(1-z_a)}{q_T^2} \biggl[(z_a \leftrightarrow z_b)\biggr] +\mathcal{O}\Bigl(\frac{q_T}{Q}\Bigr)\biggr\}
 \,,\end{align}
where 
\begin{align}
\hat{m} = \frac{m}{q_T} \, ,  \quad c = \sqrt{1+4\hat{m}^2} \, , \quad d_a = \sqrt{1+4\hat{m}^2 z_a} \, .
\end{align}

The singular fixed-order corrections for the $q_T$ spectrum at $\mathcal{O}(\alpha_s^2 T_F^2)$ consist of the primary collinear real radiation corrections in \eq{TMD_beam_coefficient} for both beam directions,
\begin{align}\label{eq:qTfo_primary}
& \frac{\df\hat{\sigma}^{(2,h)}_{gg}}{\df q_T^2 \,\df Q^2 \, \df Y}  (q_T^2,Q,m,z_a,z_b)  = 2 \,\frac{\df \hat{\sigma}^{(0)}_{Q\bar{Q}}}{\df Q^2 \,\df Y} \times \frac{\pi}{T_F} \int \df^2 p_T \, \mathcal{I}^{(1)}_{Qg}(\vec{q}_T-\vec{p}_T,m,z_a) \,  \mathcal{I}^{(1)}_{Qg}(\vec{p}_T,m,z_b) \nn \\
& =\frac{\df\hat{\sigma}^{(0)}_{Q\bar{Q}}}{\df Q^2 \,\df Y}  \, \theta(z_a) \, \theta(z_b) \, \theta(1-z_a) \, \theta(1-z_b) \, \frac{8 T_F}{q_T^2 c^4}
\biggl\{2(1 - z_a -z_b +2 z_a z_b) (z_a+z_b-2 z_a z_b) \nn \\
& \quad+ 8\hat{m}^2
[z_a(1-z_a) + z_b(1-z_b) - 3 z_a z_b (1-z_a-z_b+z_a z_b)]- 16 \hat{m}^4 z_a z_b (1 - z_a - z_b + z_a z_b) \nn \\
& \quad + \frac{1}{c} \,\ln \biggl(\frac{1+ c + 2\hat{m}^2(2+c)+2\hat{m}^4}{2\hat{m}^4}\biggr)\biggl[(1 -2 z_a +2 z_a^2) (1 -2 z_b +2 z_b^2) \nn \\
& \qquad \quad +2 \hat{m}^2 \Bigl(4 - 7 z_a(1-z_a) - 7 z_b(1-z_b) + 12 z_a z_b(1-z_a-z_b+z_a z_b)\Bigr) \nn \\
& \qquad \quad+8 \hat{m}^4 \Bigl(2 - 3 z_a(1-z_a) -3 z_b(1-z_b)  +6 z_a z_b(1-z_a-z_b+z_a z_b)\Bigr) \nn \\
& \qquad \quad+16 \hat{m}^6 z_a z_b (1 - z_a) (1 - z_b) \biggr] +\mathcal{O}\Bigl(\frac{q_T}{Q}\Bigr)
  \biggr\}\, .
\end{align}

Depending on the hierarchy between $m$ and $q_T$ and $Q$ some of the contributions in \eqs{qTfo_secondary2}{qTfo_primary} are power-suppressed and therefore only appear via nonsingular corrections in the factorization formula for the associated parametric regime in \sec{qT}. Note also that virtual corrections are reshuffled among the components of the factorization theorem, which are in addition evaluated with $\alpha_s$ in different flavor number schemes. This essentially allows for a consistent factorization and the resummation of logarithms at higher orders.

\subsection{Fixed-order result for the beam thrust spectrum}

The singular fixed-order corrections for the $\Tau$ spectrum (i.e. for $\Tau \ll Q$) at $\mathcal{O}(\alpha_s^2 C_F T_F)$ consist of the virtual (full QCD) contributions encoded in \eqs{f_QCD}{h_axial} and the secondary collinear and soft real radiation corrections contained in \eqs{Virt_Iqq2_massive}{soft_tau}. Setting common scales $\mu=\mu_H=\mu_B=\mu_S$ yields for the corrections to virtual photon production
\begin{align}\label{eq:Taufo_secondary}
 \frac{\df \hat{\sigma}^{(2,h)}_{q\bar{q}}}{\df \Tau \,\df Q^2 \, \df Y}  &=  \frac{\df \hat{\sigma}^{(0)}_{q\bar{q}}}{\df Q^2 \,\df Y} \, C_F
\biggl\{ h_{\rm virt}\Bigl(\frac{m^2}{Q^2}\Bigr) \,\delta(\Tau) \,\delta(1-z_a) \, \delta(1-z_b) \nn \\
& \quad  + \frac{1}{\Tau}\, b_{\rm{real}}^{\Tau}\Bigl(\frac{m^2}{\omega_a \Tau},z_a\Bigr)  \, \delta(1-z_b)  +\frac{1}{\Tau} \, b_{\rm{real}}^{\Tau}\Bigl(\frac{m^2}{\omega_b \Tau},z_b\Bigr) \,  \delta(1-z_a)  \nn \\
& \quad + \frac{1}{\Tau}\,s^{\Tau}_{\rm real}\Bigl(\frac{m^2}{\Tau^2}\Bigr) \, \delta(1-z_a) \, \delta(1-z_b)  +\mathcal{O}\Bigl(\frac{\Tau}{Q}\Bigr)\biggr\}\, ,
\end{align}
where $h_{\rm virt}$ and $b_{\rm{real}}^{\Tau}$ are given in eqs.~(\ref{eq:f_QCD}) and (\ref{eq:bm_realTau}), respectively, and $s_{{\rm real}}^{\Tau}$ is given implicitly by the nondistributive terms in \eq{soft_tau}. Again, for $Z$-boson production the anomalous axial current correction in \eq{h_axial} has to be included in the $\delta(\Tau)$ term.
Writing out the nontrivial terms in the spectrum we get
\begin{align}\label{eq:Taufo_secondary2}
& \frac{\df\hat{\sigma}^{(2,h)}_{q\bar{q}}}{\df \Tau \,\df Q^2 \, \df Y} (\Tau,Q,m,z_a,z_b) = \frac{\df \hat{\sigma}^{(0)}_{q\bar{q}}}{\df Q^2 \,\df Y}\,\theta(z_a) \,\theta(z_b) \,C_F\biggl\{ \delta(\Tau) \,\delta(1-z_a) \,\delta(1-z_b) \,h_{\rm virt}\Bigl(\frac{m^2}{Q^2}\Bigr) \nn \\
& \quad  + \frac{\delta(1-z_b)}{\Tau}\biggl[\frac{\theta(v_a)}{1-z_a}\biggl(-\frac{16}{9}v_a\Bigl[4-3z_a+4z_a^2 +\frac{z_a(11-21z_a+29z_a^2-15z_a^3)}{1-z_a}\,\hat{m}_a^2\Bigr] \nn \\
& \qquad  \quad-\frac{8}{3u_a} \Bigl[1+z_a^2-2\hat{m}_a^2 z_a(1+z_a^2)-4 \hat{m}_a^4 z_a^2(2-3z_a+5z_a^2)\Bigr]\ln\frac{u_a-v_a}{u_a+v_a} \biggr)\nn \\
 & \qquad  \quad+ \delta(1-z_a)  \biggl( \theta(\Tau-2m)\biggl[\frac{32}{3} \,\Li_2\Bigl(\frac{w-1}{w+1}\Bigr)
  +\frac{8}{3}\ln^2\frac{1-w}{1+w} - \frac{32}{3}\ln\frac{1-w}{1+w} \ln \hat{m}_{\Tau}
\nn \\
  & \qquad \quad -\frac{80}{9} \ln\frac{1-w}{1+w} - w\Bigl(\frac{448}{27}+\frac{128}{27}\hat{m}_{\Tau}^2\Bigr) + \frac{8\pi^2}{9}\biggr]+\frac{\Tau\,\Delta S_{\tau,m}(\Tau,m)}{2} \biggr)
 \biggr] \nn \\
 & \quad+ \frac{\delta(1-z_a)}{\Tau} \biggl[(z_a,\omega_a \leftrightarrow z_b,\omega_b)\biggr] +\mathcal{O}\Bigl(\frac{\Tau}{Q}\Bigr)\biggr\}\, ,
\end{align}
where 
\begin{align}
&\hat{m}_a = \frac{m}{\sqrt{\omega_a \Tau}} \, , \quad  \, \hat{m}_\Tau = \frac{m}{\Tau}\, , \quad u_a = \sqrt{1-4 \hat{m}_a z_a} \, , \quad v_a = \sqrt{1-\frac{4\hat{m}_a^2 z_a}{1-z_a}} \, , \quad w = \sqrt{1-4\hat{m}^2_{\Tau}}\, .
\end{align}

The singular fixed-order corrections for the $\Tau$ spectrum at $\mathcal{O}(\alpha_s^2T_F^2)$ consist of the collinear real radiation corrections in \eq{virtuality_beam_coefficient} for both beam directions,
\begin{align}\label{eq:Taufo_primary}
& \frac{\df\hat{\sigma}^{(2,h)}_{gg}}{\df \Tau \,\df Q^2 \, \df Y}(\Tau,Q,m,z_a,z_b) = 2\,\frac{\df \sigma^{(0)}_{Q\bar{Q}}}{\df Q^2 \,\df Y} \times \frac{Q^2}{T_F}\int \df \Tau' \, \mathcal{I}^{(1)}_{Qg}(\omega_a(\Tau-\Tau'),m,z_a) \,  \mathcal{I}^{(1)}_{Qg}(\omega_b \Tau',m,z_b) \nn \\
& =  \frac{\df \sigma^{(0)}_{Q\bar{Q}}}{\df Q^2 \,\df Y} \,\theta(z_a) \, \theta(z_b) \,\theta\Bigl(\Tau - \frac{m^2 z_a}{\omega_a(1-z_a)} - \frac{m^2 z_b}{\omega_b(1-z_b)} \Bigr) \,\frac{8 T_F}{\Tau} \nn \\
& \quad \times \biggl\{ \frac{2}{(1-z_a -\hat{m}_a^2 z_a)(1-z_b -\hat{m}_b^2 z_b)} \Bigl[(1 - z_a) (1 - z_b) -\hat{m}_a^2 z_a (1-z_b)-\hat{m}_b^2 z_b (1-z_a)\Bigr]\nn \\
& \qquad  \times   \Bigl[(1-z_a -z_b+ 2z_a z_b) (z_a +z_b -2z_a z_b)-\hat{m}_a^2 z_a^2(1-2z_b)^2- \hat{m}_b^2 z_b^2(1-2z_a)^2-4\hat{m}_a^2\hat{m}_b^2 z_a^2 z_b^2\Bigr] \nn \\
& \quad + \biggl(\ln\frac{ 1-z_a-\hat{m}_a^2 z_a}{z_a\hat{m}_a^2} + \ln\frac{ 1-z_b-\hat{m}_b^2 z_b}{z_b\hat{m}_b^2} \biggr) \biggl[(1 - 2 z_a + 2 z_a^2) (1 - 2 z_b + 2 z_b^2) \nn \\
& \qquad+ 2\hat{m}_a^2 z_a^2 (1 - 2 z_b + 2 z_b^2) + 2\hat{m}_b^2 z_b^2 (1 - 2 z_a + 2 z_a^2) + 8 \hat{m}_a^2 \hat{m}_b^2 z_a^2 z_b^2 \biggr]
  \biggr\}\, .
\end{align}

Depending on the hierarchy between $m$ and $\Tau$ and $Q$ some of the contributions in \eqs{Taufo_secondary2}{Taufo_primary} are power-suppressed and therefore only appear via nonsingular corrections in the factorization formula for the associated parametric regime in \sec{qT}.

\section{Plus distributions}\label{sec:distributions}
The standard plus distribution for some dimensionless function $g(x)$ is defined as
\begin{align}
 \left[\theta(x)g(x)\right]_+=\lim\limits_{\beta \rightarrow 0}\frac{\mathrm{d}}{\mathrm{d}x}\left[\theta(x-\beta)G(x)\right] \qquad \text{with} \qquad G(x)=\int_1^x\mathrm{d}x^\prime \,g(x^\prime)\;.
\end{align}
The special case used in this paper is
\begin{align}
 \mathcal{L}_n(x)=\left[\frac{\theta(x)\ln^n x}{x}\right]_+ \;.
\end{align}
The 2-dimensional plus distributions that appear in the TMD beam and soft functions are defined as
\begin{align}
 \mathcal{L}_n(\vec{p}_T,\mu)=\frac{1}{\pi\mu^{2}}\mathcal{L}_n\left(\frac{|\vec{p}_T|^2}{\mu^{2}}\right)\;.
\end{align}
For the Fourier transform we use the convention
\begin{align}
 \tilde{f}(\vec{b})=\int\mathrm{d}^2\vec{p}_T \,\mathrm{e}^{i\,\vec{b}\cdot \vec{p}_T}f(\vec{p}_T)\;.
\end{align}
The Fourier transforms of the 2-dimensional distributions required here are
\begin{align}
 \delta^{(2)}(\vec{p}_T) \;&\longleftrightarrow \; 1
 \,,\nn \\
\mathcal{L}_0(\vec{p}_T,\mu) \; &\longleftrightarrow \; -L_b
\,,\nn \\
\mathcal{L}_1(\vec{p}_T,\mu) \; &\longleftrightarrow \; \;\;\frac{L_b^2}{2}
\,,\nn \\
\mathcal{L}_2(\vec{p}_T,\mu) \; &\longleftrightarrow \; -\frac{1}{4}\Bigl(L_b^3+4\zeta_3\Bigr)\,,
\end{align}
with $L_b$ defined in eq.~\eqref{eq:Lb}.

\phantomsection
\addcontentsline{toc}{section}{References}
\bibliographystyle{../jhep}
\bibliography{../bmassDY}

\end{document}